\documentstyle[aps,prl,multicol,amssymb,eqsecnum,epsf]{revtex}

\def\mathbold{\bf}
\def\be{\begin{equation}}
\def\ee{\end{equation}}
\def\bea{\begin{eqnarray}}
\def\eea{\end{eqnarray}}
\def\br{{\mathbold r}}
\def\bq{{\mathbold q}}
\def\eps{\epsilon}
\newcommand{\corr}[1]{\langle #1\rangle}
\newcommand{\Tr}{\mathop{\rm Tr}}
\newcommand{\tr}{\mathop{\rm tr}}
\newcommand{\Trk}{\mathop{{\rm Tr}_K}}
\newcommand{\trk}{\mathop{{\rm tr}_K}}
\newcommand{\Trg}{\Tr\nolimits_\Gamma}
\newcommand{\sgn}{\mathop{\rm sgn}}
\renewcommand{\Re}{\mathop{\rm Re}}

\newcommand{\pp}{{\textstyle\frac\pi2}}
\newcommand{\ETh}{E_{\rm Th}}
\newcommand{\ac}{{\cal M}}

\def\Top#1{\vskip #1\begin{picture}(290,80)(80,500)\thinlines
\put(65,500){\line(1,0){255}}\put(320,500){\line(0,1){5}}\end{picture}}
\def\Bottom#1{\vskip #1\begin{picture}(290,80)(80,500)\thinlines
\put(330,500){\line(1,0){255}}\put(330,500){\line(0,-1){5}}\end{picture}}
\def\top{\Top{-2.8cm}}
\def\bottom{\Bottom{-2.7cm}}

\begin{document}

\draft

\title{Superconductive proximity effect in interacting disordered conductors}
\author{M. A. Skvortsov$^1$, A. I. Larkin$^{1,2}$ and M. V. Feigel'man$^1$}

\address{
$^1$L. D. Landau Institute for Theoretical Physics, Moscow
117940, Russia\\
$^2$Theoretical Physics Institute, University of Minnesota,
Minneapolis, MN 55455, USA
}

\date{\today}
\maketitle

\begin{abstract}
We present a general theory of the superconductive proximity effect
in disordered normal--superconducting (N-S) structures, based on
the recently developed~\cite{FLS} Keldysh action approach.
In the case of the absence of interaction in the normal conductor we
reproduce known results for the Andreev conductance $G_A$ at arbitrary
relation between the interface resistance $R_T$ and the diffusive
resistance $R_D$. In two-dimensional N-S systems,
electron-electron interaction in the Cooper channel
of normal conductor is shown to strongly affect
the value of $G_A$ as well as its dependence
on temperature, voltage  and magnetic field.
In particular, an unusual maximum of $G_A$ as a function of temperature
and/or magnetic field is predicted for some range of parameters
$R_D$ and $R_T$.
The Keldysh action approach
makes it possible to calculate the full statistics of charge transfer in such
structures. As an application of this method, we calculate the noise
power of an N-S contact as a function of voltage, temperature, magnetic
field and frequency for arbitrary Cooper repulsion in the N metal and
arbitrary values of the ratio $R_D/R_T$.
\end{abstract}


\begin{multicols}{2}

\section{Introduction}

The essence of the superconductive proximity effect is an existence
of Cooper correlations between electrons in the normal metal
in contact with a superconductor. The microscopic
mechanism leading to the proximity effect is Andreev
reflection~\cite{Andreev} of an electron into a hole at a
normal-metal--superconducting (N-S) interface.
The probability of Andreev reflection (as opposed to the normal reflection)
and thus the strength of the proximity
effect is determined by the transparency of the N-S interface, which
may be relatively weak due to the presence of a tunnel barrier or
mismatch between Fermi velocities of two materials.  Disorder in the
normal conductor near an N-S contact is shown
theoretically~\cite{volkov,Nazarov94,NazarovC,Carlo}
to increase considerably the effective 
probability of Andreev reflection
(see Ref.~\cite{Panne} for a recent review from the experimental viewpoint).
However, when the normal conducting region
is made of a dirty metal film, or two-dimensional electron gas with low
density of electrons, Coulomb interaction in the normal region
(neglected in~\cite{volkov,Nazarov94,NazarovC,Carlo})
gets enhanced~\cite{AA} and leads to strong quantum fluctuations
which suppress the Andreev conductance.
Several different kinds of quantum effects
are known to be relevant in low-dimensional conductors at low temperatures.
Quantum corrections to the conductivity of two-dimensional dirty conductors
grow logarithmically as temperature $T$ decreases and
become large at $\ln(1/T\tau) \sim g$
(where $g=(\hbar/e^2)\sigma$ is the dimensionless conductance, $\sigma$
is the conductance per square, and $\tau$ is the elastic scattering time).
There are two main types of these effects:
weak localization corrections~\cite{band4,GLK},
and interaction-induced corrections~\cite{AA}.
Other important quantum effects include interaction-induced
suppression of the tunneling conductance
(Coulomb zero-bias anomaly~\cite{AA,AAL}), and disorder-induced
suppression~\cite{fuku,kuroda,finkel1,finkel2,finkel3}
of the superconductive transition temperature $T_c$.
The corresponding corrections
 are of the relative order of $g^{-1}\ln^2(1/T\tau)$,
i.~e., much stronger than the weak localization and interaction corrections.

In the previous paper~\cite{FLS} we have studied the influence of the last
two effects on the Andreev conductance and the Josephson proximity coupling
in S-I-N and \mbox{S-I-N-I-S} structures
(where `I' denotes an insulating tunnel barrier).
Using the renormalization group method, we have calculated both quantities
including the region of strong fluctuations, $\ln^2(1/T\tau) \geq g$.
Cooper-channel repulsion was found to modify results of semiclassical
calculations by the power-law factors $\propto T^{1/\pi\sqrt{g}}$.
The effects of the Coulomb zero-bias anomaly (in the case of unscreened
static Coulomb interaction) were found to be even stronger, the
corresponding correction factor being of the order of
$\exp[-(1/4\pi^2 g)\ln^2(\Delta/T)]$, where $\Delta$
is the gap in the superconductive terminal.
However, results of Ref.~\cite{FLS} are limited to the lowest tunneling
approximation, i.~e., valid under the condition $R_T \gg R_D$, where
$R_T$ and $R_D$ are the resistances of the tunnel barrier in the normal state,
and of the diffusive normal conductor, correspondingly.
General semiclassical theory of \mbox{N-I-S} systems with arbitrary
ratio $R_T/R_D$ usually neglects interaction effects in the
N part of the structure~\cite{volkov,Nazarov94,NazarovC,Carlo}.
The effect of Cooper interaction
(with strength $\lambda$) on the Andreev conductance $G_A$ was studied
by Stoof and Nazarov~\cite{StNaz} for an N-S structure with the ideal
interface and 1D geometry of the normal wire, in the first order over
$\lambda$.  Relative correction to $G_A$ was found to be
small, of the order of $\lambda \ll 1$.
The effects of interaction upon noise in N-S structures
had never been studied before,
to the best of our knowledge (for the discussion
of N-S noise in the absence of interaction, see
Refs.~\cite{DeJong,Lesovik99} and the review~\cite{Blanter}).

\begin{figure}
\refstepcounter{figure} \label{F:NSisland}
\vspace{1mm}
\epsfxsize=3in
\centerline{\epsfbox{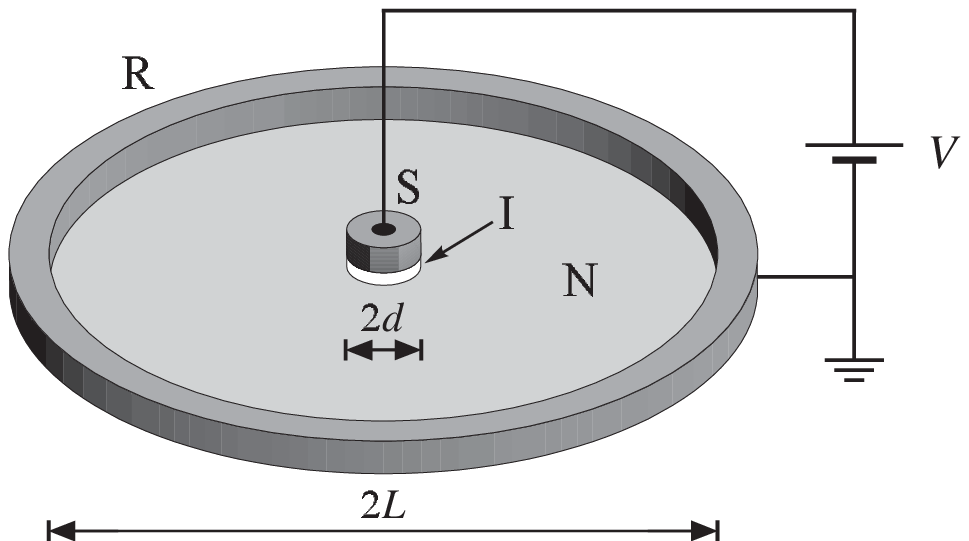}}
\vspace{3mm}
\small FIG.\ \arabic{figure}.
A small superconductive island (S) of size $2d$ connected to
a reservoir (R) through a tunnel barrier (I) and
a dirty normal film (N) of size $2L \gg 2d$.
\end{figure}

In this paper we generalize our approach~\cite{FLS}
for the case of strong proximity effect
(arbitrary relation between $R_T$ and $R_D$),
in the presence of Cooper interactions in the normal conductor.
We calculate the Andreev conductance of 2D  N-I-S structures
(shown in Fig.~\ref{F:NSisland})
at low (compared to $\Delta$) temperature and voltages, as a function
of $t=R_D/R_T$ and of the ``decoherence time"
of an electron and the Andreev-reflected hole, $\hbar/\Omega_*$,
where $\Omega_* = \max(T, eV, eDH/c)$.  We demonstrate that the
Cooper interaction effects in 2D are substantially different from the
1D case considered in Ref.~\cite{StNaz}.
In particular, at $\Omega_* \ll \ETh = \hbar D/L^2$ the lowest-order
relative correction $\delta G_A/G_A$ scales as $\lambda\ln L/d$
and grows with the size of the system $L$.
We sum up the main logarithmic terms of the order of $(\lambda\ln{L/d})^n$
and find that Cooper repulsion may lead
to nonmonotonous dependence of $G_A$ on $R_D/R_T$ and/or on
the decoherence energy scale $\Omega_*$.

Technically, our method is based on
the Keldysh functional approach for disordered superconductors~\cite{FLS}
(proposed previously for normal conductors in Ref.~\cite{KA}).
The basic object of this theory is the $8\times 8$ matrix field
$Q(t,t';\br)$ which depends on two time variables and
one space coordinate.
Its average value gives the electron Green function $G(\br,\br)$
at coincident points.
Fluctuations of the $Q$ matrix
describe slow diffuson and Cooperon modes of the electron system.
In the case of spin-independent interactions,
$Q$ reduces to a $4\times 4$ matrix.
This method seems to be more convenient than the replica functional
approach~\cite{finkel1,finkel2,finkel3}
as it does not require tedious analytic
continuation procedure and allows for the direct study of non-equilibrium
phenomena. In the limit $R_T \gg R_D$ studied in~\cite{FLS},
integration over diffuson/Cooperon modes reduces the problem to an
effective action $S_{\rm prox}[Q_S,Q_N]$ for the proximity
effect, which contains two terms only, $\Tr Q_SQ_N$ and
$\Tr(Q_SQ_N)^2$, describing
transport of single electrons and Cooper pairs, respectively
(here $Q_S$ and $Q_N$ corresponds to the superconductive
and normal boundaries of the system). At arbitrary relation
between $R_T$ and $R_D$, the proximity action
contains an infinite series of terms,
$S_{\rm prox}[Q_S,Q_N] \propto \sum_{n=1}^\infty \gamma_n \Tr (Q_SQ_N)^n$.
A set of parameters $\gamma_n$ can be conveniently parametrized via the
$2\pi$-periodic function $u(x) = \sum_{n=1}^\infty n \gamma_n \sin nx$
that encodes amplitudes
of processes of $n$-electron transfer through the system.
The evolution of $u(x)$ as a function of the ratio $R_T/R_D$
is found with the use of the functional renormalization group method
applied to the action $S_{\rm prox}[Q_S,Q_N]$.

In this paper we will not consider the weak localization and interaction
corrections assuming that the sheet conductance is relatively large,
$g \gg \ln(L/d)$.
We will also not take into account the effect of the Coulomb
zero-bias anomaly (ZBA) on the Andreev conductance.
Possibility of neglecting Coulomb ZBA effects while treating
the Coulomb-induced repulsion in the
Cooper channel~\cite{finkel1,finkel2,finkel3} is due to the fact that the
specific form of the Coulomb ZBA depends crucially on the long-range
behavior of the Coulomb potential, whereas renormalization of the
Cooper-channel interaction depends on the short-distance Coulomb amplitude
only. If long-range Coulomb forces are suppressed (i.~e., by placing a nearby
screening metal gate), the Coulomb ZBA effect may become weak, and
the main effect comes from the short-range repulsion in the Cooper channel.
This is the situation we are going to study in this paper.

Another limitation of the present discussion is that we will consider
the case of a two-dimensional (2D) geometry of the current flow
between the superconductive and normal electrodes of the structure, as
shown in Fig.~\ref{F:NSisland}. This will make possible to construct
a unified functional renormalization group treatment that takes into
account modifications of the proximity effect strength both due to
multiple Andreev reflections and due to Cooper-channel repulsion.

Finally, we restrict our present discussion to a low-frequency domain
$\omega \ll D/L^2$.
It can be shown that the frequency-dependent
Andreev conductance $G_A(\omega)$ does not contain significant
frequency dependence at scales $\omega \geq D/L^2$, due to the
long-range nature of the Coulomb interaction which makes electron liquid
effectively incompressible up to much high frequency scale
defined by the inverse time of electric field
propagation across the structure. However, the specific form of the
proximity action $S_{\rm prox}[Q_S,Q_N]$ as
a linear combination of the traces of $(Q_SQ_N)^n$ is valid only
in the low-frequency domain. General discussion of the action
becomes much more involved in the high frequency region, and
we will postpone this subject for the future studies.

The rest of the paper is organized as follows.
In Sec.~\ref{S:sigma} we introduce the Keldysh
$\sigma$-model action and describe a general scheme of expansion
in terms of diffusive slow modes.
Sec.~\ref{S:FRG} is devoted to the
derivation of the functional renormalization group (FRG) equation
for the case of non-interacting electrons in the normal conductor.
It is shown that the resulting equation for the function $u(x)$
which parametrizes the action
exactly coincides with the Euler equation for generating function of
transmission eigenvalues $T_j$ discussed in Ref.~\cite{Carlo94}.
In Sec.~\ref{S:phys} we show how physical quantities such as conductivity
and current noise can be expressed in terms of the function $u(x)$.
The main new results of the paper are presented in Sec.~\ref{S:lambda}.
In Secs.~\ref{SS:lambda-gen} and~\ref{SS:lambda-der} we derive an
additional term in the FRG equation,
that accounts the electron-electron interaction in the Cooper channel.
The solution of the full FRG equation and the corresponding results for
the Andreev conductance are discussed in Sec.~\ref{SS:lambda-cond}.
The effect of interaction upon the current noise is studied in
Sec.~\ref{SS:lambda-noise}.
Sec.~\ref{S:beyond} is devoted to the study of the dependence
of the conductance and noise
on temperature, voltage and magnetic field, which
suppress Cooperon amplitude at the energy scale
$\Omega_* = \max(T,eV,eDH/c)$.
It is shown in Sec.~\ref{SS:be_int} that repulsion in
the Cooper channel may lead to a nonmonotonous dependence of
effective interface resistance on $\Omega_*$.
Sec.~\ref{SS:conclusion} contains discussion and conclusions.
Some technical details are presented in Appendixes~\ref{A:A} and~\ref{A:B}.
Appendix~\ref{A:C} contains an analysis of the Andreev conductance in
a 2D S-I-N structure by means of the standard method of the Usadel equation in
the presence of Cooper interaction.

\section{Keldysh $\sigma$-model action}
\label{S:sigma}

Keldysh action for disordered N-S systems
derived in Ref.~\cite{FLS} can be represented as a sum of the bulk
and boundary contributions.

The bulk action, $S_{\rm bulk}$, is a functional of three fluctuating fields:
the matter field $Q(\br,t,t')$ in the film,
the electromagnetic potential $\roarrow\phi(\br,t)$
and the order parameter field $\roarrow\Delta(\br,t)$
used to decouple the quartic interaction vertex in the Cooper channel:
\bea
  S_{\rm bulk} = \frac{i\pi\nu}{4}
    \Tr \left[
      D (\nabla Q)^2
      + 4i \bigl(
        i\tau_z \partial_t + \tensor{\phi} + \tensor{\Delta}
      \bigr) Q
    \right]
\nonumber \\
   {} + \Tr \roarrow{\phi}^T (V_0^{-1}+2\nu)\sigma_x \roarrow{\phi}
  + \frac{2\nu}{\lambda} \Tr \roarrow{\Delta}^+ \sigma_x \roarrow{\Delta} ,
\label{sm}
\eea
where $D$ is the diffusion coefficient in the film,
$\nu$ is the density of states at the Fermi level per
single projection of spin, and
$\lambda$ is the dimensionless coupling constant in the Cooper channel.

Being a matrix in the time domain, $Q(\br,t,t')$ is also a $4\times4$ matrix
in the direct product $K\otimes N$ of the Keldysh and Nambu spaces.
Pauli matrices in the $K$ and $N$ spaces are denoted by $\sigma_i$
and $\tau_i$ respectively.
The field $Q$ satisfies a nonlinear constraint $Q^2=1$ and can be
parametrized as $Q=U^{-1}\Lambda U$ where $\Lambda = \Lambda_0 \tau_z$
is the metallic saddle point and
\be
\label{Lambda0}
  \Lambda_0(\eps) =
    \left( \begin{array}{cc} 1 & 2F(\eps) \\ 0 & -1 \end{array} \right)_K .
\ee
The matrix $F(\eps)$ introduced in Eq.~(\ref{Lambda0}) acts in the
Nambu space and has the meaning of a generalized distribution function.
In equilibrium $F(\eps) = \tanh(\eps/2T)$,
and its general form is $F(\eps) = f(\eps) + f_1(\eps) \tau_z$,
where $f_1(E)$ is the anomalous distribution function which
is a measure of the branch imbalance~\cite{LOreview,RamSm}.
The object $\roarrow\phi=(\phi_1, \phi_2)^T$ is a vector
in the Keldysh space, with $\phi_1, \phi_2$ being the classical
and quantum components of the field.
They are given by the symmetric and antisymmetric
linear combination of the $\phi$ fields on the forward (f)
and backward (b) branches of the Keldysh contour:
$\phi_{1,2}=(\phi_{\rm f}\pm\phi_{\rm b})/2$.
In the Keldysh technique, external (nonfluatuating) fields
(e.~g., applied voltage) have only classical component,
while physical results can be obtained by taking derivatives
with respect to the quantum component, cf.\ Sec.~\ref{S:phys}.
Fluctuating fields have both components.
$\tensor\phi$ is a shorthand notation for the matrix
$\tensor\phi = \phi_1 \sigma_0 + \phi_2 \sigma_x$ in the Keldysh space.
Similarly, $\roarrow\Delta=(\Delta_1, \Delta_2)^T$, and
$\tensor\Delta$ stands for a $4\times4$ matrix
$\tensor\Delta = [\tau_+\Delta_1 - \tau_-\Delta_1^*]\sigma_0
+ [\tau_+\Delta_2 - \tau_-\Delta_2^*]\sigma_x$,
where $\tau_\pm \equiv (\tau_x \pm i\tau_y)/2$.

In Eq.~(\ref{sm}), $V_0(q)$ is the bare static Coulomb potential
between electrons in the dirty film.
Below we will consider the situation when $V_0(q)$ is screened due to
the presence of the nearby metal gate.
In particular, if such a gate is situated at the distance $b$
from the dirty film, parallel to it,
$V_0(q) = 2\pi e^2 (1-e^{-2b q})/q \stackrel{q\to0}{\longrightarrow}
4\pi e^2 b$.
As it was discussed at length in Ref.~\cite{FLS}, the effect of
long-range fluctuations of electromagnetic potential (which lead
to the Coulomb zero-bias anomaly) is determined by the long-range
part of the bare Coulomb
interaction, and thus is suppressed once the Coulomb potential
is screened, having the relative order of
$g^{-1}\ln(8\pi\nu e^2 b)\ln(1/\Omega\tau)$.
We consider the case of a relatively short screening length $b$, and
neglect long-range electric fluctuations and the Coulomb ZBA effects.
On the other hand, Coulomb-induced repulsion in the Cooper
channel~\cite{finkel1,finkel2} does not depend on the long-range
part of $V_0(q)$ and is left unchanged by the screening gate.
It is this effect of interaction which we are going to consider in this paper.

In terms of the $\sigma$-model (\ref{sm}), diffusion-like collective
excitations of an electron system are described as slow fluctuations
of the $Q$-matrix over the manifold $Q^2=1$. Small fluctuations near
the metallic saddle point $Q=\Lambda$ can be parametrized by the
rotation matrix $U=1+W/2+\dots$, with $W$ obeying the relation
$\{W,\Lambda\}=0$. This constraint is resolved by
\be
\label{uWu}
  W = u \left( \begin{array}{cc}
     w_x \tau_x + w_y \tau_y
     & w_0 + w_z \tau_z \\
     \overline w_0 + \overline w_z \tau_z &
     \overline w_x \tau_x + \overline w_y \tau_y
  \end{array} \right)_K u,
\label{W}
\ee
where the matrix $u$ is defined as
\be
\label{u}
  u = \left( \begin{array}{cc} 1 & F \\ 0 & -1 \end{array} \right)_K .
\ee
Variables $w_i$ with $i=0,z$ couple to diagonal in the Nambu space matrices
and describe diffuson modes, while $w_j$ with $j=x,y$ couple to off-diagonal
in the Nambu space matrices and correspond to Cooperon modes.
Their propagators are given by
\bea
  \corr{ w_i(\bq)_{\eps_1\eps_2} \overline w_i(-\bq)_{\eps_2\eps_1} }
  = -\frac{1}{\pi\nu}
    \frac{1}{Dq^2 - i(\eps_1-\eps_2)},
\nonumber \\
\label{diffusons}
    \corr{ w_j(\bq)_{\eps_1\eps_2} w_j(-\bq)_{\eps_2\eps_1} }
    = -\frac{1}{\pi\nu}
      \frac{1}{Dq^2 - i(\eps_1+\eps_2)},
\\ \nonumber
    \corr{ \overline w_j(\bq)_{\eps_1\eps_2} \overline w_j(-\bq)_{\eps_2\eps_1} }
    = -\frac{1}{\pi\nu}
      \frac{1}{Dq^2 + i(\eps_1+\eps_2)}.
\eea
General contraction rules for averaging of $\corr{ \Tr AW \cdot \Tr BW }$
over $W$ are listed in Appendix~\ref{A:A}.

In two dimensions, the action (\ref{sm}) can be studied by the
renormalization group approach~\cite{finkel2,FLS}. At each step of the RG
procedure one has to eliminate fast modes with either
$\max(Dq^2,\eps_1-\eps_2)>\Omega$ (for diffusons) or
$\max(Dq^2,\eps_1+\eps_2)>\Omega$ (for Cooperons)
where $\Omega$ is the running ultraviolet RG cutoff.
There are two types of logarithmic corrections to the action (\ref{sm}).
Weak localization and interaction corrections
have the relative order of $g^{-1}\ln(1/\Omega\tau)$
and modify the conductance $g$ of the system.
Quantum corrections of the other type have the relative order of
$g^{-1}\ln^2(1/\Omega\tau)$ and emerge in renormalization
of the Cooper channel coupling $\lambda$.
They become essential at such scales where localization
corrections are still small and can be neglected.
In this case $\lambda$ is the only running parameter
of the bulk action which satisfies the RG equation~\cite{finkel2,FLS}
\be
\label{lambda-rg}
  \frac{\partial\lambda}{\partial \zeta} = - \lambda^2 + \lambda_g^2 ,
  \qquad
  \lambda_g = \frac{1}{2\pi\sqrt{g}} ,
\ee
where $\zeta=\ln(1/\Omega\tau)$ is the logarithmic variable.
The first term in the RHS of the RG Eq.~(\ref{lambda-rg})
is due to the usual BCS logarithm,
whereas the second contribution is the effect of Coulomb repulsion
between slowly diffusing electrons.

The bulk action (\ref{sm}) describes dynamics of the electron system
in the metal film. Possibility of tunneling between the
island and the film is taken into account in the lowest tunneling Hamiltonian
approximation by introducing the boundary term in the action:
\be
  S_{\rm tun} = - \frac{i\pi\gamma}{4} \Trg Q_{\rm isl} Q .
\label{tun}
\ee
Here $Q_{\rm isl}$ refers to the island, spatial integration under the trace
is taken over the area of the interface $\Gamma$,
and $\gamma$ is the dimensionless normal-state tunneling conductance
per unit area of the boundary.

Below we will consider thick enough superconductive island
so that the absolute value $|\Delta|$ of the order parameter
is not suppressed by proximity to the normal film.
The corresponding condition reads $|\Delta| \gg G_T/\nu V_{\rm isl}$,
with $V_{\rm isl}$ being the island's volume.
Also, the size of the island, $d$, is supposed
to be smaller than the superconductive coherence length so that
for $T\ll T_c$ the only relevant degree of freedom in the island is the phase
$\varphi$ of the order parameter. In the Keldysh formalism, one has to
introduce its classical, $\varphi_1$, and quantum, $\varphi_2$, components,
and arranging them into a matrix
$\tensor\varphi = \varphi_1\sigma_0 + \varphi_2\sigma_x$
one can write $Q_S$ in the subgap limit ($\eps<\Delta$) as
\be
  Q_S = - i \tau_+ e^{i\tensor\varphi} + i \tau_- e^{-i\tensor\varphi} .
\label{QS}
\ee
We will also consider below the example of a normal (non-superconductive)
island connected to the film by a tunnel barrier and biased
at some voltage with respect to it.
In that case the island's $Q$ matrix can also be formally expressed
in terms of the superconductive phase:
\be
  Q_N = e^{i\tau_z\tensor\varphi/2} \Lambda e^{-i\tau_z\tensor\varphi/2} ,
\label{QN}
\ee
with $\varphi/2$ having the meaning of the single-particle phase
which is conjugated to the number of electrons on the island
(while $\varphi$ is conjugated to the number of Cooper pairs).
In both cases the applied voltage $V(t)$ can be accounted by
the Josephson relation $d\varphi_1/dt=2eV$ for the classical
component of the phase.

\section{Functional RG: Noninteracting case}
\label{S:FRG}

\subsection{General idea}
\label{SS:idea}

In this section we will apply the renormalization group treatment
to the system consisting of a small island coupled to a diffusive
metal film by a tunnel barrier of arbitrary transparency.
The effects of the Cooper interaction in the film will be considered
later on in Sec.~\ref{S:lambda}, while now we will study
noninteracting electrons.
In this section we will restrict ourselves to the {\em zero-energy limit}
when both temperature $T$, voltage $eV$
and frequency $\omega$ are smaller that the Thouless energy $\ETh=D/L^2$,
and the perpendicular magnetic length $l_H=\sqrt{\Phi_0/H}$
($\Phi_0$ is the flux quantum, $H$ is the applied magnetic field)
is shorter than the system size $L$.
The effect of larger $T$, $V$ or $H$ is to destroy Cooperon
coherence at scales larger than $\min\{\sqrt{D/T}, \sqrt{D/eV}, l_H\}$,
and will be studied in Sec.~\ref{S:beyond}.
The effect of high-frequency ($\omega\gg\ETh$) voltage is two-fold:
it suppresses Cooperon coherence and also perturbs local
electroneutrality of electron system. We will not study this
effect in the present paper.

The method developed in the present
section can be applied both to the normal and superconductive island.
However, since the rest of the paper will be devoted mainly
to the study of charge transport in the N-S system, all formulas
will be written for the case of a superconducting island, $Q_{\rm isl} = Q_S$.
In order to translate them for the case of a normal island, one has to
substitute $Q_S$ by $Q_N$, and the word ``Cooperon'' by the word ``diffuson''.

\begin{figure}
\refstepcounter{figure} \label{F:2}
\epsfxsize=0.857in
\noindent
\centerline{\epsfbox{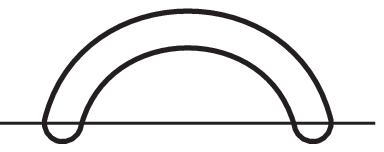}}
\par
\vspace{1mm}
\small FIG.\ \arabic{figure}.
The second order in $\gamma$ correction to the action.
The horizontal line schematically designates the N-S boundary $\Gamma$,
with the two semicircles below it denoting $Q_S$ in the island,
and the two-line object above it representing the Cooperon in the normal metal.
\end{figure}

Renormalization of the total action $S_{\rm bulk} + S_{\rm tun}$
describing an N-S contact is a complicated task.
The main problem one encounters here is that this action
does not reproduce itself under renormalization.
In the lowest order over the tunneling transparency $\gamma$
studied in Ref.~\cite{FLS}, the boundary term (\ref{tun})
generates the next-order term $\gamma_2\Tr (Q_S Q)^2$ in the action,
with $\gamma_2$ obeying the RG equation $d\gamma_2/d\zeta \propto \gamma^2$.
The physical meaning of this relation is simple.
It describes the process when a Cooper pair tunnels
from the island, coherently propagates in the normal metal
as a Cooperon and then returns back to the island.
The resulting expression is bilinear in $Q_S$ (taken at the
times of the first and second tunneling) and involves
logarithmically (in 2D) large factor due to the probability of returning
to the original point.
The corresponding diagram is shown in Fig.~\ref{F:2}.

\begin{figure}
\refstepcounter{figure} \label{F:4}
\epsfxsize=3.375in
\noindent
\centerline{\epsfbox{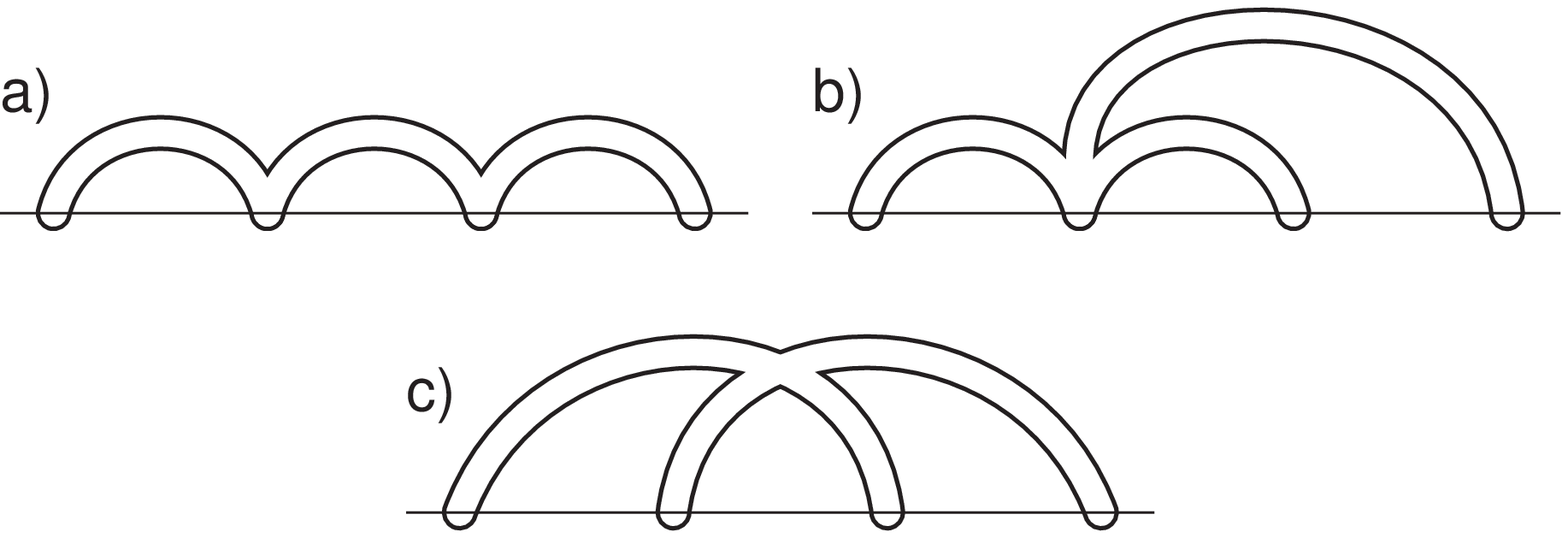}}
\par
\vspace{1mm}
\small FIG.\ \arabic{figure}.
Three leading logarithmic corrections to the action
at the fourth order in $\gamma$.
\end{figure}

In order to clear up the structure of the leading logarithmic
corrections at arbitrary interface transparencies, let us consider
the diagrams of the forth order over $\gamma$ shown in Fig.~\ref{F:4}.
All of them ought to have the {\em tree-like} structure
since any loop yields an additional small factor $\zeta/g\ll1$.
For two of them, labeled by (a) and (b), successive Andreev tunnelings
are connected by free Cooperons and diffusons, whereas the process
shown in Fig.~\ref{F:4} (c) involves 4-Cooperon interaction vertex
(Hikami box) in the bulk. Due to spatial integration, each of these
diagrams evaluates to the third power $\zeta^3$ of the logarithm,
and they exhaust the number of the most divergent diagrams of the
forth order.

Fig.~\ref{F:4} suggests an idea that the tree-like structure of
corrections to the action may be suitable for RG approach.
Indeed, for each tree one can find the branch with the smallest
momentum, say $q$. If we cut the whole tree over this branch,
it will break up into two pieces which could be calculated at
the previous steps of RG procedure since they contain only momenta
greater that $q$. The diagrams (a) and (b) conform well to this scheme,
however it is not clear how to treat the bulk nonlinearity in
the diagram (c). To overcome this problem we will employ our
freedom to choose arbitrary parametrization of the field $Q$
in terms of the matrix $W$, that can change the relative
contribution of different diagrams. For the problem in question,
the choice of the exponential parametrization, $Q=\Lambda \exp(W)$,
nullifies all diagrams with Hikami boxes within the RG precision.
To see this, we note that the leading logarithmic contribution arises
from the integration over fast momentum, so that the energy dependence
of the diffusion propagators (\ref{diffusons}) can be neglected.
To this accuracy, the matrices $W$ and $\nabla W$ commute and the term
$\Tr(\nabla Q)^2=-\Tr(\nabla W)^2$ becomes Gaussian in $W$, therefore
there are no nonlinear vertices in the bulk in the zero-energy limit.
Note also that the same
exponential parametrization is used usually for the solution
of the Usadel equations~\cite{Usadel,StNaz} for the Green functions of
disordered superconductors.

Thus, in the exponential parametrization, diagrams of the type
shown in Fig.~\ref{F:4}(c) are absent and one can carry out the RG
procedure based on the tree-like structure of the leading diagrams.
Following this approach we present the total action as a sum
of the bulk part $S_{\rm bulk}$ and the part $S_\Gamma$ which arises
from the single tunneling action (\ref{tun}) under renormalization:
\be
  S_\Gamma
  = \sum_{n=1}^\infty S_n
  = -i \pi^2 g \sum_{n=1}^\infty \gamma_n \Trg (Q_SQ)^n .
\label{sg}
\ee
The renormalized boundary action (\ref{sg}) describes processes
of multiple Andreev tunneling coupled by coherent propagation
of two-particle diffusion modes in the normal metal.
It is determined by the infinite set of coefficients (``charges'')
$\gamma_n$ and is a functional of the phase $\roarrow\varphi$
of the island's order parameter. Once $\gamma_n$'s are known,
one can easily calculate the Andreev conductance and other quantities
related to charge transfer through the system (see section~\ref{SS:cond}).
Our main purpose below will be to establish the law of transformation
of the coefficients $\gamma_n$ and to derive the corresponding
multicharge renormalization group equation.

\subsection{Derivation of the multicharge RG equation}
\label{SS:mrg-der}

According to the general approach discussed above, at each step of the RG
procedure one has to calculate a correction $\Delta S_{k,n}$ to the action
which results from connecting two pieces $S_k$ and $S_n$ of the boundary
action (\ref{sg}) by a fast diffusion mode. This gives logarithmic
contribution at scales larger than the island size $d$ and, consequently,
at energies $\Omega$ smaller than $\omega_d \equiv D/d^2$.
Therefore in this case it is convenient to define the logarithmic variable as
\be
  \zeta = \ln \frac{\omega_d}{\Omega} = 2 \ln \frac{R}{d},
\label{zeta}
\ee
where the current spatial scale $R$ is related to $\Omega$
through $\Omega=D/R^2$. For the purposes of RG,
the form of the island is of no importance, and all expressions
depend only the total normal-state interface conductance $G_T$.
Hence, one can formally consider a point-like island,
with the transparency $\gamma$ in Eq.~(\ref{tun}) becoming
$\gamma=G_T$ and $\Trg X \equiv \Tr X(\br=0)$.

Our aim is to derive an effective low-energy action of the phase
$\roarrow\varphi$ on the island, therefore we consider $Q_S$
as a slow variable and decompose only metallic $Q$ into the
fast and slow parts. Such a decomposition consistent with the
exponential parametrization adopted is given by
$Q = \tilde{U}^{-1} \Lambda \exp(W') \tilde{U}$,
where $\tilde{U}$ is a slow rotation matrix
and $W'$ is a fast variable~\cite{FLS,finkel2}.
The relevant interaction vertex reads
\be
  S_{{\rm int},n} =
  -i\pi^2g n\gamma_n \Trg (Q_S\tilde{Q})^{n-1}
  Q_S \tilde{U}^{-1} \Lambda W' \tilde{U},
\label{S_n}
\ee
where the additional factor of $n$ accounts the fact that
the fast diffusion mode can be coupled to any matrix $Q$ in $S_n$,
and $\tilde{Q} \equiv \tilde{U}^{-1} \Lambda \tilde{U}$.
The form of the vertex (\ref{S_n}) corresponds to the
tree-like structure of the diagrammatic expansion discussed
in Sec.~\ref{SS:idea}. The lack of higher-order terms in $W$
in $S_{{\rm int},n}$ amounts to neglecting closed loops
in virtue of the small parameter $\zeta/g$.

For $k\neq n$, the correction to the action is given by
$\Delta S_{k,n} = i \corr{S_{{\rm int},k}S_{{\rm int},n}}$.
This average (in the zero-energy limit) is given by Eq.~(\ref{WW1}),
with
\begin{mathletters}
\label{AB}
\bea
  A = \tilde{U}(Q_S\tilde{Q})^{k-1} Q_S \tilde{U}^{-1}\Lambda, \label{A}
\\
  B = \tilde{U}(Q_S\tilde{Q})^{n-1} Q_S \tilde{U}^{-1}\Lambda. \label{B}
\eea
\end{mathletters}%
Calculating the logarithmic integral over the fast momentum
and omitting the tilde sign over slow variables one obtains
\bea
  \Delta S_{k,n} = \frac{i\pi^2 g}2 kn \gamma_k \gamma_n \Delta\zeta
\nonumber \\
  {} \times
    \Trg \left[ (Q_SQ)^{k+n} - (Q_SQ)^{k-1} Q_S (Q_SQ)^{n-1} Q_S \right] .
\eea
Employing invariance of the trace under cyclic permutations and using
the relations $Q^2=Q_S^2=1$, we get finally for $k\neq n$
\be
  \Delta S_{k,n}
  = \frac{i\pi^2 g}2 kn \gamma_k \gamma_n \Delta\zeta
    \Trg \bigl[ (Q_SQ)^{k+n} - (Q_SQ)^{|k-n|} \bigr] .
\label{<SS>}
\ee
Thus we see that averaging of the terms $S_k$ and $S_n$ from the
action (\ref{sg}) modifies the terms $S_{k+n}$ and $S_{|k-n|}$
with $\Delta\gamma_{k+n} = - (1/2) kn \gamma_k \gamma_n \Delta\zeta$
and $\Delta\gamma_{|k-n|} = (1/2) kn \gamma_k \gamma_n \Delta\zeta$.

Analogous calculation of $\Delta S_{k,k} = (i/2) \corr{S_{{\rm int},k}^2}$
yields
\be
  \Delta S_{k,k}
  = \frac{i\pi^2 g}4 k^2 \gamma_k^2 \Delta\zeta \Trg (Q_SQ)^{2k}
\ee
that results in renormalization of the coefficient
$\Delta\gamma_{2k} = - (1/4) k^2 \gamma_k^2 \Delta\zeta$.

Taking into account all possible pairings $\Delta S_{k,n}$ between the terms
of the action (\ref{sg}), we derive the main equation that
describes evolution of the charges $\gamma_n$ under renormalization:
\be
  \frac{d\gamma_n}{d\zeta}
  = - \frac{1}{4} \sum_{k=1}^{n-1} k (n-k) \, \gamma_k \gamma_{n-k}
    + \frac{1}{2} \sum_{k=1}^\infty k (n+k) \, \gamma_k \gamma_{n+k} .
\label{mrg}
\ee
Initial conditions for these multicharge RG equations at $\zeta=0$
follow from the bare tunneling action (\ref{tun}):
\be
  \gamma_1(0) = a \equiv \frac{G_T}{4\pi g},  \qquad \gamma_{n>1}(0)=0 .
\label{initial}
\ee

Note that $Q_S^2=1$ is the only property of the island's $Q_S$ which was
used to derive the RG equations (\ref{mrg}). Therefore these equations
universally describe both normal-metal and superconductive islands
on the same footing.

Equations (\ref{mrg}) look quite complicated. Their hidden
algebraic structure becomes transparent after Fourier transformation.
To this end we introduce a $2\pi$-periodic function $u(x)$ of an auxiliary
continuous variable $x$ according to the definition
\be
  u(x) = \sum_{n=1}^\infty n \gamma_n \sin nx .
\label{u-def}
\ee
Transforming Eq.~(\ref{mrg}) into $x$-representation we obtain the
following RG equation for the function $u(x)$:
\be
  u_\zeta + u u_x = 0 .
\label{urg}
\ee
The differential equation (\ref{urg}) is nothing but the well known
Euler equation describing 1D motion of a compressible gas,
with $\zeta$ playing the role of time.
The initial condition for Eq.~(\ref{urg}) at $\zeta=0$ is given by
$u_0(x) = a\sin x$.

The Euler equation (\ref{urg}) can be easily solved with the help of
characteristics. The latter are described by the following system of
differential equations:
\be
  \frac{dx}{d\zeta} = u,
  \qquad
  \frac{du}{d\zeta} = 0.
\label{char0}
\ee
They are trivially integrated: $u=u_0(x_0)$ and $x=x_0+u_0(x_0)\zeta$.
In order to get $u(x)=u_0(x_0(x))$ one has just to find the inverse
function $x_0(x)$.
For $u_0(x)=a\sin x$, the function $u(x)$ is implicitly defined
by the relation
\be
  u(x) = a\sin(x-u(x)\zeta) .
\label{uu}
\ee

\begin{figure}
\refstepcounter{figure} \label{F:Euler}
\vspace{-1mm}
\epsfxsize=80mm
\centerline{\epsfbox{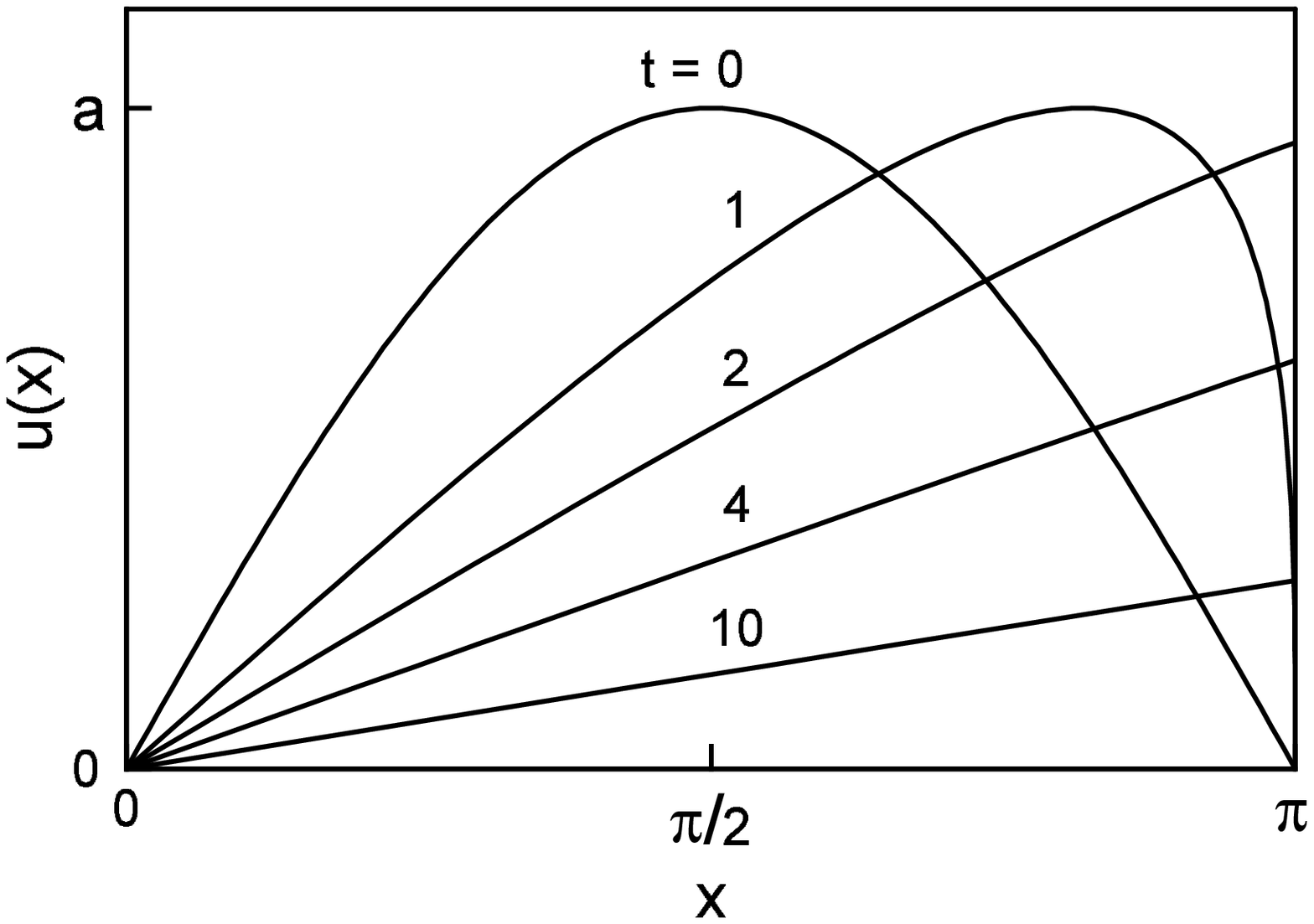}}
\vspace{2mm}
\small FIG.\ \arabic{figure}.
Solutions for the Euler equation (\protect\ref{urg})
at different values of $t=a\zeta$.
\end{figure}

The solution of the Euler equation is conveniently expressed through
the variable $t=a\zeta=G_T/G_D$ which is equal to the ratio of the tunneling
conductance of the barrier $G_T$ to the conductance of the metal film
$G_D=4\pi g/\zeta$ at the current scale.
The solution for different values of $t$ is shown in Fig.~\ref{F:Euler}.
The derivative $u_x(\pi)$ diverges at $t=1$, exactly at the point
where $G_D=G_T$.
For $t>1$, the discontinuity (``shock wave'') develops at $x=\pm\pi$.
For $t\gg1$, the solution acquires a saw-like type with
$u(x)\simeq (a/t)x = (G_D/4\pi g)x$.

\subsection{Proximity Action}
\label{SS:prox}

In the previous section we have established the transformation law
for the coefficients $\gamma_n$ of the multicharge action (\ref{sg})
under the action of the renormalization group. After eliminating
diffusion modes up to some value of the logarithmic variable $\zeta$,
the action becomes a functional of the island's $Q_S$ and $Q(R)$
in the film taken at the special scale $R \sim d\,e^{\zeta/2}$
from the island. For a finite system bounded by a perfect metallic
lead located at a distance $L$ ($L\gg d$) from the island
(see Fig.~\ref{F:NSisland}), renormalization of the action
started at the energy scale $\omega_d$ should stop at the scale
of the Thouless energy $\ETh=D/L^2$
when all diffusive electronic degrees of freedom in the dirty metal
are integrated out. The resulting action,
\be
  S_{\rm prox}[Q_S,Q_N]
  = -i \pi^2 g \sum_{n=1}^\infty \gamma_n \Tr (Q_S Q_N)^n ,
\label{prox0}
\ee
thus becomes the functional of the island's $Q_S$, Eq.~(\ref{QS}),
and the matrix $Q_N$, Eq.~(\ref{QN}), in the lead. Such an action
resulting from elimination of the modes in the diffusive conductor
will be referred to as the Proximity Action.

According to Eqs.~(\ref{QS}) and (\ref{QN}), the only degree of freedom
in the matrices $Q_S$ and $Q_N$ is the one associated with the phase
rotation. In other words, externally controlled voltage is the only
dynamical variable both in the island and in the lead.
Below we will assume that the metal lead is at zero bias,
so that $Q_N$ reduces to $\Lambda$ and the Proximity Action (\ref{prox0})
acquires the form
\be
  S_{\rm prox}[Q_S]
  = -i \pi^2 g \sum_{n=1}^\infty \gamma_n \Tr (Q_S \Lambda)^n .
\label{prox}
\ee
In this formula, $\gamma_n$ are taken at ``time'' $t = a\zeta_{\rm Th}$,
with $\zeta_{\rm Th} \equiv \ln(\omega_d/\ETh) = 2\ln(L/d)$.
It can be calculated through the integration of the FRG equation (\ref{urg})
for the function $u(x,\zeta)$ up to the scale $\zeta_{\rm Th}$.

In order to express the action in terms of the phase $\roarrow\varphi$,
one has to substitute $Q_S$ from Eq.~(\ref{QS}).
Taking the trace over the Nambu space,
we obtain that only terms with even $n$ contribute to the Proximity Action:
\be
  S_{\rm prox}[Q_S]
  = -2i \pi^2 g \sum_{n=1}^\infty (-1)^n \gamma_{2n}
    \Trk (e^{i\tensor\varphi} \Lambda_0 e^{-i\tensor\varphi} \Lambda_0)^n ,
\label{prox-s}
\ee
where the subscript `K' indicates that the
matrix trace is taken only over the Keldysh space.
As explained in Sec.~\ref{SS:mrg-der}, the action (\ref{prox}), after
substitution $Q_S \to Q_N$, also applies to the case of the normal island.
Using Eq.~(\ref{QN}) for $Q_N$ and taking the trace over the Nambu space,
we arrive at the following expression for the action:
\be
  S[Q_N]
  = -2i \pi^2 g \sum_{n=1}^\infty \gamma_{n}
    \Trk(e^{i\tensor\varphi/2} \Lambda_0 e^{-i\tensor\varphi/2} \Lambda_0)^n .
\label{prox-n}
\ee

Remarkably, our function $u(x,\zeta_{\rm Th})$ can be directly related
to the generating function of transmission coefficients $F(\phi)$
introduced by Nazarov~\cite{Nazarov94} (cf.\ also Ref.~\cite{Carlo94}).
Comparing our Eq.~(\ref{uu}) with Nazarov's Eq.~(19) one can verify that
\be
  F(\phi) = \frac{4\pi g\, u(\phi)}{\sin\phi} .
\label{F}
\ee
Note also that appearance of
discontinuity for $t>1$ goes in parallel with opening of conducting
channels with transparencies $T\to1$~\cite{Nazarov94}.
The saw-like solution for the function $u(x)$ at $t\gg1$
results\cite{Nazarov94} in the Dorokhov's\cite{Dorokhov}
bimodal distribution function ${\cal P}(T) = G_D/2T\sqrt{1-T}$
for the transmission coefficients of a diffusive system.

It might be surprising to realize equivalence of the two approaches
since Nazarov's result is valid for any geometry while our FRG
equation apparently holds only in 2D.
Such a coincidence is related to the fact that both Nazarov's derivation
and our derivation of the functional RG equation (\ref{urg})
implied the {\em zero-energy limit},
$T, eV, \omega < \ETh$.
Under this condition it is possible to derive the universal
distribution of the eigenvalues of the transmission matrix
as well as to neglect Cooperon decoherence in deriving Eq.~(\ref{urg}).
Note further that
summation of the tree-like diagrams for the action can be performed
without the use of the RG at all: one has to integrate diffusion
propagators over momentum independently for each branch of the tree.
But in the zero-energy limit, such integration is expressed in terms
of the Green function of the Laplace operator, thus reducing to the total
resistance of the system. Therefore, in this case Eq.~(\ref{urg})
rewritten in terms of $t=G_T/G_D$ remains valid for any geometry of the
disordered normal region.

Power of the Keldysh multicharge Proximity Action will be
crucial if one has to go beyond the zero-energy limit
and, especially, for the problems with interaction.
The corresponding FRG equations replacing the zero-energy-limit Euler
equation (\ref{urg}) will be derived in Secs.~\ref{S:beyond} and
\ref{S:lambda}.

\section{Physical quantities}
\label{S:phys}

Now we will discuss how the action (\ref{prox})
together with the FRG equation (\ref{urg})
valid in the zero-energy limit can be used to
calculate physically observable quantities.
In this manuscript we will focus on the conducting properties
of the system between the island and the lead.

Once the Proximity Action $S_{\rm prox}[Q_S]$ is known,
the system conductance as well as current statistics
can be calculated following the method of Ref.~\cite{FLS}.
We will suppose that the island is connected to the voltage source
by an ideal conductor so that phase fluctuations are suppressed
(the effect of phase fluctuations will be studied elsewhere\cite{qpt}).
In this case the general expression
$Z=\int D\varphi_1 D\varphi_2 \exp\{iS_{\rm prox}[\varphi_1,\varphi_2]\}$
for the Keldysh partition function reduces to
$Z=\exp\{iS_{\rm prox}[\varphi_{s1},\varphi_{s2}]\}$
where $\roarrow\varphi_s$ is the island's phase controlled by the source.
Dynamics of its classical component is governed by the applied voltage
through the Josephson relation:
\be
  \frac{d\varphi_{s1}}{dt} = 2eV(t) ,
\ee
while the current operator is given by the derivative with respect
to the quantum component of the phase:
\be
  \hat{I}(t) = ie \frac{\delta}{\delta\varphi_{s2}(t)} .
\label{I}
\ee
In order to obtain the expectation value of currents taken at different
moments of time one should apply the operators (\ref{I}) to $\ln Z$:
\be
  \corr{I(t_1) \dots I(t_k)}
  = \hat{I}(t_1) \dots \hat{I}(t_k) \ln Z[\varphi_{s1},\varphi_{s2}]
    \bigr|_{\varphi_{s2}=0} .
\ee

Below we will employ these relations to calculate the conductance
and noise for the case of the normal and superconductive islands.

\subsection{Normal junction: conductance and noise}
\label{SS:N}

To illustrate the formalism, we start by considering the conductance
of the system for the case of the normal island. We will trace how
addition of resistances $R_T$ and $R_D$ is realized within the multicharge
action approach and calculate the current noise power.

The current response $\corr{I(t)} = \sum_{n=1}^\infty \corr{I(t)}_n$
to the biased voltage $V(t)$ is given by the sum of the
partial contributions arising from the terms of the action (\ref{prox-n}):
\be
  \corr{I(t)}_n = 2ie\pi^2g\, \gamma_n
    \, \frac{\delta
      \Trk(e^{i\tensor\varphi/2} \Lambda_0 e^{-i\tensor\varphi/2} \Lambda_0)^n
    }{\delta\varphi_{2}(t)}
  \biggr|_{\varphi_{2}=0}
\label{ii1}
\ee
(here and below we omit the source phase subscript `s').
Now using Eq.~(\ref{A:B:1n}) we calculate the current $I=(e^2/\hbar) G\,V$,
where the dimensionless conductance $G$ is given by
\be
  G = 4 \pi g \sum_{n=1}^\infty n^2 \, \gamma_{n} .
\label{Gng}
\ee
In terms of the function $u(x)$ it reads
\be
  G = 4\pi g\, u_x (0).
\label{Gnu}
\ee

According to the Euler equation (\ref{urg}), $G$ obeys
\be
  \frac{\partial(1/G)}{\partial\zeta} = \frac{1}{4\pi g} ,
\label{1/Gn}
\ee
that amounts to the anticipated addition of resistances:
\be
  G = \frac{G_T}{1+t} = \frac{G_TG_D}{G_T+G_D} ,
\label{G}
\ee
where
\be
  G_D = \frac{2\pi g}{\ln(L/d)}
\label{GN}
\ee
is the metal film conductance between the island and the lead.

Eq.~(\ref{G}), though derived in the low-frequency limit, $\omega\ll D/L^2$,
can be shown to remain valid for much larger frequencies,
up to $\omega_{\rm max} = 2\pi\sigma/L$ (for the bare Coulomb interaction)
or $\omega_{\rm max} = 4\pi\sigma/b$ (if the
Coulomb interaction is screened by the presence of a gate situated at the
distance $b$ from the film, see Sec.~\ref{S:sigma}).
This frequency determines the time scale $\omega_{\rm max}^{-1}$ at which
the electroneutrality of the system sets in.

After this lesson let us turn to a more involved calculation
of the current-current correlator at a fixed constant bias $V$.
Now one has to apply two current operators to the action (\ref{prox-n}).
Similarly to the above consideration we present the Fourier transformed
correlator as $\corr{I_{\omega}I_{-\omega}} =
\sum_{n=1}^\infty \corr{I_{\omega}I_{-\omega}}_n$ with
\be
  \corr{I_{\omega}I_{-\omega}}_n = -2e^2\pi^2g\, \gamma_n
  \frac{\delta^2
    \Trk(e^{i\tensor\varphi/2} \Lambda_0 e^{-i\tensor\varphi/2} \Lambda_0)^n
  }{\delta\varphi_2(\omega)\, \delta\varphi_2(-\omega)}
  \biggr|_{\varphi_{2}=0} \! .
\label{ii2}
\ee
The corresponding functional derivative is calculated in Appendix~\ref{A:B}.
Substituting it from Eq.~(\ref{A:B:2n}) and employing Eq.~(\ref{Gng})
one obtains
\bea
  \corr{I_{\omega}I_{-\omega}} &=&
  \frac{e^2G}{3\hbar} \,
  \Bigl\{
    (3-P_N(t)) \, \Psi(\omega)
\nonumber \\
  {} &+&
    \frac12 P_N(t) \, [\Psi(\omega-eV) + \Psi(\omega+eV)]
  \Bigr\} ,
\label{IIn}
\eea
with the function $\Psi(\omega)$ defined as
\be
  \Psi(\omega) \equiv \int_0^\infty [1 - F(E_+)F(E_-)] dE
    = \omega \coth \frac{\omega}{2T} ,
\label{psi}
\ee
where $E_\pm=E\pm\omega/2$.
The function $P_N(t)$ is given by
\be
  P_N(t) =
    1 + \frac{2 \sum_{n=1}^\infty n^4 \gamma_n}{\sum_{n=1}^\infty n^2 \gamma_n}
    = 1 - \frac{2u_{xxx}(0)}{u_x(0)}.
\ee

The third derivative, $u_{xxx}(0)$,
can be easily calculated with the help of characteristics.
The one reaching the point $x=0$ is obviously $x(\zeta)=0$
with $x_0=0$. Writing $u(x)=u_0(x_0)$ and using the  relation
$\partial_x = [1+t\cos x_0]^{-1} \partial_{x_0}$ at $x_0=0$
we obtain
\be
  u_{xxx}(0)
  = -\frac{a}{(1+t)^4}
  = -\frac{1}{4\pi g} \frac{G^4}{G_T^3},
\label{uxxxn}
\ee
and, therefore,
\be
  P_N(t) = 1 + 2 \frac{G^3}{G_T^3}.
\label{PN}
\ee

Eqs.~(\ref{IIn}) and (\ref{PN}) provide a general description
of the noise power for arbitrary relation between the frequency
(which should be smaller than $\ETh$),
voltage and temperature, and arbitrary $R_T/R_D$.
Though it was derived in the zero-energy limit,
the condition $(T,eV)<\ETh$ can be relaxed since
for a normal island, charge propagation in the diffusive conductor
is described in terms of diffusons which are insensitive to decoherence.

In the limit $V\to0$, Eq.~(\ref{IIn}) reduces
to the Nyquist-Johnson equilibrium thermal noise:
\be
  \corr{I_{\omega}I_{-\omega}}_{\rm Nyq}
  = \frac{e^2G}{\hbar} \, \omega \coth \frac{\omega}{2T} .
\label{Nyquist}
\ee
In the limit $eV \gg (\omega, T)$, one obtains for the shot noise power:
\be
  \corr{II}_{\rm shot}
  = \frac{e^2G}{3\hbar} P_N(t) \, eV ,
\ee
that coincides with Nazarov's
result~\cite{Nazarov94} obtained via
the distribution function of transmission coefficients.
The function $P_N(t)$ is plotted in Fig.~\ref{F:P(t)} by the dashed line.
For $t\gg1$, when the system resistance is dominated by the diffusive
conductor, the shot noise power is three times smaller than its Poisson value.
This result was first obtained in Ref.~\cite{1/3}
with the help of the Dorokhov's bimodal distribution\cite{Dorokhov},
and in Ref.~\cite{Nagaev}, in the framework of the classical
Boltzmann equation with Langevin sources.

The Proximity Action (\ref{prox}) together with the FRG equation
(\ref{urg}) allows, in principle, for the calculation of higher momenta of
current fluctuations and even the full statistics of transmitted
charge (cf.\ Refs.~\cite{LeviLes,ALYa}).
We leave this problem for future investigation.

\subsection{Andreev conductance}
\label{SS:cond}

In this section we will turn to the case of a superconductive island.
We will assume that the island is not very small so that its order
parameter satisfies $\Delta>\omega_d$. This means that at the scales
$\Omega<\omega_d$ relevant for RG, quasiparticle reflection from the
island is of Andreev type and one can use the subgap expression
(\ref{QS}) for the island's $Q_S$.

Similarly to the N-island case, we write the current as a sum
of the partial contributions $\corr{I(t)}_n$ from different
terms of the action (\ref{prox-s}):
\be
  \corr{I(t)}_{n} = (-1)^n 2ie\pi^2g\, \gamma_{2n}
    \, \frac{\delta
      \Trk(e^{i\tensor\varphi} \Lambda_0 e^{-i\tensor\varphi} \Lambda_0)^n
    }{\delta\varphi_{2}(t)}
  \biggr|_{\varphi_{2}=0} .
\label{ii3}
\ee
Calculating the derivative with the help of Eq.~(\ref{A:B:1s}) we
find for the dimensionless Andreev conductance:
\be
  G_A = 16 \pi g \sum_{n=1}^\infty (-1)^{n} n^2 \, \gamma_{2n} ,
\label{Gsn}
\ee
or, in terms of $u(x)$:
\be
  G_A = 4\pi g\, u_x (\pp) .
\label{Gsu}
\ee
Note that Eqs.~(\ref{Gnu}) and (\ref{Gsu}) for the case of the normal
and superconducting island look very similar. The only difference
is at the point where the derivative should be taken.

Now we will use the general expression (\ref{Gsu}) to calculate $G_A$
in the noninteracting case described by Eq.~(\ref{urg}).
The simplest way to calculate $u_x(\pp)$ is to use characteristics
of the Euler equation. To do this, one has to find the one which leads
to $x=\pi/2$ at ``time'' $t$. Writing its initial point as
$x_0=\pi/2-\Theta(t)$ we obtain the following equation for the function
$\Theta(t)$:
\be
  \Theta(t) = t \cos \Theta(t) .
\label{Theta-def}
\ee
For small $t$, $\Theta(t\ll1) = t + O(t^3)$ while in the limit
of large $t$, $\Theta(t\gg1) = \pi/2 - \pi/2t + O(t^{-2})$.
Now, applying $\partial_x = [1+t\cos x_0]^{-1} \partial_{x_0}$
to $u(x)=u_0(x_0)$ we obtain
\be
  u_x(\pp) = \frac{a\sin\Theta(t)}{1+t\sin\Theta(t)}
\label{uxs}
\ee
and, finally,
\be
  G_A
  = G_T \frac{\sin\Theta(t)}{1+t\sin\Theta(t)}
  = G_D \frac{t\sin\Theta(t)}{1+t\sin\Theta(t)} ,
\label{GA}
\ee
where $G_D$ is defined in Eq.~(\ref{GN}).

In terms of the resistance, the zero-energy result (\ref{GA}) can be written as
\be
  R_A = R_D + R_{T,\rm eff},
\label{R+R}
\ee
where $R_{T,\rm eff}(t) = R_T/\sin\Theta(t)$ has the meaning of an
effective interface resistance connected in series with
the metal-film resistance $R_D = (\hbar/e^2)G_D^{-1}$.
The former decreases with the increase of $R_D$ due to disorder-induced
enhancement of Andreev reflection determined by the probability
of returning to the origin.

Equation (\ref{GA}) for the Andreev conductance had been
previously derived by Nazarov both with the help of the Usadel
equation\cite{NazarovC} and the distribution function $F(\phi)$
of transmission eigenvalues\cite{Nazarov94}.
His latter approach, though formulated in a very different language,
appears to be directly connected to our multicharge action approach.
Using the relation between $F(\phi)$ and $u(x)$ established
in Eq.~(\ref{F}), one can verify that Nazarov's result\cite{Nazarov94}
$G_A=(\partial F/\partial\phi)|_{\phi=\pi/2}$ exactly
coincides with our Eq.~(\ref{Gsu}).

\subsection{Noise of NS current}
\label{SS:noise}

Decomposing the current-current correlator
into the sum of the partial contributions,
$\corr{I_{\omega}I_{-\omega}}_{n}$, similarly to Eqs.~(\ref{ii3}) and
(\ref{ii2}), using Eq.~(\ref{A:B:2s}), and employing Eq.~(\ref{Gsn})
we obtain
\bea
  \corr{I_{\omega}I_{-\omega}} &=&
  \frac{e^2G_A}{3\hbar} \,
  \Bigl\{
    (3-P_S(t)) \, \Psi(\omega)
\nonumber \\
  {} &+&
    \frac12 P_S(t) \, [\Psi(\omega-2eV) + \Psi(\omega+2eV)]
  \Bigr\} ,
\label{IIs}
\eea
where $\Psi(\omega)$ is defined in Eq.~(\ref{psi}).
The superconductive noise function $P_S(t)$ is given by
\be
  P_S(t) =
    1 + \frac{2 \sum_{n=1}^\infty (-1)^n n^4 \gamma_{2n}}
      {\sum_{n=1}^\infty (-1)^n n^2 \gamma_{2n}}
    = 1 - \frac{u_{xxx}(\pp)}{2u_x(\pp)}.
\label{PS0}
\ee
Eq.~(\ref{IIs}) has the same form as its analogue for the N island
with the replacement $P_N(t) \to P_S(t)$ and $eV \to 2eV$.

\begin{figure}
\refstepcounter{figure} \label{F:P(t)}
\epsfxsize=80mm
\centerline{\epsfbox{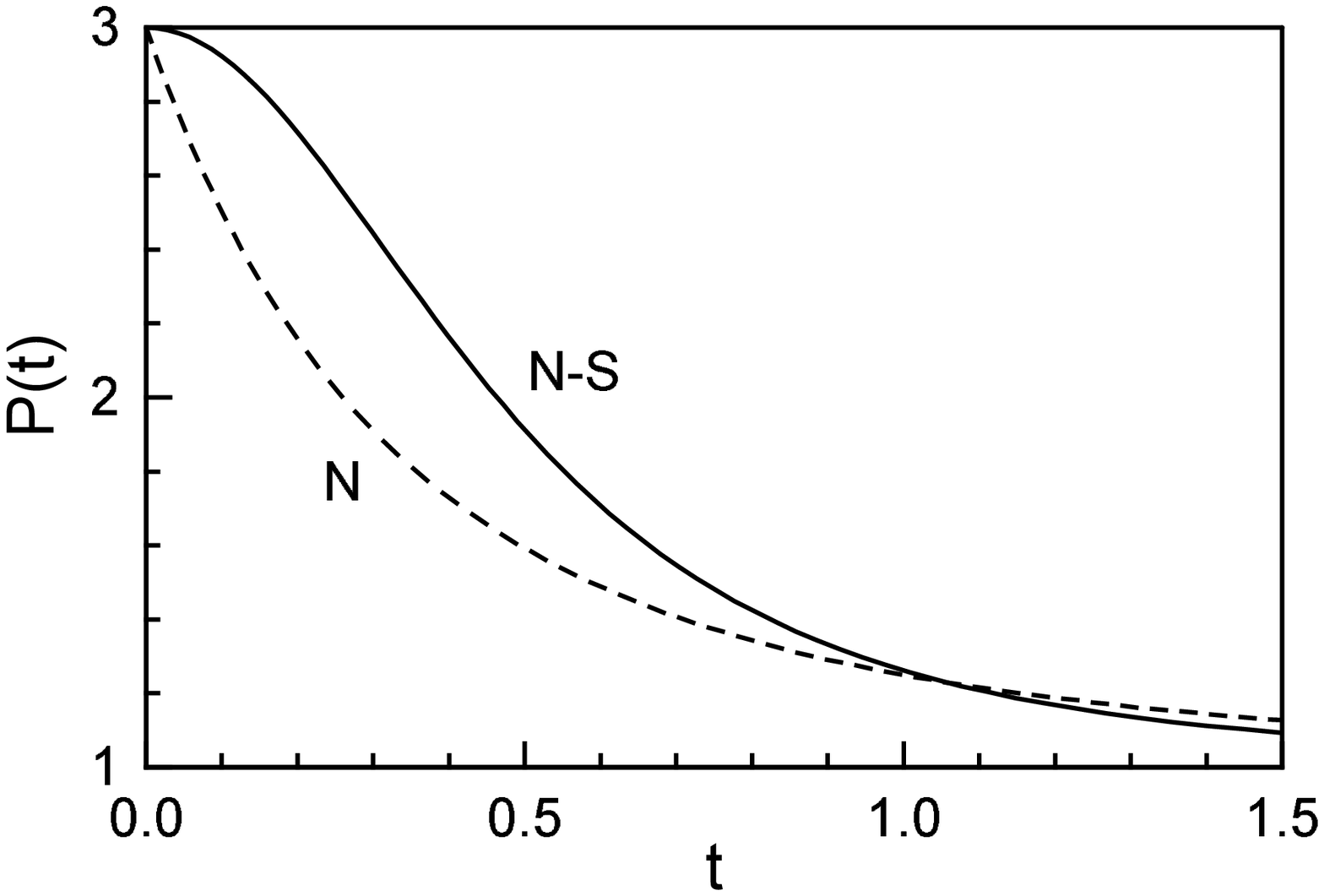}}
\vspace{2mm}
\small FIG.\ \arabic{figure}.
Noise functions $P_N(t)$ (dashed line) and $P_S(t)$ (solid line)
for normal and N-S junctions, correspondingly, vs.\
the ratio $t=G_T/G_D$.
\end{figure}

For the noninteracting system, the value of $u_x(\pp)$ was calculated in
Eq.~(\ref{uxs}), and calculating $u_{xxx}(\pp)$ in the same way we obtain
\be
  u_{xxx}(\pp) =
  - \frac{a\sin\Theta(t)}{(1+t\sin\Theta(t))^4}
  - \frac{3at\cos^2\Theta(t)}{(1+t\sin\Theta(t))^5} .
\label{uxxxs}
\ee
Thus, we get
\be
  P_S(t) = 1 +
  \frac{1+\Theta\tan\Theta+3\Theta\cot\Theta}{2(1+\Theta\tan\Theta)^4} ,
\label{PS}
\ee
where $\Theta\equiv\Theta(t)$ is defined in Eq.~(\ref{Theta-def}).

In the limit of vanishing bias,
Eq.~(\ref{IIs}) with $P_S(t)$ from Eq.~(\ref{PS})
reproduces the Nyquist-Johnson thermal noise given by Eq.~(\ref{Nyquist}),
with $G$ being substituted by $G_A$, Eq.~(\ref{GA}).
In the opposite limit of large bias, $eV \gg (\omega, T)$,
it gives the power of the shot noise between
the normal and superconducting terminals~\cite{DeJong}:
\be
  \corr{II}_{\rm shot}
  = \frac{2e^2G_A}{3\hbar} P_S(t) \, eV .
\label{sshot}
\ee
The function $P_S(t)$ is plotted in Fig.~\ref{F:P(t)}
together with its analogue for the normal junction $P_N(t)$.
On increasing $t$, both $P_N(t)/3$ and $P_S(t)/3$ evolve
from the value of 1 describing the Poisson noise at a tunnel contact
to $1/3$ corresponding to the diffusive conductor.
An excess factor of 2 in Eq.~(\ref{sshot}) accounts the
fact that in an N-S system charge is transferred by Cooper pairs.

Results obtained in Secs.~\ref{SS:cond} and \ref{SS:noise} refer
to the zero-energy limit.  In Sec.~\ref{S:beyond} we will extend our theory
to the case of large temperature or voltage, $\max(T,eV)>\ETh$.

\section{Effect of interaction in the Cooper channel}
\label{S:lambda}

\subsection{General idea}
\label{SS:lambda-gen}

Here we calculate the effect of repulsion in the Cooper channel
on the renormalization of the effective boundary action (\ref{sg}).
In the lowest order over the interface transparency, the RG equation
for $\gamma_1$ was derived in Ref.~\cite{FLS}. We will extend this
result to account the whole set of charges $\{\gamma_n\}$.

\begin{figure}
\refstepcounter{figure} \label{F:4l1}
\epsfxsize=3.375in
\noindent
\centerline{\epsfbox{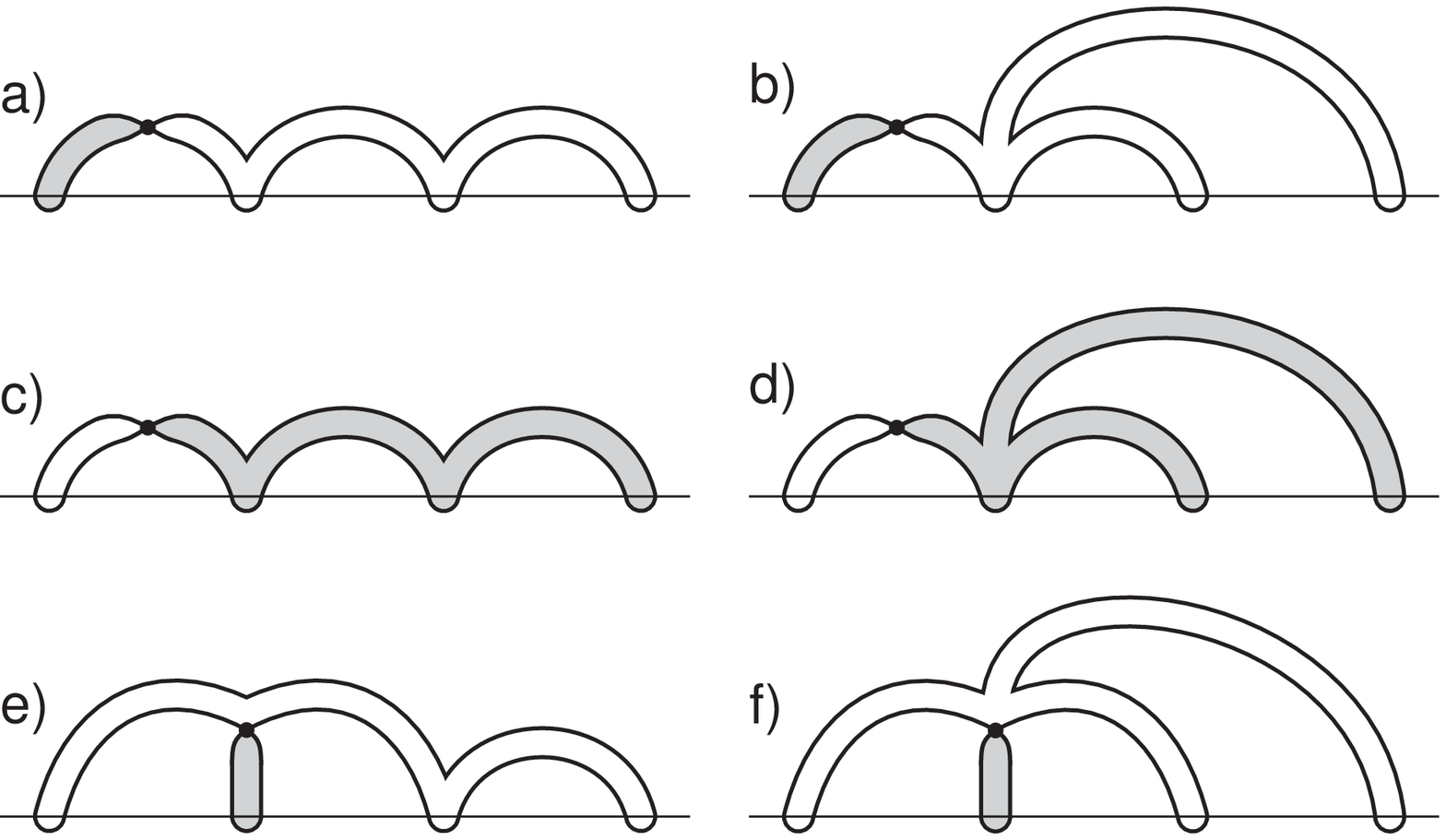}}
\par
\vspace{3mm}
\small FIG.\ \arabic{figure}.
Leading diagrams in the forth order in $\gamma$ and the first order
in $\lambda$. Shadowed is the piece of the diagram
where the logarithmic contribution of the Cooperon adjacent
to the $\lambda$-vertex results from the energy integration.
\end{figure}

As in Sec.~\ref{S:FRG}, we first turn to the analysis of the lowest
order perturbative corrections to the action. The Cooper channel
interaction is described by the following vertex in the bulk action:
\be
  S_\lambda
  = \frac{\pi^2\nu\lambda}4
    \int d\br\, dt \tr \sigma_x [ Q^2 - (\tau_z Q)^2 ] ,
\label{S_lambda}
\ee
obtained from the $\sigma$-model action (\ref{sm}) by eliminating
the fluctuating $\roarrow\Delta$ field~\cite{FLS}.
The tree-like diagrams for the action of noninteracting system
should be modified now by all possible insertions of the vertexes $S_\lambda$.
In the lowest order in the interface transparency $\gamma$
(cf.\ Fig.~\ref{F:2}), that was carried out in Ref.~\cite{FLS}.
The diagrams of the forth order in $\gamma$ and the first order in
$\lambda$ are shown in Fig.~\ref{F:4l1}. They are obtained from
the diagrams of Fig.~\ref{F:4} by insertions of the Cooper
vertex denoted by a dot. The latter cuts a diagram into two
parts with independent energy integration variables.

The leading logarithmic contribution of a diagram from Fig.~\ref{F:4l1}
is due to the {\em momentum} integration in each diffusion mode
except for one of the Cooperons adjacent to the vertex $S_\lambda$,
whose contribution is logarithmic due to the {\em energy} integral $\int dE/E$.
The corresponding part of the diagram containing such a Cooperon is
shadowed in Fig.~\ref{F:4l1}. Thus, each of the
diagrams shown in Fig.~\ref{F:4l1} evaluates to the forth power
of the logarithm, $\propto \lambda\gamma^4\zeta^4$.

To achieve this situation, the energy $E$ and momenta in the
shadowed part of the diagram must obey some inequalities.
First, the momentum of the Cooperon adjacent to the
$\lambda$-vertex, $q$, must satisfy the relation $Dq^2<E$.
Second, momenta of all other shadowed Cooperons, $q_i$, must
satisfy the relation $Dq_i^2>E$. Only under these conditions the
diagram evaluates to the maximal power of the logarithmic variable $\zeta$.

Were all diffusion modes in the shadowed region of the diagram Cooperons,
the energy $E$ would be the slowest variable in this region.
Then, on the level of the RG, the shadowed region would be taken into account
by the pairing of the vertex $S_\lambda$ with the multicharge action
$S_\Gamma$ [Eq.~(\ref{sg})] obtained at the previous steps of the RG procedure.
Since $E$ would be the slowest variable in the shadowed part,
all metallic $Q$'s entering the action $S_\Gamma$ paired to $S_\lambda$
should be set to $\Lambda$ indicating that no further pairings with slower
variables are allowed to the shadowed region.
As a result, as will be shown in Sec.~\ref{SS:lambda-der},
the shadowed region will give a correction to the term
$S_1 \propto \gamma_1 \Trg Q_SQ$.
However, the shadowed region might contain diffusons as well.
Their momenta $q_i$ are not restricted by the inequality $Dq_i^2>E$
and may be ``slower'' than the energy $E$.
Were this the case, the multicharge action (\ref{sg}) would not
reproduce itself upon pairing with $S_\lambda$
since one would not be able to perform the integral over $E$
as the result would depend on next steps of the RG.

Fortunately, diffuson contribution in the shadowed part of a diagram
can be disregarded. This can be seen already from the diagrams
of the fourth order shown in Fig.~\ref{F:4l1}.
Among them only the diagram (c) contains a shadowed diffuson.
However, direct calculation shows that its contribution to the action
vanishes identically~\cite{why?}.
This statement can be generalized to arbitrary order.
One can prove that any diagram containing a diffuson in the shadowed
region is exactly zero. Physically, the shadowed part of a diagram accounts
the correction to the Usadel spectral angle $\theta$
(cf.\ Appendix~\ref{A:C}) which is determined solely by Cooperon modes.

\subsection{Functional RG equation}
\label{SS:lambda-der}

To calculate the correction to the action due to the pairing
$S_{n,\lambda}=i\corr{S_{{\rm int},n}S_{{\rm int},\lambda}}$
one should extract one fast $W'$ from $S_\lambda$:
\be
  S_{{\rm int},\lambda} =
    \frac{\pi^2\nu\lambda}4
    \Tr \{ \sigma_x, \tilde Q \} \Lambda (W' - \tau_z W' \tau_z) ,
\ee
and another one from $S_n$ according to Eq.~(\ref{S_n}).
The only fast variable to be integrated out at this step of the RG
procedure is the energy $E$ running over the Cooperon propagator
$\corr{W'W'}$. This fast energy runs also over all $\tilde Q$'s
under the trace in $S_{{\rm int},n}$.
Substituting $\tilde Q = \Lambda(E)$ into Eq.~(\ref{S_n}) as explained above
and performing the averaging over the fast Cooperon in
$\corr{S_{{\rm int},n}S_{{\rm int},\lambda}}$
with the use of Eq.~(\ref{WW}), we obtain that only the terms with odd $n$
give a nonzero contribution to the action
(hereafter the tilde sign over slow variables is omitted):
\bea
  S_{n,\lambda} = &&
  (-1)^{(n-1)/2} \frac{i \pi^3 g n\gamma_n \lambda}{4}
  \int_\Gamma d\br
  \int \frac{dt\, dE'\, dE}{(2\pi)^2\,E}
\nonumber \\
  &&
  \times \tr \{ (Q_S \Lambda_0(E))^n + (\Lambda_0(E) Q_S)^n, \sigma_x \} \, Q .
\label{S_nl}
\eea
Since $E$ is a fast variable, all $Q_S(t)$ are taken at the same time $t$,
and $Q=\int e^{i\omega t} Q_{E'-\omega/2,E'+\omega/2} (d\omega/2\pi)$.
The sign in the above equation comes from commutation of $\tau_z$
contained in $\Lambda=\Lambda_0\tau_z$ with $Q_S$ in order to transform
$(Q_S\Lambda)^n$ to $(Q_S\Lambda_0)^n$.

Let us calculate
\be
  \ac_n = \{ (Q_S \Lambda_0)^n + (\Lambda_0 Q_S)^n, \sigma_x \} ,
\label{ac} 
\ee
which enters Eq.~(\ref{S_nl}).
It can be written as $\ac_n = \{ X^n + X^{-n}, \sigma_x \}$,
where $X=Q_S \Lambda_0$.
According to Eq.~(\ref{QS}), $Q_S$ contains
the classical and quantum parts: $Q_S=p+q\sigma_x$.
Then it is easy to show that $\ac_1 = 4 Q_S F(E)$.
Now we will show by induction that for all odd $n$, $\ac_n=\ac_1$.
Consider the difference $\ac_{n+2}-\ac_n$.
It contains
$X^{n+2}-X^{n}-X^{-n}+X^{-n-2}
= (X^{n+1}-X^{-n-1})(X-X^{-1})
= {\cal F}(X) (X-X^{-1})^2$.
Within the RG accuracy,
the distribution function $F(E)=\sgn E$ so that $F^2=1$.
Under this condition,
$(X-X^{-1})^2 \equiv [Q_S,\Lambda_0]^2=0$
that proves the statement $\ac_n=\ac_1$.

Substituting
\be
   \{ (Q_S \Lambda_0(E))^n + (\Lambda_0(E) Q_S)^n,  \sigma_x \} = 4 Q_S \sgn E
\ee
into Eq.~(\ref{S_nl}), we get
\be
  S_{n,\lambda} =
  (-1)^{(n-1)/2} i \pi^3 g n\gamma_n \lambda \Delta\zeta \Trg Q_S Q.
\ee
Therefore, only $S_1$ is subject to renormalization by the $\lambda$-term:
\be
  \left(
    \frac{d\gamma_1}{d\zeta}
  \right)_{\mbox{\scriptsize due to $\lambda$}} =
  - \lambda \sum_{l=0}^\infty (-1)^l (2l+1)\gamma_{2l+1} .
\label{g1lambda}
\ee
As a result, an additional term will appear in the RHS of the Euler
equation~(\ref{urg}):
\be
  u_\zeta + u u_x = - \lambda(\zeta)\, u(\pp) \sin x .
\label{urgl}
\ee

Here $\lambda(\zeta)$ is described by the bulk RG equation (\ref{lambda-rg})
(tunnel coupling between the superconductive island and the film,
though induces finite $\Delta$ in the bulk, does not have logarithmic
effects on the Cooper interaction).
Its solution~\cite{finkel2,FLS} is given by
\be
\label{lambda-res}
  \lambda(\zeta) =
    \frac{\lambda_d+\lambda_g\tanh\lambda_g \zeta}
      {\displaystyle 1+\frac{\lambda_d}{\lambda_g}\tanh\lambda_g \zeta} ,
\ee
with $\lambda_d$ defined at the energy scale $\omega_d$
(cf.\ the definition (\ref{zeta}) of the logarithmic variable $\zeta$):
\be
  \lambda_d =
    \frac{\displaystyle
       \lambda_n+\lambda_g\tanh \left(\lambda_g \ln\frac1{\omega_d\tau}\right)}
      {\displaystyle 1+
  \frac{\lambda_n}{\lambda_g}\tanh
    \left(\lambda_g \ln\frac1{\omega_d\tau}\right) } ,
\label{lambda_d}
\ee
and $\lambda_n$ is the interaction constant at the energy scale $\hbar/\tau$.

\subsection{Andreev conductance}
\label{SS:lambda-cond}

Here we consider the effect of repulsive interaction in the Cooper channel
of a normal conductor on the Andreev conductance, in two limiting cases
of weak and strong interaction. Of course, we always assume that
dimensionless coupling constant $\lambda \ll 1$.
We will see below that the limit of strong interaction
means $\lambda \gg G_T/4\pi g$, and thus can be realized at
relatively poor transparency of the interface.
However, we will see also that even the weak-$\lambda$ solution produces
the interaction-induced correction to $G_A$ which grows logarithmically
with the spatial scale $L$ and eventually crosses over
to the strong-coupling limit.

\subsubsection{First order correction}
\label{SSS:first}

In the limit of small $\lambda(\zeta)$, it can be taken into
account perturbatively.
The equations for the characteristics of Eq.~(\ref{urgl}) have the form:
\be
  \frac{dx}{d\zeta} = u ,
  \qquad
  \frac{du}{d\zeta} = - \lambda(\zeta)\, u(\pp,\zeta) \sin x .
\label{char}
\ee
The presence of $u(\pp,\zeta)$ in Eq.~(\ref{char}) which
is determined by the trajectory reaching $x=\pi/2$ in ``time'' $\zeta$
introduces a nonlocal in $\zeta$ coupling between different trajectories.
In the lowest order over $\lambda(\zeta)$, we search for the solution
in the form $x(\zeta) = x_0 + a \zeta \sin x_0 + \delta x(\zeta)$ and
$u(\zeta) = a\sin x_0 + \delta u(\zeta)$,
where $\delta x$ and $\delta u$ are proportional to $\lambda$
and vanish at $\zeta=0$. Substituting this into Eqs.~(\ref{char})
and keeping only terms linear in $\lambda$, we obtain
\end{multicols}
\top
\be
  \frac{d\delta x}{d\zeta} = \delta u ,
  \qquad
  \frac{d\delta u}{d\zeta}
  = - \lambda(\zeta)\, a \cos [\Theta(a\zeta)] \sin [x_0 + a\zeta\sin(x_0) ] .
\ee
These equations can be easily integrated as
\bea
  \delta x(\zeta) =
  - a \int_0^\zeta (\zeta-\eta)
    \lambda(\eta) \cos [\Theta(a\eta)] \sin [x_0 + a\eta\sin x_0 ] d\eta,
\label{char-dx}
\\
  \delta u(\zeta) =
  - a \int_0^\zeta
    \lambda(\eta) \cos [\Theta(a\eta)] \sin [x_0 + a\eta\sin x_0 ] d\eta.
\label{char-du}
\eea

Thereby, we know the characteristics of Eq.~(\ref{urgl})
in the first order over $\lambda(\zeta)$.
As we have already seen in Sec.~\ref{SS:cond}, in order to
calculate the Andreev conductance $G_A=4\pi g u_x(\pp)$, one has to
search for the characteristic which leads to $x=\pi/2$ in ``time'' $\zeta$.
Resolving equation $x(\zeta)=\pi/2$ we find that the initial point of
such a trajectory, $x_0 = \pi/2 - \Theta(a\zeta) + \delta x_0$,
gets shifted by
\be
  \delta x_0 =
    \frac{a}{1+a\zeta\sin\Theta(a\zeta)}
    \int_0^\zeta (\zeta-\eta)
    \lambda(\eta) \cos [\Theta(a\eta)]
    \cos \left[
      \Theta(a\zeta) \left( 1 - \frac{\eta}{\zeta} \right)
    \right] d\eta
\label{x0l}
\ee
compared to the noninteracting case.
A nonzero $\lambda$ influences the derivative
$u_x(\pp) = (\partial u/\partial x_0)/(\partial x/\partial x_0)$
which determines the Andreev conductance in two ways.
The first correction, $\delta^{(1)}u_x(\pp)$,
to the quasiclassical result (\ref{uxs})
is related to the change of the initial point $x_0$ of the trajectory.
The second correction, $\delta^{(2)}u_x(\pp)$,
is due to the modification of the functional
dependences of $x(\zeta)$ and $u(\zeta)$ versus $x_0$
by the terms (\ref{char-dx}) and (\ref{char-du}).
A straightforward calculation yields
\be
  \delta^{(1)}u_x(\pp)
  = - \frac{a\cos\Theta(a\zeta)} {[1+a\zeta\sin\Theta(a\zeta)]^2} \delta x_0
\label{du1}
\ee
and
\be
  \delta^{(2)}u_x(\pp)
  = - \frac{a} {[1+a\zeta\sin\Theta(a\zeta)]^2}
    \int_0^\zeta
    \lambda(\eta) \cos [\Theta(a\eta)]
    \cos \left[
      \Theta(a\zeta) \left( 1 - \frac{\eta}{\zeta} \right)
    \right]
    [1+a\eta\sin\Theta(a\zeta)]^2
    d\eta.
\label{du2}
\ee

Finally, with the help of Eq.~(\ref{Gsu}) we obtain for the Andreev
conductance in the lowest order over the Cooper-channel interaction
$\lambda(\zeta)$:
\bea
  \frac{G_A}{G_D}
  = \frac{t\sin\Theta(t)}{1+t\sin\Theta(t)}
  - && \frac{\Theta(t)} {[1+t\sin\Theta(t)]^3} \,
    \zeta
    \int_0^1 \lambda(x\zeta) \, (1-x)\, \frac{\Theta(xt)}{x}
    \cos[\Theta(t) (1-x)] \,
    dx
\nonumber \\
  {}
  - && \zeta
    \int_0^1 \lambda(x\zeta)
    \, \frac{[1+xt\sin\Theta(t)]^2}{[1+t\sin\Theta(t)]^2} \,
    \frac{\Theta(xt)}{x} \sin[\Theta(t) (1-x)] \,
    dx .
\label{dGa}
\eea
\bottom
\begin{multicols}{2}

It is interesting to compare Eq.~(\ref{dGa}) with the result of the
standard (cf., e.~g., Ref.~\cite{StNaz}) calculation of $G_A$ by means of
direct solution of the Usadel equations, presented in Appendix~\ref{A:C}.
While both Eqs.~(\ref{dGa}) and (\ref{ga/gn}) contain corrections linear
in $\lambda$, their meaning is somewhat different.
Whereas Eq.~(\ref{dGa}) contains, in general, the renormalized
coupling constant $\lambda(\zeta)$ which depends on the energy
scale through Eq.~(\ref{lambda-res}), the result (\ref{ga/gn}) is
expressed in terms of the bare coupling constant $\lambda_d$ defined
at the scale $\omega_d$, cf.\ Eq.~(\ref{lambda_d}).
Note that the Usadel equation contains, in principle,
logarithmic dependence of $\lambda(\zeta)$ which
follows from the first term of Eq.~(\ref{lambda-rg}):
$\lambda(\zeta) = \lambda_d/ (1 + \lambda_d \zeta)$.
It emerges as a result of the selfconsistent determination of the
proximity-induced $\Delta$ in the normal metal, cf.\ Eq.~(\ref{delta}).
However, in order to take these effects into account within the
approach presented in Appendix~\ref{A:C}, it would be necessary
to solve the whole problem with higher-order accuracy in $\lambda_d$,
that seems to be an extremely complicated task.
The main complication is the appearance of the supercurrent flowing
in the normal region that leads to the rotation of the phase
$\varphi({\bf r})$ of the
anomalous Green function and also mixes both components, $f$ and $f_1$,
of the distribution function.
As a consequence, the problem becomes hardly tractable.
Staying within first-order accuracy over $\lambda_d$,
we identify $\lambda(\zeta) \equiv \lambda_d$ and find exact agreement
between Eqs.~(\ref{dGa}) and (\ref{ga/gn}).
Moreover, two contributions to the correction to $G_A$
given by Eqs.~(\ref{du1}) and (\ref{du2})
have their direct counterparts within the Usadel method:
the first one is due to the modification of the spectral angle
$\theta(r=d)$ [cf.\ Eq.~(\ref{theta_d})],
whereas the second one is, physically, due to the correction
to the distribution function $f_1(r=L)$ by the presence of
the induced $\Delta$ [cf.\ Eq.~(\ref{I_N})].

Another principal limitation of the quasiclassical approach based on the
Usadel equation is that the latter does not allow for an account of
fluctuation corrections, such as, e.~g., Finkel'stein correction to
$\lambda(\zeta)$. This is related to the fact that the Usadel equations
are nothing more than the mean field approximation for the Keldysh
action (\ref{sm}). Therefore, our approach seems to be the only tool
for studing the effect of the Cooper channel interaction on the conducting
properties of 2D N-S systems.

A somewhat cumbersome expression (\ref{dGa}) can be simplified in the
limiting cases of small and large $t$.
For $t\ll1$, the noninteracting Andreev conductance
$G_A^{(0)} \approx G_T^2/G_D$,
and one has for the interaction correction $\delta G_A \equiv G_A-G_A^{(0)}$:
\begin{mathletters}
\label{dGA-lim}
\be
  \frac{\delta G_A}{G_A} = - 2 \zeta
    \int_0^1 \lambda(x\zeta) (1-x) \, dx.
\label{dGA-t<1}
\ee
In the opposite limit, $t\gg1$, the conductance is determined by the
diffusive conductor, $G_A^{(0)} \approx G_D$, and
\be
  \frac{\delta G_A}{G_A} = - \frac{\pi}{2} \zeta
    \int_0^1 \lambda(x\zeta) \, x \cos\frac{\pi x}{2} \, dx.
\label{dGA-t>1}
\ee
\end{mathletters}%
An important consequence of Eqs.~(\ref{dGA-lim}) is that the relative value of
the $\lambda$-correction to the subgap conductance is proportional to
$\lambda\zeta$ at any value of $G_T/G_D$, including the case of
a completely transparent interface with $G_T^{-1}=0$.
Thus, our result differs from the one obtained by Stoof and
Nazarov~\cite{StNaz} who considered the same kind of problem in the 1D
geometry and found a weak regular correction $\sim \lambda$ only.
The growth of the interaction-induced correction in 2D
indicates that qualitative changes in the behavior of the function $u(x)$
do occur at large enough scales $\zeta$ and one has to sum the leading
logarithmic terms, $(\lambda\zeta)^n$.

In the most general case of arbitrary $\lambda(\zeta)$,
the FRG Eq.~(\ref{urgl}) should be solved numerically.
However, in the limiting cases of {\em strong} and {\em weak}
Cooper interaction, an analytical solution can be obtained.
The former refers to the situation when the first correction,
(\ref{dGA-t<1}), becomes of the order of one already in the limit of weak
transparency of the barrier, $t\ll1$. The latter applies when the first
correction, (\ref{dGA-t>1}), becomes important only in the diffusive
regime, $t\gg1$.
The boundary between the limits of strong and weak repulsion in the
Cooper channel slightly depends on the relation between $\lambda_d$
and $\lambda_g$.
In the rest of this section we will consider the case when
$\lambda_d\sim\lambda_g$.
Then repulsion is strong (weak) provided that $\lambda_g\gg a$
($\lambda_g\ll a$).
We want to emphasize that the notion of strong/weak interaction refer
not to the value of $\lambda$ itself but to the ratio
$\lambda/a = 4\pi g\lambda/G_T$.
In both cases the interaction correction can be either small or large
depending on the value of $\lambda\zeta = 4\pi g\lambda/G_D$.

The limiting cases of strong and weak repulsion will be considered
Secs.~\ref{SSS:strong} and \ref{SSS:weak},
and then (in Sec.~\ref{SSS:arb}) we will turn to the discussion
of the general case of arbitrary $\lambda$.

Eqs.~(\ref{dGA-lim}) describe the interaction correction to the total
resistance of the system. It is instructive to study also the correction
to the effective interface resistance defined in Eq.~(\ref{R+R}).
In the limit of strong repulsion, this correction
is determined by Eq.~(\ref{dGA-t<1}).
In the limit of weak repulsion,
 $R_{T,\rm eff} = R_T/\sin\Theta(t) \approx R_T$
is only a small part of $R_A$ at $t\gg1$.
The $\lambda$ term is also a small correction if $t\ll a/\lambda_g$.
Expanding Eq.~(\ref{dGa}) in the region $1\ll t\ll a/\lambda_g$
and assuming for simplicity that
$\lambda(\zeta)=\lambda_g$, one obtains:
\be
  \frac{R_{T,\rm eff}}{R_T} \approx
  1 + \frac{\pi^2}{8t^2}
  + \left( 1 - \frac{2}{\pi} \right) \frac{\lambda_g}{a} t^2 ,
\label{Rl}
\ee
where the first two terms come from expansion of $1/\sin\Theta(t)$.
According to Eq.~(\ref{Rl}), $R_{T,\rm eff}(t)$ first decreases with $t$
up to $t\sim(a/\lambda_g)^{1/4}$ and then starts to increase. This increase
is a  qualitatively new feature appeared due to the interaction.
At the upper boundary of applicability of Eq.~(\ref{dGa}),
at $t\sim a/\lambda_g$, one has $R_{T,\rm eff}(t)/R_T \sim a/\lambda_g \gg1$.
Thus, even within the range where the first
correction is still small, weak Cooper repulsion changes the
dependence of the effective interface resistance
which starts to grow with the increase of $t$.
The behavior of $R_{T,\rm eff}(t)$
at $t\geq a/\lambda_g$ will be studied in Sec.~\ref{SSS:weak}.

\subsubsection{Strong repulsion}
\label{SSS:strong}

If the Cooper interaction is strong in the sense that
\be
  \lambda_g \gg a \equiv \frac{G_T}{4\pi g},
\label{cond-str}
\ee
then the initial stage of the evolution of $u(x)$ is better represented
in terms of the RG equations for the coefficients $\gamma_1$ and $\gamma_2$,
cf.\ Eqs.~(\ref{mrg}) and (\ref{g1lambda}):
\be
  \frac{d\gamma_1}{d\zeta} = -\lambda(\zeta) \gamma_1 ,
  \qquad
  \frac{d\gamma_2}{d\zeta} = -\frac14 \gamma_1^2 ,
\label{2-charge}
\ee
where $\lambda(\zeta)$ is given by Eq.~(\ref{lambda-res}).
Simple calculation gives
%
\begin{mathletters}
\label{two-ch-sol}
\bea
  \gamma_1(\zeta) &=&
  \frac{a\lambda_g}
    {\lambda_g\cosh\lambda_g\zeta + \lambda_d\sinh\lambda_g\zeta} ,
\\
  \gamma_2(\zeta) &=&
  - \frac{a^2/4}
    {\lambda_g\coth\lambda_g\zeta + \lambda_d} .
\eea
\end{mathletters}%
At $\zeta \gg \zeta_1 \equiv \lambda_g^{-1}$,
the first harmonics $\gamma_1$ vanishes,
while the second one saturates at
$\gamma_2^{(1)} = - a^2/4(\lambda_d+\lambda_g)$.
The corresponding subgap conductance is then given by
\be
  G_A^{(1)} = - 16\pi g \gamma_2^{(1)} =
  \frac{G_T^2}{4\pi g(\lambda_d+\lambda_g)} .
\label{GA1}
\ee
At the scale $\zeta \geq \zeta_1$ higher harmonics $\gamma_{n>2}$
are still much smaller than $\gamma_2$ once the condition (\ref{cond-str})
is fulfilled, so that
\be
  u(x) = - \tilde{a} \sin 2x , \qquad
  \tilde{a} = \frac{a^2}{2(\lambda_d+\lambda_g)} .
\label{u1_ini}
\ee

{} From now on further evolution of $u(x,\zeta)$ with the increase of $\zeta$
is given by the solution of the Euler equation (\ref{urg}) with the initial
condition (\ref{u1_ini})
[a somewhat similar situation will be discussed in
Sec.~\ref{SS:beyond-gen}].
Change of variables $u(x,\zeta) = v(y,\eta)$ with $y = 2x-\pi$ and
$\eta = 2\zeta$ makes the solution for $v(y,\eta)$ identical
(up to the replacement $ a\to \tilde{a} $) to the
solution for $u(x,\zeta)$ discussed in Sec.~\ref{SS:mrg-der}.
In particular, the ``shock wave" develops in the solution for $u(x)$ at
the scale $\zeta_2 = 1/2\tilde{a} = (\lambda_g +\lambda_d)/a^2$ that
is much larger than $\zeta_1$ under the condition~(\ref{cond-str}).

To find the subgap conductance of the system with $\ln(L/d) \gg \zeta_1$,
one should calculate, according to the general rule (\ref{Gsu}),
the quantity $u_x(\pp,\zeta)$ with $\zeta=2\ln(L/d)$.
This calculation is simplified by the fact that $u(\pp,\zeta) \simeq 0$ at
$\zeta \gg \zeta_1$.
A similar situation had been encountered in Sec.~\ref{SS:N}
where the normal-state conductance was considered, cf.\ Eq.~(\ref{1/Gn}).
Differentiating Eq.~(\ref{urgl}) over $x$ at $x =\pi/2$
under the condition $u(\pp)=0$, one gets for the Andreev
conductance the same Eq.~(\ref{1/Gn}), with
the initial condition (\ref{GA1}).
As a result, one obtains
\be
  \frac{1}{G_A} \approx
  \frac1{G_A^{(1)}} + \frac{1}{2\pi g}\ln\frac{L}{d} ,
\label{GAint}
\ee
where the first term is defined by Eq.~(\ref{GA1}), and the second term
(which can be of arbitrary magnitude compared to the first one)
is the normal-state resistance of the film (in units of $\hbar/e^2$).
The first term in Eq.~(\ref{GAint}) has the meaning of the effective
interface resistance defined in Eq.~(\ref{R+R}). In the leading order
over $\lambda_g/a$, $R_{T,\rm eff}$ is given by $1/G_A^{(1)}$.
Retaining also the next-to-leading contribution, one can write
$R_{T,\rm eff}$ as
\be
  \frac{e^2}{\hbar} R_{T,\rm eff} =
  \frac{4\pi g(\lambda_g+\lambda_d)}{G_T^2} -
  \frac{C(\lambda_d/\lambda_g)}{\sqrt{g}}
  + O\left( \frac{G_T^2}{g^{3/2}} \right),
\label{RTeff-strong}
\ee
where the function $C(\lambda_d/\lambda_g)$ is
of the order of one for $\lambda_d\sim\lambda_g$.
The second term in the RHS of Eq.~(\ref{RTeff-strong}) is small compared
to both terms in Eq.~(\ref{GAint}) once the conditions (\ref{cond-str})
and $\ln(L/d) \gg \zeta_1$ are fulfilled.

\subsubsection{Effect of weak repulsion at large scales}
\label{SSS:weak}

Here we study the solution of the FRG equation (\ref{urgl})
in the limit of weak repulsion.
For the sake of simplicity we will consider the case of
scale-independent $\lambda(\zeta)=\mbox{const}$
corresponding to the Finkelstein's fixed point $\lambda=\lambda_g$.

\begin{figure}
\refstepcounter{figure} \label{F:ulambda}
\vspace{-1mm}
\epsfxsize=80mm
\centerline{\epsfbox{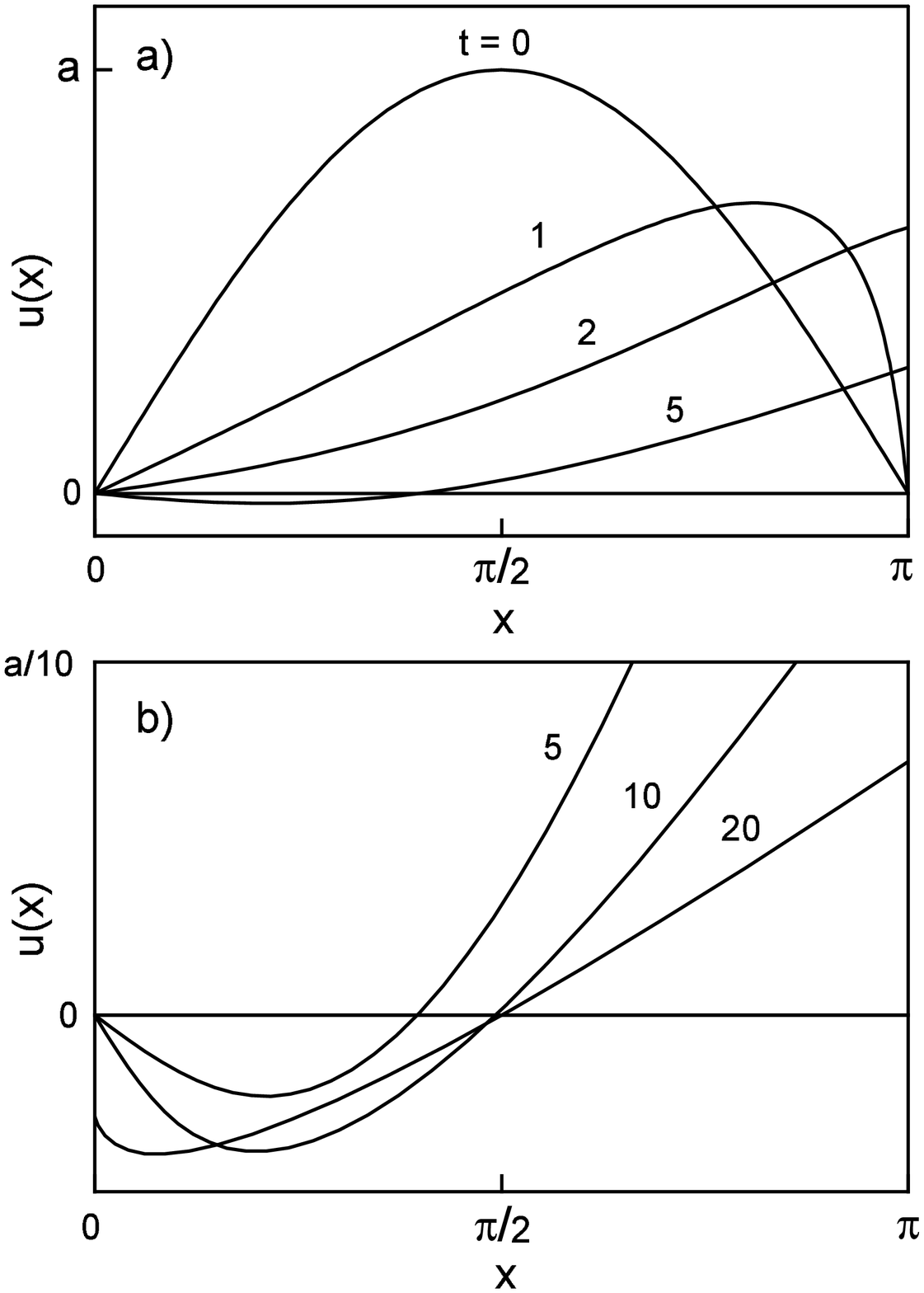}}
\small FIG.\ \arabic{figure}.
Solutions for Eq.~(\protect\ref{urgl})
with 
$\lambda/a=1/2$
for different values of $t$:
a) initial stage of evolution;
b) solution for large $t$.
\end{figure}

Consider first the qualitative effect of weak repulsion, $\lambda_g\ll a$,
on the evolution of the function $u(x)$.
At small $t\ll a/\lambda_g$, it leads to a small reduction
of the amplitude of the solution compared to the noninteracting case.
At $t\sim a/\lambda_g\gg1$, when this correction
becomes of the order of the solution itself, $u(x)$ changes its sign
and becomes negative on some part of the interval $0<x<\pi/2$.
For even larger $t\gg a/\lambda_g$, $u(\pp)$ quickly approaches zero
so that $u(x<\pp)<0$ and $u(x>\pp)>0$.
Then the characteristics (\ref{char})
with $x>\pi/2$ keep going on the right, while the characteristics
with $x<\pi/2$ go on the left, to the direction of negative $x$.
As a result, shock wave will appear also at $x=0$ that, however,
does not influence any physical results given by derivatives
of $u(x)$ at $x=\pi/2$.
The numerical solution for $\lambda(\zeta)=\lambda_g=a/2$ is
shown in Fig.~\ref{F:ulambda}.

To analyze Eq.~(\ref{urgl}) in the limit $\lambda_g\ll a$,
it is convenient to make the following rescaling:
\be
  s = \lambda_g\zeta, \quad
  v = u/\lambda_g.
\label{s-def}
\ee
Then, in terms of the function $v(x,s)$, Eq.~(\ref{urgl}) takes the form
\be
  v_s + v v_x = - v(\pp) \sin x,
\label{vrgl}
\ee
with the initial condition $v_0(x) = (a/\lambda_g) \sin x$.

\begin{figure}
\refstepcounter{figure} \label{F:R/Runiv}
\vspace{-1mm}
\epsfxsize=80mm
\centerline{\epsfbox{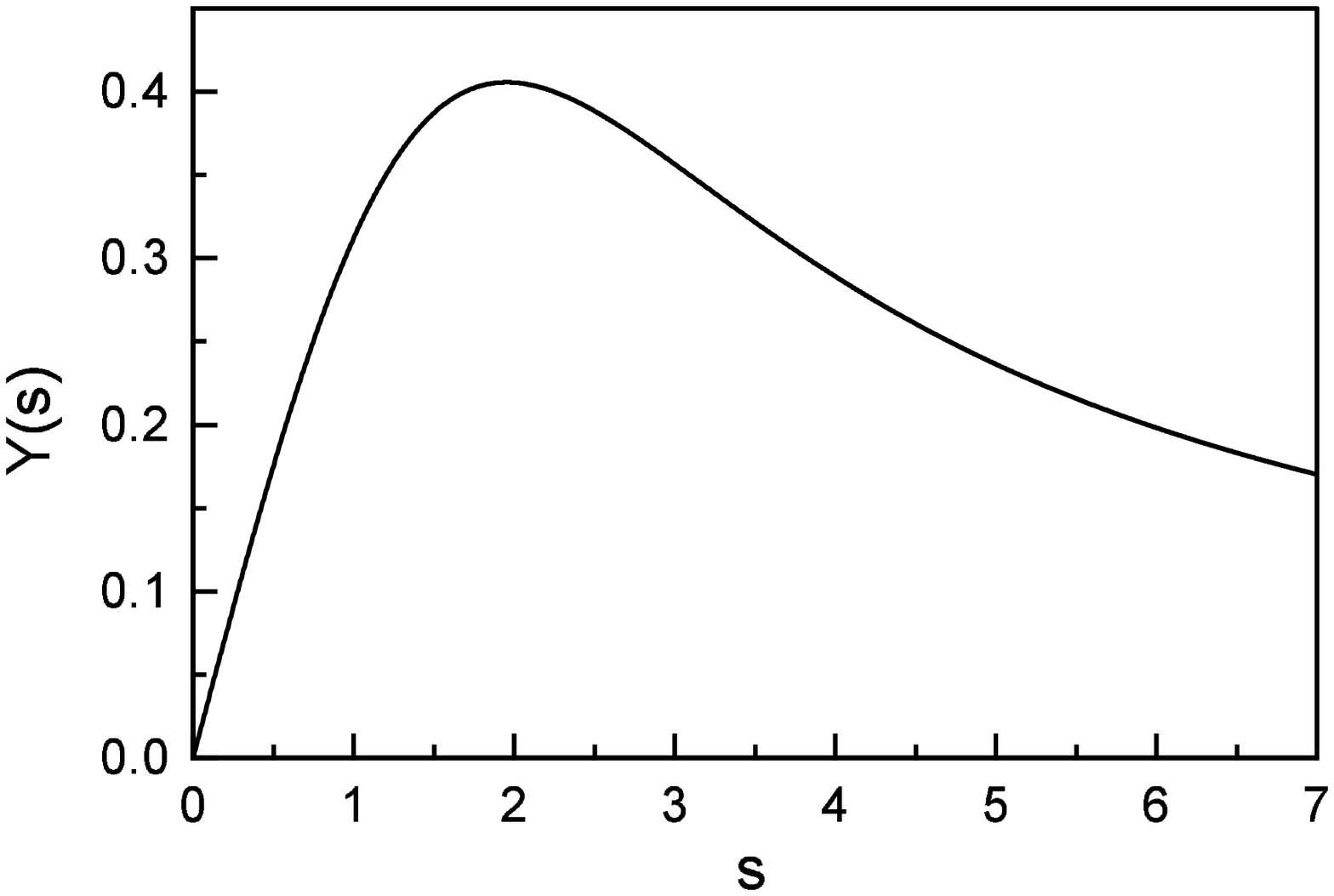}}
\vspace{2mm}
\small FIG.\ \arabic{figure}.
The universal function $\Upsilon(s)=R_{T,\rm eff}/R_D$ vs.\
$s=(\ln L/d)/\pi\sqrt{g}$.
\end{figure}

Since the magnitude of this initial condition is much larger than 1,
$v(x,s)$ acquires a saw-like behavior at $s\sim \lambda_g/a\ll1$
(i.~e., at $t\sim 1$).
Therefore, the details of $v_0(x)$ are irrelevant for the study
of the solution at large scales, $s\sim1$,
and one can formally consider the initial condition $v_0(x)=Ax$
with $A\to\infty$.
The solution for Eq.~(\ref{vrgl}) obtained with such $v_0(x)$
does not depend on $\lambda_g$ and describes a universal behavior
of the system for $t\gg1$ in the limit of weak repulsion.
The corresponding dependence $\Upsilon(s)=R_{T,\rm eff}(s)/R_D$
of the effective interface resistance
obtained by numerical solution of Eq.~(\ref{vrgl}) and normalized to $R_D$
is shown in Fig.~\ref{F:R/Runiv}.
The function $\Upsilon(s)$ has a maximum $\Upsilon=0.406$ at $s=1.95$,
and in the limiting cases it is given by
\be
  \Upsilon(s) = \cases{
    (1-2/\pi)\,s, & for $s\ll1$; \cr
    1.19/s, & for $s\gg1$.
  }
\label{Upsilon}
\ee

\subsubsection{Arbitrary $\lambda$ and $t$}
\label{SSS:arb}

\begin{figure}
\refstepcounter{figure} \label{F:R/R1}
\epsfxsize=80mm
\centerline{\epsfbox{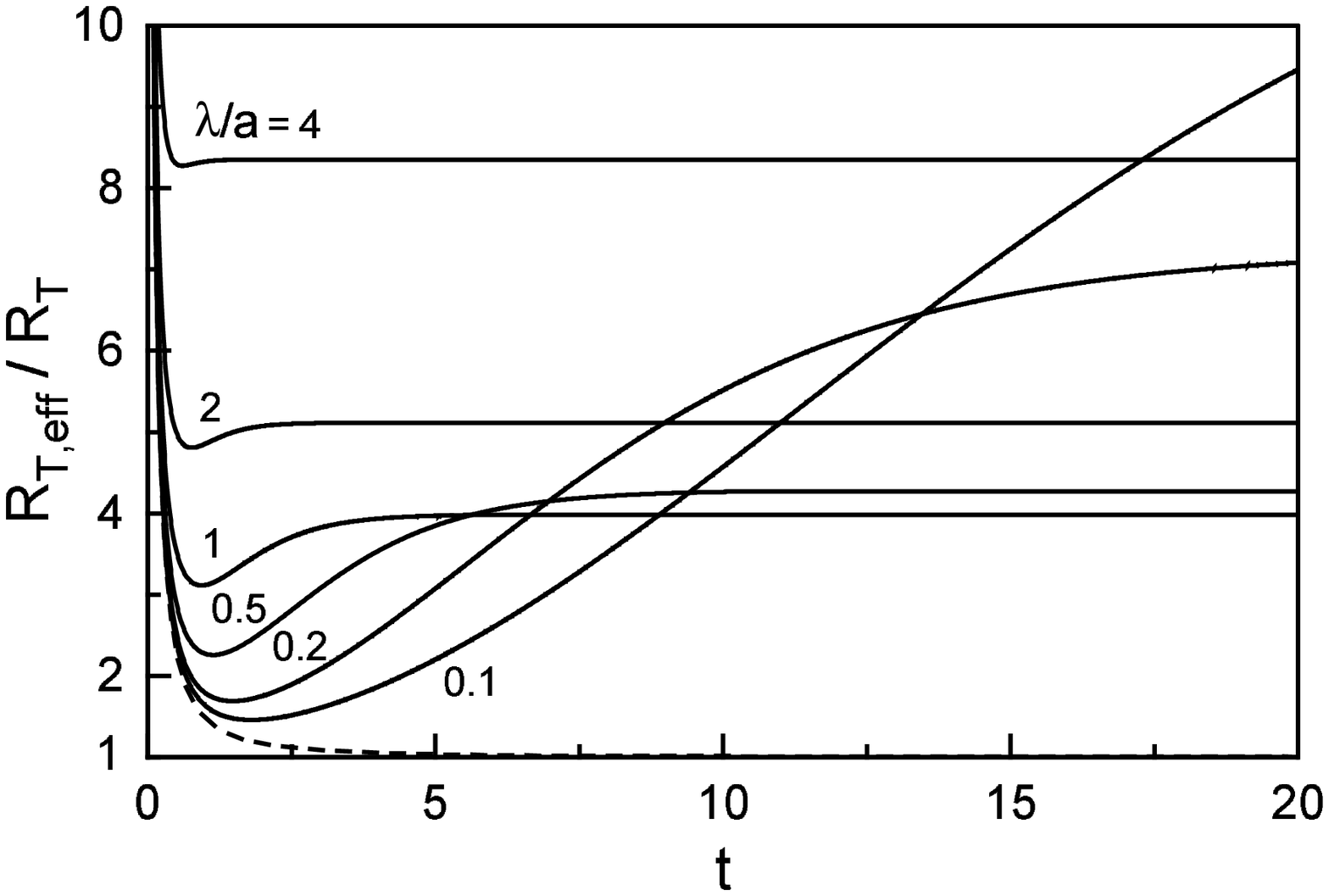}}
\vspace{2mm}
\small FIG.\ \arabic{figure}.
Dependence of the effective interface resistance $R_{T,\rm eff}(t)/R_T$
vs.\ $t=R_D/R_T$ for different values of $\lambda/a$ obtained by numerical
solution of Eq.~(\protect\ref{urgl}) for $\lambda(\zeta)=\mbox{const}$.
The dashed line corresponds to the noninteracting case, $\lambda=0$.
\end{figure}

Solution of the FRG equation (\ref{urgl}) for arbitrary $\lambda$ and $t$
should be obtained numerically.
The effective interface resistance $R_{T,\rm eff}$
normalized to the tunneling resistance $R_T$
as a function of $t=R_D/R_T$ is plotted in Fig.~\ref{F:R/R1}
(as in Sec.~\ref{SSS:weak},
we consider $\lambda(\zeta)=\lambda_g=\mbox{const}$).
The dashed line shows $R_{T,\rm eff}/R_T=1/\sin\Theta(t)$ for the
noninteracting case [cf.\ Eq.~(\ref{GA})].

\begin{figure}
\refstepcounter{figure} \label{F:R-infty}
\epsfxsize=80mm
\centerline{\epsfbox{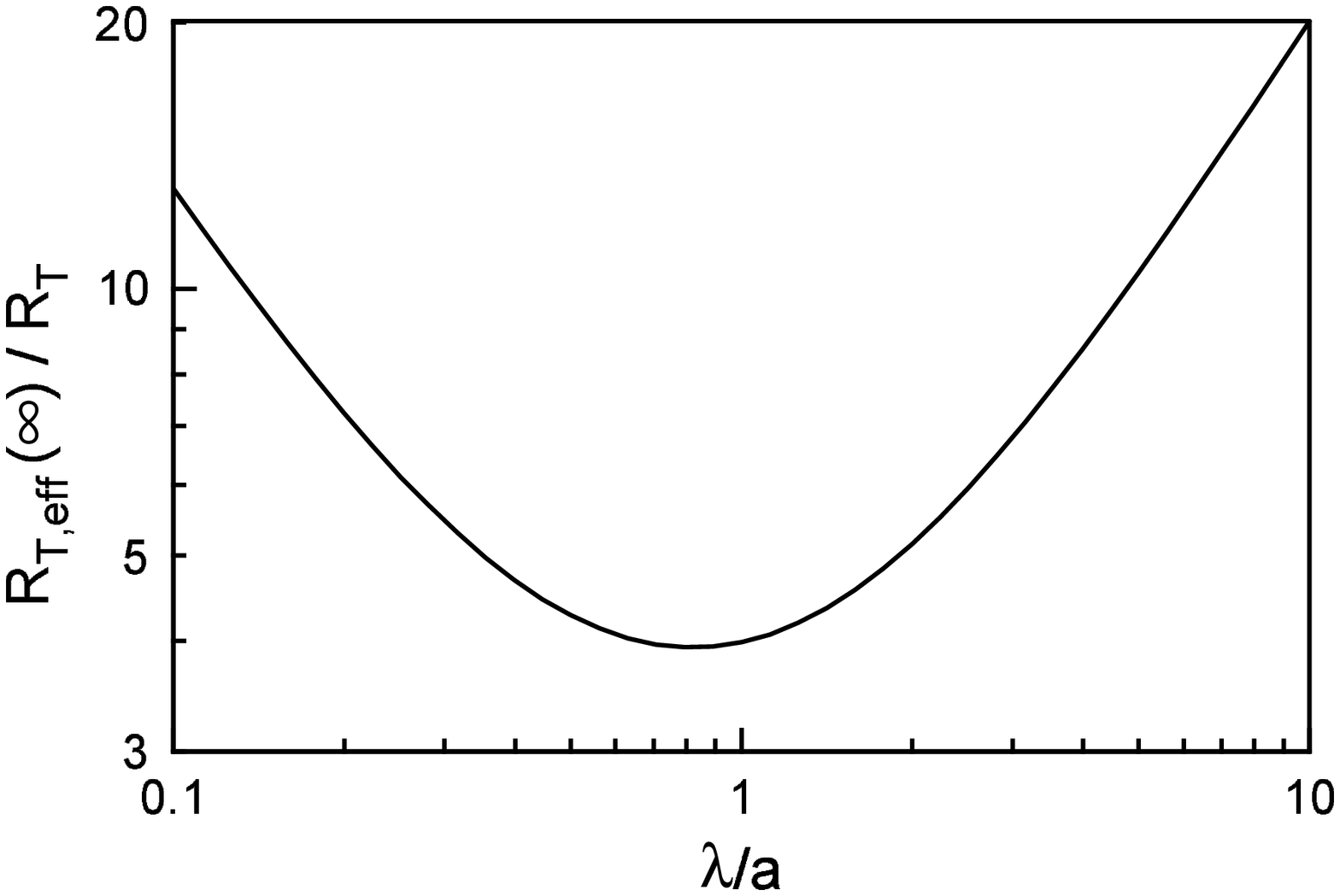}}
\vspace{2mm}
\small FIG.\ \arabic{figure}.
The limiting value $R_{T,\rm eff}(t=\infty)/R_T$ vs.\ $\lambda/a$
(for $\lambda(\zeta)=\mbox{const}$).
\end{figure}

For the case of strong repulsion, $\lambda_g\gg a$,
$R_{T,\rm eff}(t)$ very quickly (at $t\sim a/\lambda_g\ll1$)
reaches its asymptotic value
and saturates at $R_{T,\rm eff}(t=\infty) \approx(2\lambda_g/a)R_T$,
cf.\ Eq.~(\ref{RTeff-strong}).
The limiting value $R_{T,\rm eff}(\infty)$ decreases with
the decrease of $\lambda_g/a$ up to $\lambda_g/a \sim 1$.
For small $\lambda_g/a$, corresponding to the case of weak repulsion,
it starts to grow again with
$R_{T,\rm eff}(\infty)/R_T \approx 1.19a/\lambda_g$,
according to Eq.~(\ref{Upsilon}).
Note that in this limit $R_{T,\rm eff}(t)$ reaches its asymptotic value
at large scale $t\sim a/\lambda_g$.
The dependence of $R_{T,\rm eff}(\infty)$ as a function of
$\lambda_g/a$ is shown in Fig.~\ref{F:R-infty}.

The most striking feature of Fig.~\ref{F:R/R1} is a nonmonotonous
dependence of $R_{T,\rm eff}(t)$,
especially pronounced for weak repulsion, $\lambda_g\ll a$.
In this limit, the effective interface resistance
significantly exceeds its noninterating value, $R_T$,
at large scales, $t\geq(a/\lambda_g)^{1/2}$.
Such a nonmonotonuos dependence arises even within applicability
of the first-order correction (\ref{dGa}) as discussed in the end
of Sec.~\ref{SSS:first}.
Another important feature of Fig.~\ref{F:R/R1} is that
the limits $\lambda\to0$ and $R_D\to\infty$ {\em do not commute}.
Indeed, for any small but finite $\lambda_g$,
$R_{T,\rm eff}(t)/R_T$ will eventually (though, at very large $t$)
deviate from the noninteracting dependence
(dashed line in Fig.~\ref{F:R/R1}) and become large.

We will see in Sec.~\ref{S:beyond} that such a nonmonotonous dependence
of $R_{T,\rm eff}(t)$ will manifest itself in a nonmonotonous temperature
and voltage behavior of the subgap conductance.

\begin{figure}
\refstepcounter{figure} \label{F:R/R2}
\epsfxsize=80mm
\centerline{\epsfbox{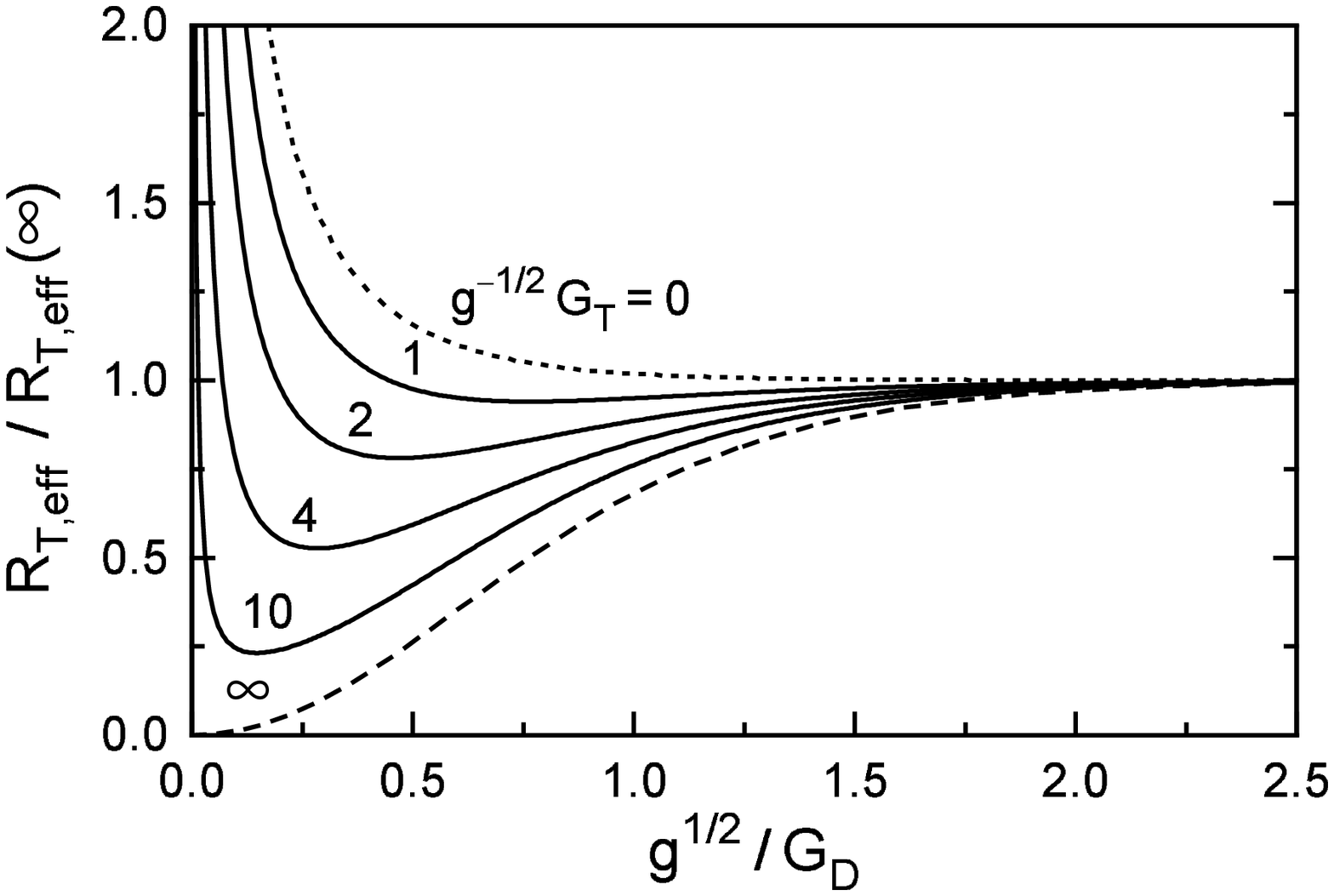}}
\vspace{2mm}
\small FIG.\ \arabic{figure}.
The effective interface resistance $R_{T,\rm eff}$ normalized to its limiting
value at $G_D^{-1}=\infty$ (see Fig.~\ref{F:R-infty}) as a function
of $\sqrt{g}/G_D = \lambda_g\zeta/2$ for various values of $G_T/\sqrt{g}$.
The plot corresponds to the Finkelstein's fixed point
$\lambda=\lambda_g\equiv1/(2\pi\sqrt{g})$.
\end{figure}

In this section we assumed that $\lambda(\zeta)$ had already reached
its fixed-point value $\lambda_g=1/(2\pi\sqrt{g})$, so that
$\lambda_g/a = 2\sqrt{g}/G_T$.
Thus, the limits of strong and weak interaction are separated by
$G_T\sim \sqrt{g}$. For $G_T\ll\sqrt{g}$, interaction is strong,
and for $G_T\gg\sqrt{g}$ (and, in particular, for the case of a completely
transparent interface, $G_T=\infty$), interaction is weak.
It is then natural to measure all dimensionless
conductances in units of $\sqrt{g}$.
The effective interface resistance $R_{T,\rm eff}$ normalized
to its limiting value $R_{T,\rm eff}(\infty)$ at large spatial scales
(cf.\ Fig.~\ref{F:R-infty}) as a function of
$\sqrt{g}/G_D = s/2 = \lambda_g\zeta/2$ is shown in Fig.~\ref{F:R/R2},
with different curves corresponding to different values of $G_T/\sqrt{g}$.
The asymptotic curve (dashed line) for the transparent interface,
$G_T\to\infty$, is given by
$R_{T,\rm eff} = \Upsilon(2\sqrt{g}\,(e^2R_D/\hbar))\, R_D$,
cf.\ Eq.~(\ref{Upsilon}).
In this limit, for $R_D\gg\hbar/e^2\sqrt{g}$ one has
\be
  \frac{e^2}{\hbar} R_A
  = \frac{0.6}{\sqrt{g}}
  + \frac{1}{2\pi g} \ln \frac{L}{d} .
\ee

\subsection{Noise}
\label{SS:lambda-noise}

\begin{figure}
\refstepcounter{figure} \label{F:diagram}
\epsfxsize=70mm
\centerline{\epsfbox{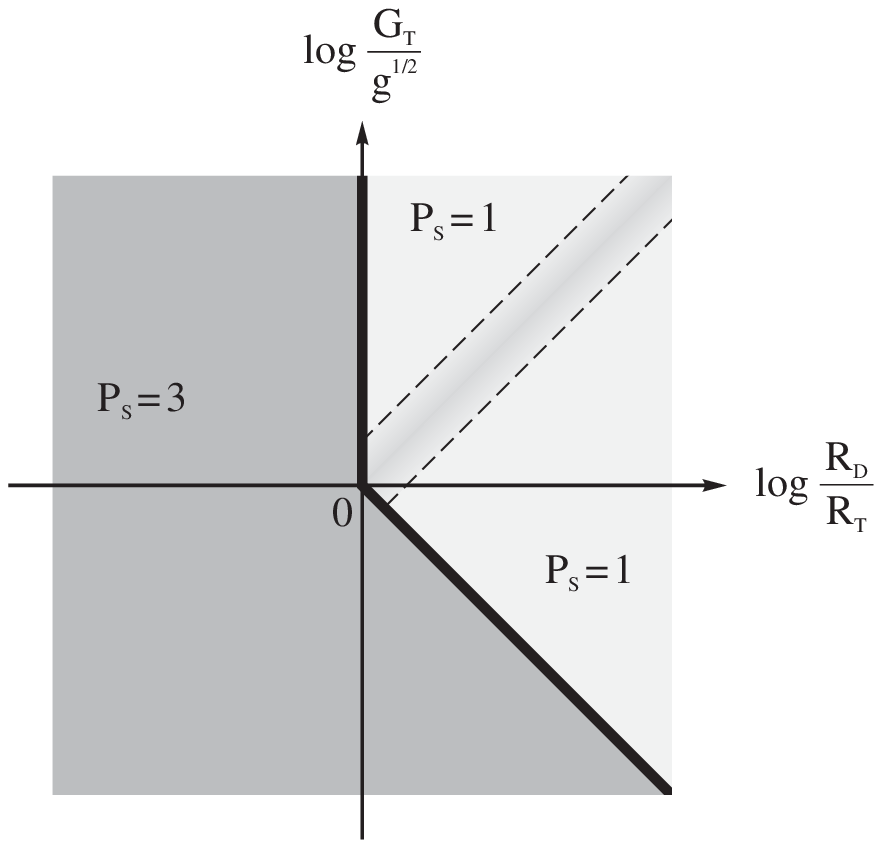}}
\small FIG.\ \arabic{figure}.
Schematic map of the noise coefficient $P_S$
as a function of the tunnel and diffusive resistances in the presence
of Cooper repulsion $\lambda=\lambda_g$. Dark area refers to
the tunnel limit $P_S\approx 3$, whereas light regions correspond to
the diffusive regime with $P_S \approx 1$.
\end{figure}

In this section we will analyze the effect of interaction on the noise
of N-S current. As shown in Sec.~\ref{SS:noise}, the current-current
correlator is determined by the noise function $P_S(t)$,
see Eqs.~(\ref{IIs}) and (\ref{PS0}).
Qualitatively, $P_S$ can be estimated by comparing the effective
tunneling resistance, $R_{T,\rm eff}$, with the diffusive resistance, $R_D$.
If $R_{T,\rm eff} \gg R_D$ then $P_S \approx 3$, while in the opposite case
($R_{T,\rm eff} \ll R_D$) $P_S \approx 1$.
As in Sec.~\ref{SSS:weak}, we will assume here the $\lambda(\zeta)$ had
already reached its fixed point $\lambda=\lambda_g$.
Then $P_S$ becomes a function of two parameters:
$t=R_D/R_T$ and $\lambda_g/a=2\sqrt{g}/G_T$.
Summarizing results of Sec.~\ref{SS:lambda-cond}, one can sketch
the boundaries between regions with $P_S=3$ and $P_S=1$
on the plane ($\log t$, $\log(G_T/\sqrt{g})$), see Fig.~\ref{F:diagram}.

In the case of strong repulsion, $G_T\ll\sqrt{g}$,
the function $u(x)$ very quickly (at $t\sim G_T/\sqrt{g}$)
reduces to the second harmonic, Eq.~(\ref{u1_ini}),
and $R_{T,\rm eff}$ 
saturates at
$R_{T,\rm eff} = R_A^{(1)} \equiv (4\sqrt{g}/G_T)R_T$
[cf.\ Eq.~(\ref{GA1})].
As explained in Sec.~\ref{SSS:strong}, further evolution of $u(x)$
is analogous (after a proper rescaling and shift of variables)
to the evolution of $u(x)$ for the noninteracting case
considered in Sec.~\ref{SS:mrg-der}.
The property $u(\pp)=0$ makes the calculation
of the noise function $P_S$ analogous
to the calculation of the function $P_N$ for the case of the {\em normal}\/
island, see Sec.~\ref{SS:N}.
So, one obtains $P_S^{\rm int} = P_N(R_D/R_A^{(1)})$.
Thus, in the limit $G_T\ll\sqrt{g}$,
the crossover between the tunnel ($P_S=3$) and diffusive ($P_S=1$) character
of noise is shifted to $t\sim \sqrt{g}/G_T \gg 1$, see Fig.~\ref{F:diagram}.

In the limit of weak repulsion, $G_T\gg\sqrt{g}$, the situation is more
interesting. For $t\sim 1$, interaction corrections can be neglected
and $P_S$ is given by the noninteracting expression (\ref{PS}).
So, at $t\sim 1$, $P_S$ decreases from 3 to 1, the corresponding
boundary being shown in Fig.~\ref{F:diagram}.
Later, at $t\sim G_T/\sqrt{g}\gg1$
(when $R_D\sim\hbar/e^2\sqrt{g}$)
interaction corrections become relevant.
In this region, $R_{T,\rm eff}$ is of the order of $R_D$
(cf.\ Fig.~\ref{F:R/Runiv}),
and one may anticipate that $P_S$ will deviate from 1.
For even larger $t$ when resistance is dominated by the diffusive conductor,
$P_S$ will eventually reduce down to 1.
This crossover region
is marked in Fig.~\ref{F:diagram} by the dashed lines.

\begin{figure}
\refstepcounter{figure} \label{F:lam_shot}
\epsfxsize=72mm
\centerline{\epsfbox{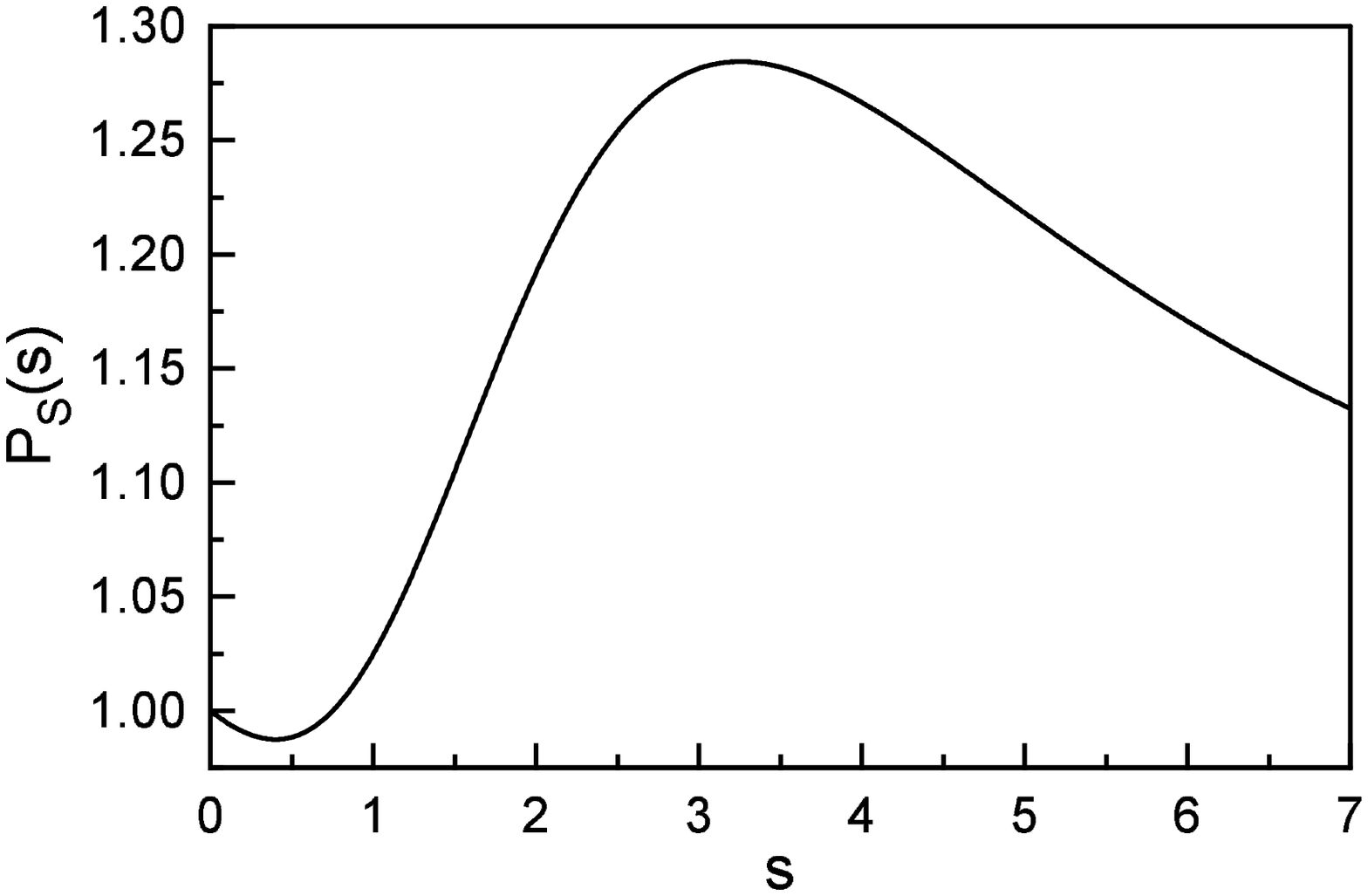}}
\small FIG.\ \arabic{figure}.
Noise function $P_S(s)$ vs.\ $s=2\sqrt{g}/G_D$
for the case of weak interaction, $G_T\gg\sqrt{g}$.
\end{figure}

The behavior of $P_S$ in the crossover region $t\sim G_T/\sqrt{g}\gg1$
can be obtained by numerical solution of Eq.~(\ref{vrgl}).
The resulting $P_S(s)$ is plotted as a function of
$s=\lambda_g\zeta=(\ln L/d)/\pi\sqrt{g} = 2(e^2/\hbar) R_D \sqrt g$
[cf.\ Eq.~(\ref{s-def})]
in Fig.~\ref{F:lam_shot}.
$P_S(s)$ has a mimimum $P_S=0.99$ at $s=0.40$
and a maximum $P_S=1.28$ at $s=3.25$.

The negative derivative of $P_S(s)$ for $s\ll1$ can be obtained analytically.
Seeking the solution of Eq.~(\ref{vrgl}) as a series over $s$,
one finds
\be
  v(x,s) = \frac xs - \frac{\pi(1-\cos x)}{2x} + O(s),
\ee
where the first term is the usual saw function, while the other terms
are due to the RHS of Eq.~(\ref{vrgl}). Now, with the help of
Eq.~(\ref{PS0}) one gets for $s\ll1$
\be
  P_S(s)
    = 1 - \left( \frac12 - \frac{12}{\pi^2} + \frac{24}{\pi^3} \right) s
    = 1 - 0.058 \, s .
\ee

\section{Temperature, voltage and magnetic field effects}
\label{S:beyond}

\subsection{General analysis}
\label{SS:beyond-gen}

In this section we consider the case of a superconductive island
in the situation when temperature and/or voltage
are high compared to the Thouless energy,
or the perpendicular magnetic length
$l_H=\sqrt{\Phi_0/H}$ is shorter than the system size $L$.
Then Cooperon coherence is destroyed at the energy
scale $\Omega_* > \ETh$,
where $\Omega_* = \max(T,eV,eDH/c)$.
At the same time we will assume that frequency $\omega$ of external voltage
(or current) is smaller than the Thouless energy.

According to the general approach explained in Sec.~\ref{SS:prox},
we have to determine an effective Proximity Action via the RG
procedure accomplished down to the energy scale $\ETh$.
Now the whole energy interval of the RG, $\omega_d>\Omega>\ETh$,
can be divided into two regions with different FRG equations.
In the region $\omega_d > \Omega > \Omega_*$, renormalization of
the action (\ref{sg}) is described by Eq.~(\ref{urg}) derived in
Sec.~\ref{S:FRG} for the noninteracting case, or its analogue (\ref{urgl})
in the presence of interaction.
In the region $\Omega_* > \Omega > \ETh$, Cooperons are suppressed
while diffuson modes still contribute to the renormalization
of the action (\ref{sg}).
Thus, at large scales, $\Omega < \Omega_*$,
the charges $\gamma_n$ depend on two RG ``time'' arguments,
$\zeta$ and $\zeta_* \equiv \ln(\omega_d/\Omega_*)$.
For $\zeta > \zeta_*$, the bulk matrix $Q$ is purely diagonal
in the Nambu space [cf.\ Eq.~(\ref{uWu})].
On the other hand, according to Eq.~(\ref{QS}), the island's matrix
$Q_S$ is off-diagonal in the Nambu space, due to the Andreev
nature of the subgap tunneling across the interface.
As a result, $\tr(Q_SQ)^n$ with odd $n$ vanishes,
and so $\gamma_n$'s with odd $n$ do not contribute to the
multicharge RG equation (\ref{mrg}) at scales $\zeta > \zeta_*$.

To describe evolution of $\gamma_{n}$'s at $\zeta > \zeta_*$,
one should modify the derivation presented in Sec.~\ref{SS:mrg-der}
for the case of even $n$. The difference is that now only the
first line of Eq.~(\ref{WW}) corresponding to diffuson pairing
contributes to the average
$\Delta S_{k,n} = i \corr{S_{{\rm int},k}S_{{\rm int},n}}$.
But for even $k$ and $n$, and at $\zeta > \zeta_*$
the matrices $A$ and $B$ defined in Eq.~(\ref{AB}) commute with $\tau_z$.
Then the zero-energy limit of the first line of Eq.~(\ref{WW}) reproduces
Eq.~(\ref{WW1}) used in the derivation of Eq.~(\ref{<SS>}).
Hence, we conclude that evolution of $\{\gamma_{2m}\}$
at $\zeta > \zeta_*$ is described by the same Eq.~(\ref{mrg}),
with all $\gamma_{2m+1}$ set to zero.

To describe evolution of the set $\{\gamma_n\}$ at $\zeta > \zeta_*$,
we find it convenient to introduce Fourier transformation with respect
to even $n$:
\be
  \tilde u(x) = \sum_{n=1}^\infty 2n \gamma_{2n} \sin 2nx
  = \frac12 [u(x)-u(\pi-x)] .
\label{t-u-def}
\ee
Then the FRG equation for the function $\tilde u(x)$ acquires the form
of the Euler equation:
\be
  \tilde u_\zeta + \tilde u \tilde u_x = 0.
\label{u-tilde}
\ee
The function $\tilde u(x,\zeta;\zeta_*)$ that determines physical
quantities (see below) is then given by the solution of
Eq.~(\ref{u-tilde}) with the initial condition
\be
  \tilde u(x,\zeta_*;\zeta_*) = \frac12 [u(x,\zeta_*)-u(\pi-x,\zeta_*)] ,
\label{ini-tilde}
\ee
where $u(x,\zeta_*)$ is the solution for the FRG equation
(\ref{urgl}) at $\zeta_*=\ln(\omega_d/\Omega_*)$.
We want to emphasize that the reduction $u(x)\to\tilde u(x)$
at $\zeta=\zeta_*$ describes the crossover from the $\Omega_*\ll\ETh$
to $\Omega_*\gg\ETh$ regimes only with logarithmic accuracy.
The number in the correct cutoff of logarithm
is beyond the RG precision.

Below we apply the described scheme to the calculation of
physical quantities.
To determine the Andreev conductance at large $\Omega_* > \ETh$
in terms of the function $\tilde u(x)$ we should use a generalized
version of Eq.~(\ref{ii3}). Namely, we should take into account that
for $\epsilon > \ETh$, parameters $\gamma_{2n}$ depend on the energy
argument $\epsilon$ running under the trace [cf.\ Eq.~(\ref{B:M})],
since it is just the value
of $\epsilon$ which determines the Cooperon coherence scale.
Thus, $\gamma_{2n}(\epsilon)$ should be put under the sign of
$\Tr$ in Eq.~(\ref{ii3}).  The trace operator contains an integral over
$\epsilon$ those main contribution comes from $\epsilon \sim \Omega_*$.
As a result, we obtain the following expression for the (nonlinear) current:
\be
  I(V) = \frac{e^2}{\hbar} G_A(\Omega_*) V ,
\label{IVA}
\ee
where the value of $G_A(\Omega_*)$ is determined to logarithmic
accuracy as
\be
  G_A(\Omega_*)
  \equiv G_A(t,t_*)
  = 4\pi g\, \tilde u_x (\pp,\zeta;\zeta_*) .
\label{Gsu-E}
\ee
Calculation of $\tilde u_x (\pp)$ is very simple since
$\tilde u(\pp)=0$. Hence, similarly to Eq.~(\ref{1/Gn}) one gets
\be
  \left.\frac{\partial(1/G_A)}{\partial\zeta}\right|_{\zeta >\zeta_*} =
  \frac{1}{4\pi g} .
\label{1/Gs}
\ee

Eqs.~(\ref{Gsu-E}), (\ref{1/Gs}) can be naturally interpreted with the help
of the effective interface resistance $R_{T,\rm eff}$
introduced in Eq.~(\ref{R+R}). An important property of
this quantity is that it is formed by Cooperons only
which are taken into account by the FRG equation for $u(x)$.
At scales $t>t_*\equiv a\zeta_*$,
Cooperon coherence is lost, and $R_{T,\rm eff}(t)$
saturates at the constant level $R_{T,\rm eff}(t_*)$
becoming independent of $R_D$ any more.
As a result,
\be
  R_A(t,t_*) = R_D + R_{T,\rm eff}(t_*),
\label{RA*}
\ee
where $R_D=R_Tt=\ln(L/d)/2\pi\sigma$ is the energy-independent resistance
of the normal film.

Consideration that lead to Eq.~(\ref{Gsu-E}) can be generalized to
higher correlators of current as well.
To logarithmic accuracy, the noise power is given by
the zero-energy expressions (\ref{IIs}) and (\ref{PS0})
with $u(x,\zeta)$ replaced by $\tilde u(x,\zeta;\zeta_*)$.
The quantities $\tilde u_x(\pp)$ and $\tilde u_{xxx}(\pp)$ entering
this expression can be calculated similarly to Eq.~(\ref{uxxxn}):
\bea
  \tilde u_x(\pp,\zeta;\zeta_*)
  &=& \frac{u_x(\pp,\zeta_*)} {1+(\zeta-\zeta_*)u_x(\pp,\zeta_*)} ,
\\
  \tilde u_{xxx}(\pp,\zeta;\zeta_*)
  &=& \frac{u_{xxx}(\pp,\zeta_*)}{[1+(\zeta-\zeta_*)u_x(\pp,\zeta_*)]^4} .
\eea
As a result, the current-current correlation function can be written
in the form similar to Eq.~(\ref{IIs}):
\bea
  \corr{I_{\omega}I_{-\omega}} = && {}
  \frac{e^2G_A(t,t_*)}{3\hbar} \,
  \Bigl\{
    (3-P_S(t,t_*)) \, \Psi(\omega)
\nonumber \\
  {} + \frac12 P_S && (t,t_*) \, [\Psi(\omega-2eV) + \Psi(\omega+2eV)]
  \Bigr\} ,
\label{IIs2}
\eea
where $G_A(t,t_*)$ is given by Eq.~(\ref{RA*}),
and the noise function $P_S(t,t_*)$ depends now on two RG ``time'' variables.
Using Eqs.~(\ref{Gsu}), (\ref{PS0}), (\ref{Gsu-E}) and (\ref{RA*}),
one obtains
\be
  P_S(t,t_*) = 1 + \frac{G_A^3(t,t_*)}{G_A^3(t_*,t_*)} [ P_S(t_*)-1 ] .
\label{PStt0}
\ee
Here both $G_A(t_*,t_*)\equiv G_A(t_*)$ and $P_S(t_*)$
are given by the zero-energy results at $t=t_*$.

\subsection{Noninteracting case}
\label{SS:beyond-zero}

\subsubsection{Andreev conductance}

To calculate the Andreev conductance we
substitute $R_{T,\rm eff}(t_*)=R_T/\sin\Theta(t_*)$ into Eq.~(\ref{R+R}):
\be
  G_A(\Omega_*)
  = G_T \frac{\sin\Theta(t_*)}{1+t\sin\Theta(t_*)}
  = G_D \frac{t\sin\Theta(t_*)}{1+t\sin\Theta(t_*)} ,
\label{GA2}
\ee
that can be obtained from the zero-energy result (\ref{GA})
by replacement $\Theta(t) \to \Theta(t_*)$.

Below we apply Eq.~(\ref{GA2}) to the analysis of two specific examples.

\begin{figure}
\refstepcounter{figure} \label{F:GA1}
\epsfxsize=80mm
\centerline{\epsfbox{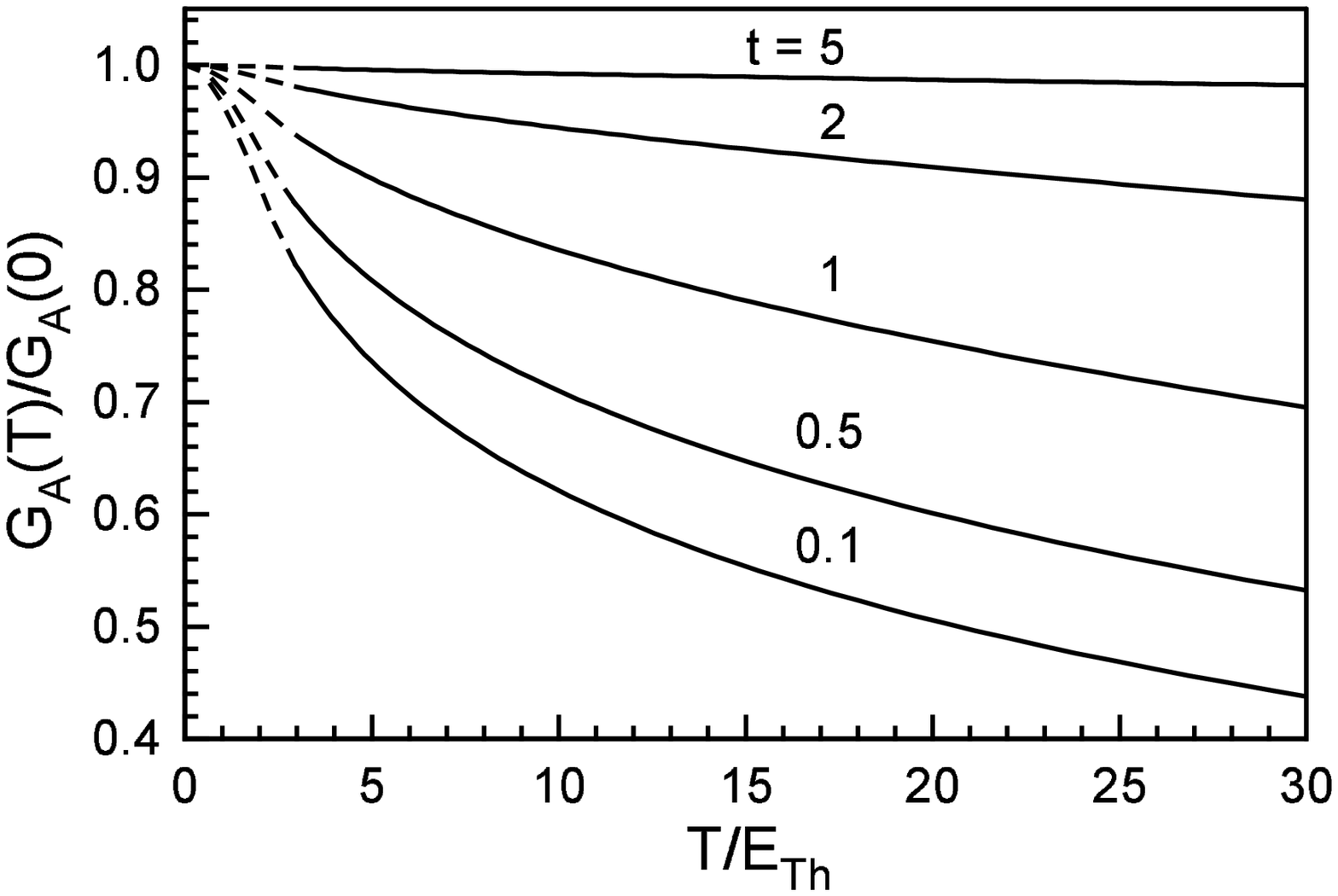}}
\small FIG.\ \arabic{figure}.
The linear Andreev conductance $G_A(T)$ normalized to its zero-temperature
value vs.\ $T/\ETh$ for different values of the ratio $t=G_T/G_D$.
The plot corresponds to the geometry shown in Fig.~\protect\ref{F:NSisland}
with $L/d=20$, i.~e., $\omega_d/\ETh=400$.
In the intermediate region $T\sim\ETh$ (sketched by dashed line),
corrections are nonlogarithmic and cannot
be taken into account within the RG approach.
\end{figure}

We start from the case of the {\em linear} conductance as a function of
temperature. The corresponding curves $G_A(T)$ for several values
of the ratio $t=G_T/G_D$ are presented in Fig.~\ref{F:GA1}.
In the limit $G_T \gg G_D$ any dependence on $T$ disappears and
N-S conductance is equal to the diffusive conductance $G_D$.
This result is in disagreement with calculations~\cite{Art,StNaz} which
predict, for an ideal N-S structure in the 1D geometry,
a conductance maximum at $T \sim \ETh$ with the relative magnitude
about $10\%$ (the so-called finite-bias anomaly).
This discrepancy is due to the limited precision of our calculation scheme,
which is equivalent to the summation of the main logarithmic terms.
Finite-bias anomaly is due to the energy-dependence of the effective
diffusion constant $D(\epsilon)$ at $\epsilon \sim \ETh$, and this effect
is beyond the main logarithmic approximation.  As one can see from
Fig.~\ref{F:GA1}, the zero-bias anomaly is stronger than the finite-bias
anomaly for small enough values of $t$.

Consider now the {\em nonlinear} subgap conductance at high temperature
$T \gg \ETh$ and arbitrary relation between $T$ and $eV$.
At $eV \ll T$ we are back to the linear conductance case
with $t_* = t_T \equiv (G_T/4\pi g)\ln(\omega_d/T)$,
whereas at large voltage $eV \gg T$ we have
$t_* = t_V \equiv (G_T/4\pi g)\ln(\omega_d/eV)$.

To find the behavior of $G_A(T,V)$ in the crossover region
$eV \sim T \gg \ETh$ we need to improve the logarithmic
accuracy of Eq.~(\ref{GA2}).
To this end, we perform a more accurate calculation of the energy
integral under the trace in the generalized version of Eq.~(\ref{ii3})
taking into account the energy dependence of $\gamma_n(E)$:
\bea
  \frac{G_A(T,V)}{G_D} &=&
  \frac1{2eV} \int_0^\infty dE
  \left[\tanh\frac{E_+}{2T} - \tanh\frac{E_-}{2T}\right]
\nonumber \\
  && {} \times
  \frac{t\sin\Theta(t_E)}{1+t\sin\Theta(t_E)} ,
\label{integral}
\eea
where $E_\pm = E \pm eV$, and $t_E=(G_T/4\pi g)\ln(\omega_d/E)$.
The logarithmic factor $\ln(\omega_d/E)$ comes from the integration
over 2D Cooperon modes, $\int^{1/d}d^2{\bf q}/(Dq^2\pm2iE)$, which determines
the Cooperon amplitude at coinciding point
(i. e., the probability of return).
In the presence of transverse magnetic field, Cooperon modes are quantized,
so integration over momenta is substituted by the summation over
Landau levels, see, e.~g., Ref~\cite{AKLL}.
Then the effect of magnetic field upon the subgap conductance
can be accounted by the replacement of $t_E$ in Eq.~(\ref{integral}) by
\be
  t_E(H) = \frac{G_T}{4\pi g}
  \left[
    \ln\frac{\hbar c}{2eHd^2} -
    \Re\psi\left(\frac12 - i\frac{Ec}{2eDH}\right)
  \right] ,
\label{tEH}
\ee
where $\psi(x) = d\ln\Gamma(x)/d x$ is the digamma function.

In the limiting case of weak interface transparency, $t\ll 1$, the Andreev
conductance is given by
\be
  \frac{G_A(T,V,H)}{G_T^2/4\pi g} =
  \cases{
    \displaystyle \ln\frac{\omega_d}{eV} + 1,
      & for $eV \gg (T, \frac{eDH}{c})$; \cr
    \displaystyle \ln\frac{2\omega_d}{\pi T} + \gamma, \rule{0pt}{18pt}
      & for $T \gg (eV, \frac{eDH}{c})$; \cr
    \displaystyle \ln\frac{2\omega_d c}{eDH} + \gamma, \rule{0pt}{18pt}
      & for $\frac{eDH}{c} \gg (T, eV)$;
  }
\label{int_as}
\ee
where $\gamma = 0.577\dots$ is the Euler constant. Comparing
these asymptotics, we conclude that crossover from the voltage
to temperature-dominated effective interface resistance occurs at
$2eV \approx \pi e^{1-\gamma} T \approx 4.8\, T$. Similarly,
crossover from the temperature- to magnetic-field dominated
resistance occurs at $ H \approx  \pi c T/e D$.

\begin{figure}
\refstepcounter{figure} \label{F:GA2}
\epsfxsize=80mm
\centerline{\epsfbox{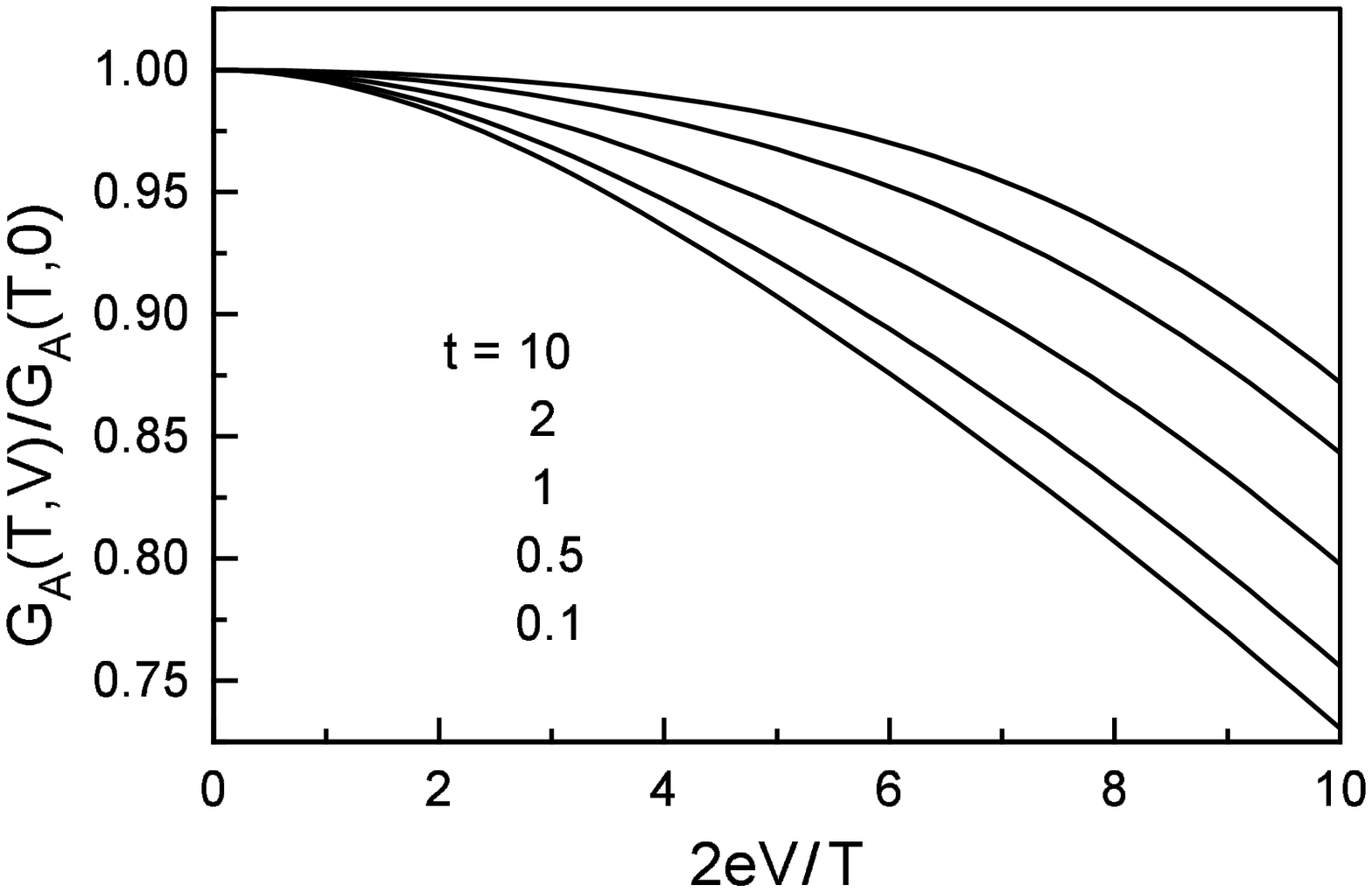}}
\small FIG.\ \arabic{figure}.
The nonlinear Andreev conductance $G_A(T,V)$ normalized to
the linear conductance $G_A(T,0)$ vs.\ the ratio $2eV/T$ for $T/\ETh=10$.
As in Fig.~\protect\ref{F:GA1}, $\omega_d/\ETh=400$.
Different curves correspond to different ratio $t=G_T/G_D$:
0.1 (bottom), 0.5, 1, 2, 10 (top).
\end{figure}

For arbitrary $t$, the behavior of $G_A$ in the crossover region
should be obtained by numerical integration in Eq.~(\ref{integral}).
The corresponding plots of $G_A(T,V)$ as a function of $2eV/T$
for different values of $t$ are presented in Fig.~\ref{F:GA2}.
Note, finally, that the difference between the nonlinear conductance
$G_A(V)\equiv I_A/V$ and the differential conductance $dI_A/dV$
should be neglected in the main logarithmic approximation.

\subsubsection{Current fluctuations}

In order to calculate the noise function $P_S(t,t_*)$
at $\Omega_*\gg\ETh$ we substitute Eqs.~(\ref{PS}) and (\ref{GA2})
into Eq.~(\ref{PStt0}). As a result, we obtain
\be
  P_S(t,t_*) = 1 +
  \frac{1+\Theta_*\tan\Theta_*+3\Theta_*\cot\Theta_*}
  {2(1+\Theta_*\tan\Theta_*)(1+t\sin\Theta_*)^3} ,
\label{Ptt}
\ee
where $\Theta\equiv\Theta(t)$, $\Theta_*\equiv\Theta(t_*)$,
and $G_A(\Omega_*)$ is given by Eq.~(\ref{Gsu-E}).

\begin{figure}
\refstepcounter{figure} \label{F:N}
\epsfxsize=80mm
\centerline{\epsfbox{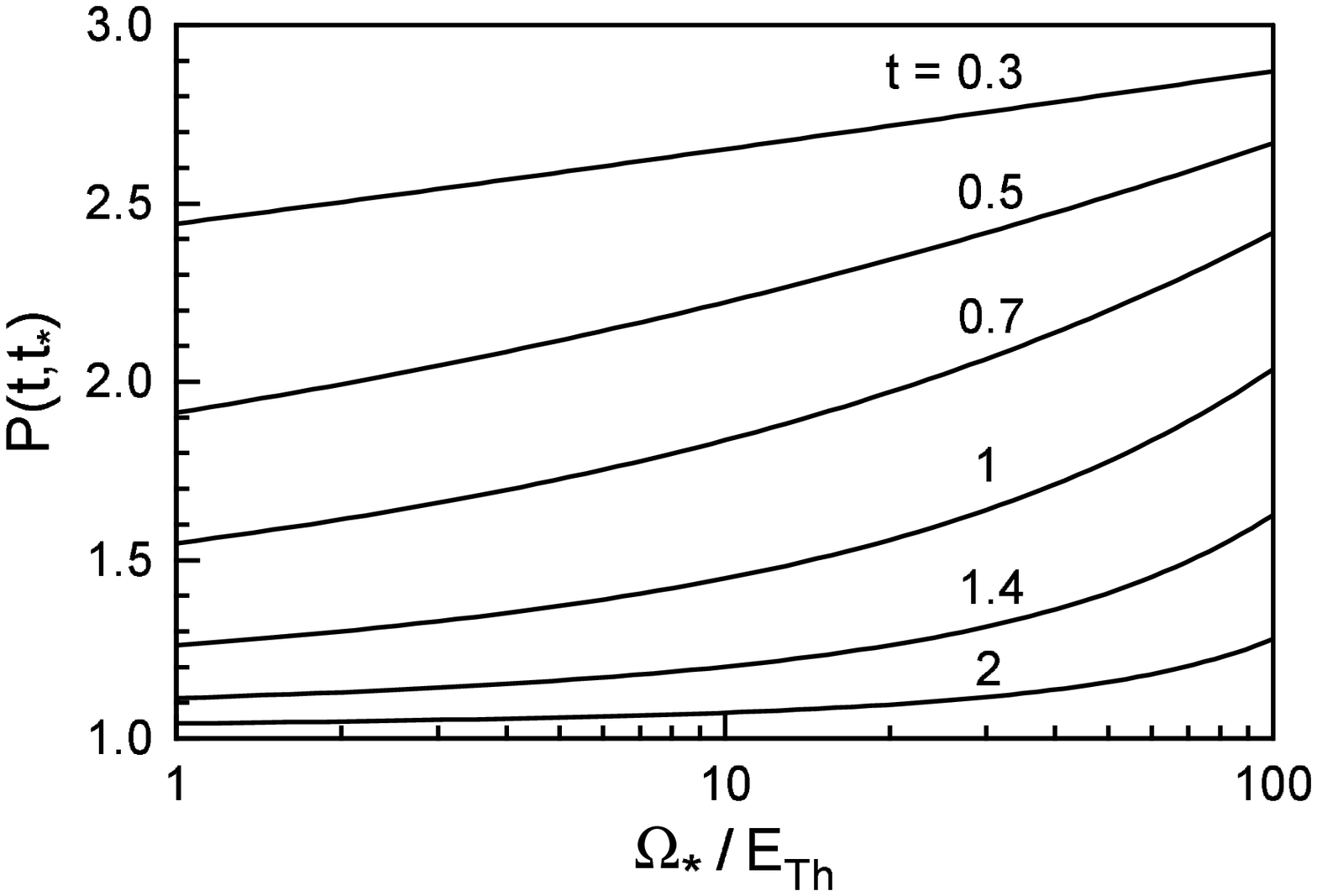}}
\small FIG.\ \arabic{figure}.
Noise function $P_S(t,t_*)$ vs.\ $\Omega_*/\ETh$ for different
values of $t=G_T/G_D$. As in Fig.~\protect\ref{F:GA1},
the ratio $\omega_d/\ETh=400$.
\end{figure}

Eqs.~(\ref{IIs2}) and (\ref{Ptt}) determine fluctuations of \mbox{N-S}
current as a function of temperature, voltage, magnetic field and
frequency ($\omega\ll\ETh$) at arbitrary values of the ratio $t = G_T/G_D$.
Dependence of noise upon
$T$, $V$ and $\omega$ comes in two different ways: via the functions
$\Psi(\omega\pm 2eV) = \Omega\coth((\omega\pm 2eV)/2T)$ and via the Cooperon
cutoff scale $\Omega_* =\max(T,eV,\omega , eDH/c)$ which determines
$t_* = (G_T/4\pi g)\ln(\omega_d/\Omega_*)$.
In the limit $t \to \infty$, the function $P_S(t,t_*)/3$ approaches 1/3
and the expression for noise reduces to the standard form~\cite{Lesovik99}
for purely diffusive N-S junction at $\Omega_* < \ETh$~\cite{noiseexp}.
However, the range of $t$ where such a universal behavior sets in, does
depend upon $\Omega_*$: the increase of $\Omega_*$ leads to the decrease of
the proximity angle $\Theta_*$, which, in turn, increases the factor
$P_S(t,t_*)$. A number of curves characterizing the behavior of $P_S(t,t_*)$
as a function of $\Omega_*/\ETh$ at different values of $t$ is presented
in Fig.~\ref{F:N}.

\subsection{Effects of interaction}
\label{SS:be_int}

This section contains the main application of our theory since
variation of temperature, voltage, or magnetic field is a natural
tool to study system properties. Here all effects discussed
in the body of the paper come into play altogether.
Therefore, we will repeat briefly the main concepts the reader
could gain from the above discussion.

At scales smaller than the Cooperon coherence length $\sqrt{D/\Omega_*}$,
i.~e., at $\zeta<\zeta_*\equiv\ln(\omega_d/\Omega_*)$,
the system is described by the function $u(x,\zeta)$
which evolves according to the FRG equation (\ref{urgl}).
At larger scales, $\zeta>\zeta_*$, one should introduce a
``two-time'' function $\tilde u(x,\zeta;\zeta_*)$.
It is obtained from $u(x,\zeta)$ by the reduction (\ref{ini-tilde})
and obeys the FRG equation (\ref{u-tilde}).
The conductance and noise power are given by Eqs.~(\ref{Gsu}) and (\ref{PS0}),
with $u(x,\zeta)$ being substituted by $\tilde u(x,\zeta;\zeta_*)$,
and $\zeta=\ln(\omega_d/\ETh)=2\ln(L/d)$.
The $\Omega_*$ dependence of the conductance can be easily expressed
with the help of the effective interface resistance according to
Eq.~(\ref{RA*}).

\begin{figure}
\refstepcounter{figure} \label{F:ga_t_int}
\epsfxsize=80mm
\centerline{\epsfbox{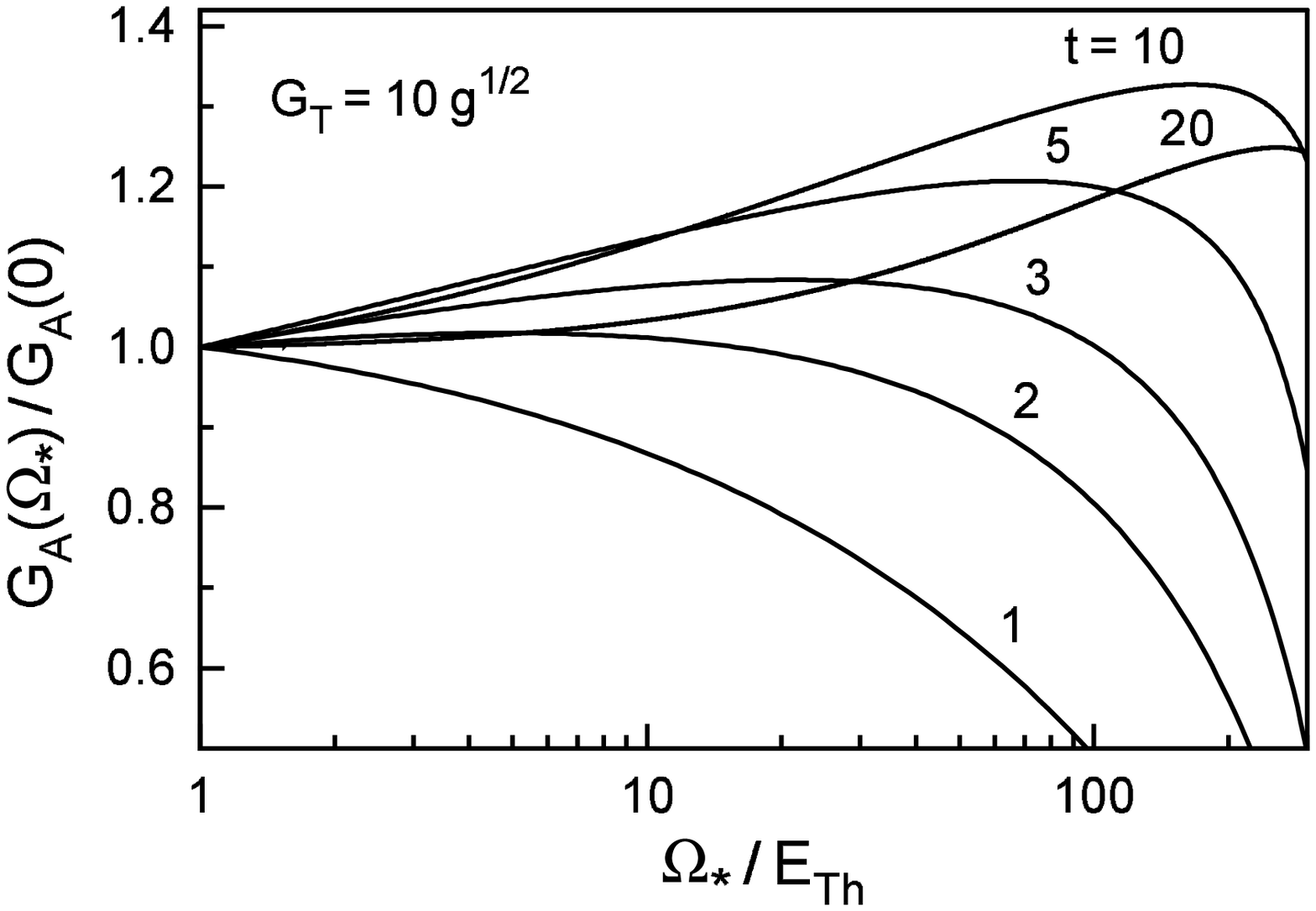}}
\small FIG.\ \arabic{figure}.
Dependence of the Andreev conductance $G_A(\Omega_*)$ (normalized to
the zero-energy value) on the ratio $\Omega_*/\ETh$
for different values of $t$. For all plots, $\omega_d/\ETh=400$,
$\lambda=\lambda_g$, and $G_T=10\sqrt{g}$.
\end{figure}

In the noninteracting approximation the effect of large $\Omega_* \gg \ETh$
was to decrease the Andreev conductance compared to the zero-energy
limit, $\Omega_* \ll \ETh$, due to the increase of $R_{T,\rm eff}(t_*)$.
The drastic change introduced by the Cooper interaction is that
$R_{T,\rm eff}(t_*)$ shown in Fig.~\ref{F:R/R1}
is no longer a monotonous function of $t_*$.
Therefore, depending on the relation between parameters of the problem,
the sub-gap conductance may either decrease or increase with the
increase of $\Omega_*$. This unusual enhancement of conductivity with
the increase of the decoherence energy scale $\Omega_*$ is most
pronounced in the limit of weak repulsion, $G_T\gg2\sqrt{g}$
(we again assume that $\lambda(\zeta)$ had reached the Finkelstein's
fixed point $\lambda_g$).
In this case, according to the result of Sec.~\ref{SSS:first},
$R_{T,\rm eff}(t_*)$ decreases with $t_*$ at
$t_* \ll (G_T/\sqrt{g})^{1/4}$, and increases with $t_*$ at
 $t_* \gg (G_T/\sqrt{g})^{1/4}$, cf.\ Eq.~(\ref{Rl}).
Hence, the total conductance decreases with the growth of $\Omega_*$
for $\Omega_*\gg\Omega_{\rm cr}$ and increases
for $\Omega_*\ll\Omega_{\rm cr}$,
where $\Omega_{\rm cr} = \omega_d\exp(-cg^{7/8}/G_T^{3/4})$,
and $c=2\pi^{7/4}(\pi-2)^{-1/4} \approx 14.3$.
Since the total resistance is the sum of $R_{T,\rm eff}$ and $R_D$,
cf.\ Eq.~(\ref{RA*}), the magnitude of the effect is determined
by the ratio $R_{T,\rm eff}/R_D$ which, according to Fig.~\ref{F:R/Runiv},
has a maximum at $t\approx G_T/\sqrt{g}$.
An example of such a nonmonotonous
dependence of $G_A(\Omega_*)$ is shown in Fig.~\ref{F:ga_t_int}.
The curves differ by the ratio $t=R_D/R_T$ and correspond to
$G_T=10\sqrt{g}$ (i.~e., $\lambda_g/a=0.2$).

In the opposite case of strong repulsion, $G_T\ll2\sqrt{g}$,
the $\Omega_*$ dependence of $G_A$ is absent for $t_*\gg G_T/\sqrt{g}$
when $R_{T,\rm eff}(t_*)$ is $t_*$ independent.
The reason is that strong repulsion makes Cooperons ineffective at
scales $\zeta\gg\zeta_1\equiv1/\lambda_g$ when all $\gamma_n$'s with odd
$n$ vanish, see Sec.~\ref{SSS:strong}.
Therefore, the reduction $u(x)\to \tilde u(x)$
at the scale $\zeta_*$ leaves the function $u(x)$ intact indicating that
physical results are $\Omega_*$ independent.
They become $\Omega_*$ dependent at relatively large scales when
$\zeta_* < \zeta_1$, i.~e., at $\ln(\omega_d/\Omega_*) < 2\pi\sqrt{g}$.
The principal effect of $\Omega_*$ is then to decrease $G_A$
with the increase of $\Omega_*$.

\begin{figure}
\refstepcounter{figure} \label{F:pss}
\epsfxsize=80mm
\centerline{\epsfbox{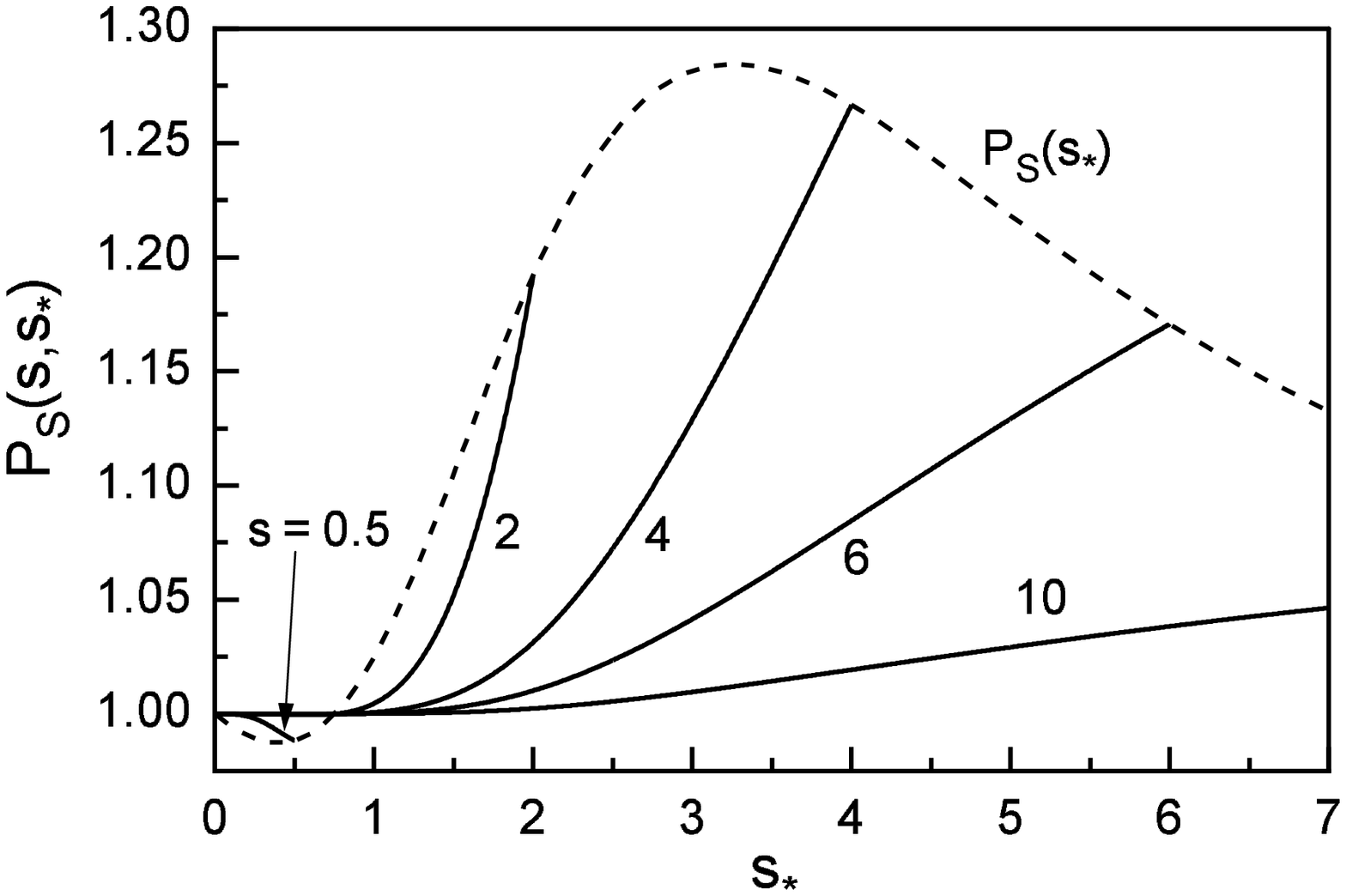}}
\small FIG.\ \arabic{figure}. Noise function $P_S(s,s_*)$ vs.\ $s_*$
for different values of $s$. The dashed line shows
$P_S(s_*,s_*)=P_S(s_*)$.
\end{figure}

Finally, we will dwell on the $\Omega_*$ dependence of the noise power.
According to Eq.~(\ref{IIs2}), the latter is described by the function
$P_S(t,t_*)$ which now depends on two RG ``time'' arguments.
Eq.~(\ref{PStt0}) relates $P_S(t,t_*)$ to the zero-energy expressions
for $G_A(t_*)$ and $P_S(t_*)$.

We start from the case of the weak interaction. The crossover from
the tunnel ($P_S=3$) to diffusive ($P_S=1$) character of noise at
$t,t_*\sim1$ is well described within the noninteracting approximation
investigated above. To study the noise function in the vicinity of
the interaction-induced peak in $P_S$ at $t,t_*\sim G_T/\sqrt{g} \gg1$
(cf.\ Fig.~\ref{F:lam_shot}) we substitute the functions $\Upsilon(s)$
and $P_S(s)$ shown in Figs.~\ref{F:R/Runiv} and \ref{F:lam_shot},
respectively, into Eq.~(\ref{PStt0}). The resulting dependence of
$P_S(s,s_*)$ as a function of $s_*$ is plotted in Fig.~\ref{F:pss}
for different values of $s$. We remind that $P_S(s,s_*)$ is defined
at $s_*<s$ and reduces to the zero-energy result at coincident ``times'':
$P_S(s_*,s_*)=P_S(s_*)$. The latter function is sketched in Fig.~\ref{F:pss}
by the dashed line.
Taking into account that
$s_* = \lambda_g\zeta_* = (1/2\pi\sqrt{g})\ln(\omega_d/\Omega_*)$,
one obtains from Fig.~\ref{F:pss} that $P_S$ decreases with the increase
of the Cooperon decoherence energy scale $\Omega_*$ (at $t,t_* \gg 1$),
as if the system becomes more diffusive.
This trend is opposite to what one has in the noninteracting case
when the increase of $\Omega_*$ drives the system toward the tunnel
limit, thus, increasing $P_S$.
Note also that, contrary to $P_S(s)$ (cf.\ Fig.~\ref{F:lam_shot}),
$P_S(s,s_*)$ shown in Fig.~\ref{F:pss}
is a monotonous function of $s_*$ at a fixed $s$.

In the limit of strong repulsion, $G_T\ll\sqrt{g}$, the zero-energy
noise function $P_S(t)$ exhibits a crossover from the tunnel to diffusive
regimes at $t\sim \sqrt{g}/G_T\gg1$.
Nevertheless, $P_S(t,t_*)$ remains $t_*$ independent upto much smaller
$t_*\sim G_T/\sqrt{g}\ll1$ corresponding to relatively large
energy scales, $\ln(\omega_d/\Omega_*) \simeq 2\pi\sqrt{g}$.
This effect is of the same origin as the above-mentioned $\Omega_*$
independence of $G_A$ in the limit of strong repulsion.

\section{Conclusions}
\label{SS:conclusion}

We have shown in this paper that electron transport through
mesoscopic N-I-S structures can be fully described
in terms of the Keldysh-space Proximity Action
$S_{\rm prox}[Q_S,Q_N]$, as defined in Eq.~(\ref{prox0}).
This action is a functional of two matrices, $Q_S$ and $Q_N$,
corresponding to the superconductive and normal leads.
Throughout the paper, we choose the gauge with the normal lead being in
equilibrium, with $Q_N=\Lambda$.  The superconductive terminal
is characterized by the matrix $Q_S$ which contains both the classical
($\varphi_1$) and quantum ($\varphi_2$) components
of the order parameter phase.
Physical response and correlation
functions of any order can be determined from the Proximity Action by
calculating the derivatives with respect to the quantum component
at a given value of the classical component of the phase,
as explained in the beginning of Sec.~\ref{S:phys}.
In principle, the Proximity Action functional contains information
about full charge transfer statistics (cf.\ Refs.~\cite{LeviLes,ALYa}).

The Proximity Action functional is known once the set of ``charges"
$\gamma_n$ or, equivalently, the periodic function $u(x)$,
Eq.~(\ref{u-def}), is specified.
It was explained in Sec.~\ref{SS:prox} that the function
$u(x)$ is directly related to the generating functional for the
transmission coefficients, introduced in Ref.~\cite{Nazarov94}.
Actually, the Proximity Action approach bears an obvious
analogy with the scattering matrix approach~\cite{Carlo,Nazarov94} as
both describe transport properties in terms of the characteristics
of the terminals (stationary-state Green functions of the terminals $Q_{S,N}$
in the former versus asymptotic scattering states in the latter approach).
The Proximity Action method is well-suited for the
treatment of interaction effects in the contact region, the task which
is out-of-reach for the standard scattering-matrix technique.

Actual calculation of the function $u(x)$ that determines the
Proximity Action is accomplished by the Functional Renormalization
Group (FRG) method.  In the simplest case (no interaction in the N region,
and all relevant energies are much below the Thouless energy scale
$\ETh = \hbar D/L^2$) the FRG equation reduces to the Euler
equation (\ref{urg}) for 1D motion of a compressible gas.
Although we derived this equation for the normal conductor of 2D geometry,
its solution, Eq.~(\ref{uu}), is applicable (being expressed in terms of
the ratio $t= G_T/G_D$ of the tunneling to diffusive conductances) to any
coherent conductor.
If higher energy scales $\Omega \geq \ETh$
or electron-electron interactions are involved the applicability
of our FRG procedure is limited to 2D diffusive conductors.
The generalized FRG equation that takes into account interaction constant
$\lambda$ in the Cooper
channel of the N conductor is given by Eq.~(\ref{urgl}).
We derive this equation and analyze its solution in Sec.~\ref{S:lambda}
analytically in two limiting cases of weak interaction of arbitrary sign
and of strong repulsion. We also provide the results of numerical
solution of Eq.~(\ref{urgl}) in the intermediate region of moderate
repulsion.
The main feature of the interaction
effect in 2D is that it leads to the infrared-growing corrections
of the order of $\lambda \ln(L/d)$, that should be summed up
non-perturbatively for the system of sufficiently large size $L$,
for any value of the ratio $G_T/G_D$.
This is in contrast with the 1D case (as treated in Ref.~\cite{StNaz}
for the case of the fully transparent interface, $G_T^{-1}=0$),
where the interaction correction to the resistance was found to be small,
of the order of $\lambda$ itself.

Physical consequences of Cooper repulsion in a 2D proximity system
depend on relation between $\lambda$ and the ratio $G_T/4\pi g$.
For clarity, we assumed that interaction constant had reached the
Finkelstein's fixed point $\lambda = 1/2\pi\sqrt{g}$.
Thus, strong repulsion corresponds to a relatively weak tunneling
conductance $G_T \ll 2\sqrt{g}$, and vice versa.
It is convenient to characterize N-I-S transport in terms
of an effective interface resistance $R_{T,\rm eff}$, so that
the total resistance is equal to $R_D + R_{T,\rm eff}$.
If $G_T \ll 2\sqrt{g}$, then
$R_{T,\rm eff} \approx (4\sqrt{g}/G_T) R_T \gg R_T$.
This asymptotic result is valid at $t \geq G_T/\sqrt{g}$.
At longer length scales, such that $\ln(L/d) \geq \pi \sqrt{g}$,
Cooper-channel correlations are irrelevant and the form of the
Proximity Action coincides with that of an N-I-N structure,
up to the replacements of charge quantum and the tunneling resistance:
$e \to 2e$ and $R_T \to R_{T,\rm eff}$.
As a particular example of this property,
in Sec.~\ref{SS:lambda-noise} we calculated the N-I-S noise intensity.
In the opposite case, $G_T \gg 2\sqrt{g}$, the evolution of
$R_{T,\rm eff}$ with the growth of $t$ is rather unexpected
(cf.\ Fig.~\ref{F:R/R2}):
first it decreases with $t$, reaches a minimum value of the order of $R_T$
at $t \sim (G_T/\sqrt{g})^{1/4}$, and then grows up to the asymptotic
value $0.6\,\hbar/e^2\sqrt{g}$ at $t \geq G_T/\sqrt{g}$.
Non-monotonous behavior with a maximum of relative height about 30$\%$
is demonstrated by the noise power as well, cf.\ Sec.~\ref{SS:lambda-noise}.

At finite temperature, voltage or magnetic field Cooperon coherence
is destroyed at the time of the order of $\hbar/\Omega_*$, where
$\Omega_* = \max(T, eV, eDH/c)$.  We calculated the Andreev conductance
and shot noise power in the presence of these phase-breaking effects
in Sec.~\ref{SS:beyond-zero} for a non-interacting normal conductor.
The role of phase-breaking effects in the presence of repulsion in
the normal conductor is discussed in Sec.~\ref{SS:be_int}.
In the case of highly transparent interface, $G_T \gg 2\sqrt{g}$,
and at $R_D \gg R_T$, an unusual ``finite-bias'' maximum
of the Andreev conductance $G_A(\Omega_*)$ is predicted,
cf.\ Fig.~\ref{F:ga_t_int}, which is of the consequence of
non-monotonous behavior of $R_{T,\rm eff}(t)$ mentioned above.
Contrary to the well-known~\cite{Art,StNaz} finite-bias anomaly
at $(T, eV) \sim \ETh$ in the non-interacting case, this new anomaly
is expected at much larger value of $\Omega_*/\ETh$.
Moreover, it should be seen as a function of magnetic field as well.
In simple terms the origin of this new effect can be understood as follows:
repulsion in the normal metal produces a superconductive
``gap function'' in the normal conductor, $\Delta_N$, with the negative
(compared to $\Delta_S$ in a superconductor) sign. Due to its opposite
sign, $\Delta_N $ decreases the conductance of the structure, therefore any
decoherence that reduces $\Delta_N $ leads to the increase of the conductance.

We did not consider in this paper the weak localization (WL) effect and
the Coulomb zero-bias anomaly (ZBA). The reason to neglect them, as explained
in the Introduction, is that relative magnitudes of these effects
scale as $g^{-1}|{\ln\Omega|}$. However, both of these effects have
high-frequency cutoff at the inverse elastic scattering time $1/\tau$,
and their contributions to the tunneling and diffusive conductances from
the energy scales $\omega_d < \Omega < 1/\tau$ are not necessarily small.
Since logarithmic corrections to the Proximity Action we have studied are
coming from much lower energy scales $\Omega \leq \omega_d$, these effects
can be considered separately: high-frequency fluctuations
can be accounted via the replacement of the bare conductances
$G_T$ and $g$ by their values corrected by the WL and ZBA effects
at the energy scales $\omega_d < \Omega < 1/\tau$,
provided that the renormalized value of $g$ is still large.

In this paper we consider the low-frequency case $\omega \ll \ETh$ only.
This limitation is due to technical complications: at higher frequencies
the effects of dynamic screening should be taken into account,
which makes the expression for the Proximity Action much more involved.
However, it is possible to show that the results obtained for the
Andreev conductance are still valid for frequencies up to much higher
frequency scale $\omega_{\rm max}$ [defined after Eq.~(\ref{GN})]
at which the electroneutrality of the system sets in.
Experimentally, N-S noise in the presence of high-frequency radiation
($\omega \gg \ETh$) was studied in Ref.~\cite{Kozhevnikov}.
We leave theoretical consideration of this problem for the future.

\acknowledgements

We are grateful to G.~B.~Lesovik and Yu.~V.~Nazarov for useful discussions.
This research was supported by the NSF grant DMR-9812340 (A.~I.~L.),
NWO-Russia collaboration grant, Swiss NSF-Russia collaboration
grant, RFBR grant 98-02-16252, and by the Russian Ministry of Science
within the project ``Mesoscopic electron systems for quantum computing"
(M.~V.~F. and M.~A.~S.).

\end{multicols}

\appendix

\section{Contraction Rules}
\label{A:A}

The contraction rule for averaging over $W$ is given by~\cite{FLS}
\be
  \corr{ \Tr AW \cdot \Tr BW } =
  \frac{1}{2\pi\nu} \int \frac{d\bq\, d\eps_1 d\eps_2}{(2\pi)^{4}}
  X(\bq,\eps_1,\eps_2) ,
\ee
where
\bea
  X =
    \frac{
      \tr \left[
        A_{12} \Lambda_2 B_{21}\Lambda_1
        + A_{12} (\Lambda_0)_2 B_{21} (\Lambda_0)_1
        - A_{12} \tau_z B_{21} \tau_z
        - A_{12} B_{21} \right]
        \left[ Dq^2 + i(\eps_1-\eps_2)(\Lambda_0)_1 \right]
    }{(Dq^2)^2+(\eps_1-\eps_2)^2}
\nonumber
\\ {}
  +
    \frac{
      \tr \left[
        A_{12} \Lambda_2 B_{21}\Lambda_1
        - A_{12} (\Lambda_0)_2 B_{21} (\Lambda_0)_1
        + A_{12} \tau_z B_{21} \tau_z
        - A_{12} B_{21} \right]
        \left[ Dq^2 + i(\eps_1+\eps_2)(\Lambda_0)_1 \right]
    }{(Dq^2)^2+(\eps_1+\eps_2)^2} .
\label{WW}
\eea
Here $A_{12}\equiv A(\bq,\eps_1,\eps_2)$, $B_{21}\equiv B(-\bq,\eps_2,\eps_1)$,
$\Lambda_m\equiv\Lambda(\eps_m)$, and $(\Lambda_0)_m\equiv\Lambda_0(\eps_m)$.
The first (second) line of Eq.~(\ref{WW}) corresponds to diffuson (Cooperon)
pairing.

In the zero-energy limit, $Dq^2\gg\eps_1,\eps_2$,
this expression can be simplified as
\be
  \corr{ \Tr AW \cdot \Tr BW } =
    \frac{2}{\pi g} \int \frac{d\bq\, d\eps_1 d\eps_2}{(2\pi)^{4}}
    \frac{ \tr ( A_{12}\Lambda_2 B_{21}\Lambda_1 - A_{12}B_{21} ) } {q^2} .
\label{WW1}
\ee

\bottom

\begin{multicols}{2}

\section{Contribution of a single term from the multicharge action
to the conductance and noise}
\label{A:B}

Here we calculate the quantity which describes the contribution
of a single term from the action (\ref{prox-s}) to the DC conductance
of the system in the zero-energy limit ($\Omega_*\ll\ETh$):
\be
  \frac{\delta}{\delta\varphi_2}
  \left.
    \Trk(e^{i\tensor\varphi} \Lambda_0 e^{-i\tensor\varphi} \Lambda_0)^n
  \right|_{\varphi_{2}=0} ,
\label{A:L0}
\ee
with $\varphi_1$ obeying the Josephson relation $\dot\varphi_1 = 2eV$.
The derivative can act either on $e^{i\tensor\varphi}$ or
on $e^{-i\tensor\varphi}$ so that (\ref{A:L0}) reduces to
\be
  in \int_{-\infty}^\infty \frac{d\eps}{2\pi}
  \trk \sigma_x [ M^n(V) - M^n(-V) ] ,
\label{A:L1}
\ee
where $M$ is diagonal in the energy space matrix:
\be
  M_{\eps\eps}(V) = \Lambda_0(\eps_+) \Lambda_0(\eps_-)
  =
  \left( \begin{array}{cc}
    1 & 2[F(\eps_-)-F(\eps_+)] \\ 0 & 1
  \end{array} \right)_K ,
\label{B:M}
\ee
and $\eps_\pm=\eps\pm eV$.
Calculating $M^n$, integrating over $\epsilon$ and tracing with
$\sigma_x$ one finds
that both terms in Eq.~(\ref{A:L1}) yield the same contributions, and
\be
  \frac{\delta}{\delta\varphi_2}
  \left.
    \Trk(e^{i\tensor\varphi} \Lambda_0 e^{-i\tensor\varphi} \Lambda_0)^n
  \right|_{\varphi_{2}=0}
  = \frac{8i}{\pi}\, n^2 eV .
\label{A:B:1s}
\ee

In a similar way one can obtain
\be
  \frac{\delta}{\delta\varphi_2}
  \left.
    \Trk(e^{i\tensor\varphi/2} \Lambda_0 e^{-i\tensor\varphi/2} \Lambda_0)^n
  \right|_{\varphi_{2}=0}
  = \frac{2i}{\pi}\, n^2 eV .
\label{A:B:1n}
\ee

Now we turn to another expression emerging in calculation of the
current-current correlator from the action (\ref{sg}):
\be
  L_n(t,t';V) =
  \frac{\delta^2}{\delta\varphi_2(t)\, \delta\varphi_2(t')}
  \left.
  \Trk(e^{i\tensor\varphi} \Lambda_0 e^{-i\tensor\varphi} \Lambda_0)^n
  \right|_{\varphi_{2}=0} .
\label{A:B1}
\ee

Here there are several possibilities depending on where
the derivatives act on. First of all we note that they cannot
act on the same exponent $e^{i\tensor\varphi}$ (or $e^{-i\tensor\varphi}$).
Indeed, in this case Eq.~(\ref{A:B1}) would reduce to
$\Tr M^n(V) \propto \Tr\openone$ that would give zero
according to the rules of the Keldysh $\sigma$-model~\cite{KA}.
Therefore there are four different contributions (coinciding time indices
$t,t'$ are omitted):
\be
  L_n(V) = L_n^{a}(V) + L_n^{a}(-V) + L_n^{b}(V) + L_n^{b}(-V) ,
\label{A:B7}
\ee
where
\end{multicols}
\top
\bea
  L_n^{a}(t,t';V) &=&
  n \sum_{p,q\geq0 \atop p+q=n-1}
    \Trk e^{2ieVt}\, \Lambda_0\, M^p(-V)\, \sigma_x\,
        e^{-2ieVt'}\, \Lambda_0\, M^q(V)\, \sigma_x
\nonumber \\
  L_n^{b}(t,t';V) &=&
  - n \sum_{r,s\geq0 \atop r+s=n-2}
    \Trk e^{2ieVt}\, \Lambda_0\, e^{-i\varphi_1}\,
      \Lambda_0\, M^r(V)\, \sigma_x\,
    e^{2ieVt'}\, \Lambda_0\, e^{-i\varphi_1}\,
      \Lambda_0\, M^s(V)\, \sigma_x
\eea
Performing Fourier transformation to the frequency domain one obtains
\bea
  L_n^{a}(\omega;V) &=&
  n \sum_{p,q\geq0 \atop p+q=n-1}
    \int_{-\infty}^\infty \frac{d\eps}{2\pi}
    \trk \, ( \Lambda_0\, M^p(-V)\, \sigma_x )_{\eps+\omega+2eV}
        \, ( \Lambda_0\, M^q(V)\, \sigma_x )_{\eps} ,
\nonumber \\
  L_n^{b}(\omega;V) &=&
  - n \sum_{r,s\geq0 \atop r+s=n-2}
    \int_{-\infty}^\infty \frac{d\eps}{2\pi}
    \trk \, \Lambda_0(\eps+\omega+2eV)
        \, (\Lambda_0\, M^r(V)\, \sigma_x )_{\eps+\omega}
        \, \Lambda_0(\eps+2eV) \, \Lambda_0\, M^s(V)\, \sigma_x )_{\eps} .
\eea
Evaluating matrix exponents and computing traces we obtain
\bea
  L_n^{a}(\omega;V) &=&
  4 n \sum_{p,q\geq0 \atop p+q=n-1}
  \int_{-\infty}^\infty
  \left[
    (p+1) F(\eps+\omega+2eV) - p F(\eps+\omega)
  \right]
  \left[
    (q+1) F(\eps) - q F(\eps+2eV)
  \right]
  \frac{d\eps}{2\pi} ,
\nonumber \\
  L_n^{b}(\omega;V) &=&
  - 4 n \sum_{r,s\geq0 \atop r+s=n-2}
  (r+1) (s+1)
  \int_{-\infty}^\infty
  \left[
     F(\eps+\omega) - F(\eps+\omega+2eV)
  \right]
  \left[
    F(\eps) - F(\eps+2eV)
  \right]
  \frac{d\eps}{2\pi} .
\eea
Now integrating over energy with the help of Eq.~(\ref{psi}), we arrive at
\bea
  L_n^{a}(\omega;V) &=&
  \frac{4n}\pi \sum_{p,q\geq0 \atop p+q=n-1}
  \bigl\{
    2(p+1)q \Psi(\omega) - pq \Psi(\omega-2eV) - (p+1)(q+1) \Psi(\omega+2eV)
  \bigr\} ,
\nonumber \\
  L_n^{b}(\omega;V) &=&
  \frac{4n}\pi \sum_{r,s\geq0 \atop r+s=n-2}
  (r+1) (s+1)
  \left[
    2\Psi(\omega) - \Psi(\omega-2eV) - \Psi(\omega+2eV)
  \right] .
\label{A:B11}
\eea
Evaluating the sums with the help of
\be
  \sum_{p,q\geq0 \atop p+q=n-1} pq = {n\choose3}, \quad
  \sum_{p,q\geq0 \atop p+q=n-1} (p+1)q = \!\!\!
  \sum_{r,s\geq0 \atop r+s=n-2} (r+1)(s+1) = {n+1\choose3}, \quad
  \sum_{p,q\geq0 \atop p+q=n-1} (p+1)(q+1) = {n+2\choose3},
\ee
we get
\be
  L_n^{a}(\omega;V) + L_n^{b}(\omega;V) =
  \frac{2n^2}{3\pi}
  \Bigl\{
    4(n^2-1) \Psi(\omega)
    - (n-1)(2n-1) \Psi(\omega-2eV) - (n+1)(2n+1) \Psi(\omega+2eV)
  \Bigr\} .
\label{A:B13}
\ee

Finally, substituting Eq.~(\ref{A:B13}) into Eq.~(\ref{A:B7}), we obtain
\be
  \frac{\delta^2}{\delta\varphi_2(\omega)\, \delta\varphi_2(-\omega)}
  \left.
  \Trk(e^{i\tensor\varphi} \Lambda_0 e^{-i\tensor\varphi} \Lambda_0)^n
  \right|_{\varphi_{2}=0}
  = - \frac{4n^2}{3\pi}
  \Bigl\{
    (2n^2+1) \bigl[ \Psi(\omega-2eV) + \Psi(\omega+2eV) \bigr]
    - 4(n^2-1) \Psi(\omega)
  \Bigr\} .
\label{A:B:2s}
\ee
Analogously,
\be
  \frac{\delta^2}{\delta\varphi_2(\omega)\, \delta\varphi_2(-\omega)}
  \left.
  \Trk(e^{i\tensor\varphi/2} \Lambda_0 e^{-i\tensor\varphi/2} \Lambda_0)^n
  \right|_{\varphi_{2}=0}
  = - \frac{n^2}{3\pi}
  \Bigl\{
    (2n^2+1) \bigl[ \Psi(\omega-eV) + \Psi(\omega+eV) \bigr]
    - 4(n^2-1) \Psi(\omega)
  \Bigr\} .
\label{A:B:2n}
\ee

\bottom
\begin{multicols}{2}

\section{Solution of the Usadel equation in the first order over $\lambda$}
\label{A:C}

We will generalize here the method used in Ref.~\cite{StNaz}
for the calculation of $G_A$ in the presence of interaction.
Since we consider general case of an arbitrary interface transparency,
the first step will be to find the interaction-induced correction
to the ``spectral angle" $\theta_E(r)$
that parametrizes~\cite{NazarovC,StNaz} the retarded semiclassical Green
function in the N conductor,
\be
\label{thetaR}
  \hat{G}^R(E) =
    \left( \begin{array}{cc}
      \cos\theta & ie^{i\varphi}\sin\theta \\
      -ie^{-i\varphi}\sin\theta & -\cos\theta
  \end{array} \right) ,
\ee
where the proximity angle $\theta$ and the order parameter phase $\varphi$
depend on the energy $E$ and the space coordinate ${\bf r}$.
In N-S systems with a single superconductive terminal physical quantities
do not depend on the phase on it, $\varphi(0)$, so
below we put $\varphi(0) = -\pi/2$.  In general, when interaction in the
N  conductor is present, normal current flowing through the N-S structure
induces a supercurrent ${\bf j}_s \propto \nabla\varphi$,
so that the phase $\varphi({\bf r})$ acquires some nontrivial distribution.
However, this effect appears in the second order of expansion over
Cooper interaction constant $\lambda_d$. Below we solve the Usadel equation
within the first order over $\lambda_d$, neglecting, therefore,
the effects of supercurrent.

\subsection{Spectral angle}

The Usadel equation for the spectral angle $\theta_E(r)$ has the form
\be
  D\nabla^2\theta_E + 2iE \sin\theta_E + 2\Delta\cos\theta_E = 0 ,
\label{usadel}
\ee
where the selfconsistency equation for the order parameter reads
\be
  \Delta(r) = - \lambda_d \int_0^{\infty} dE
    \tanh\frac{E}{2T} \mathop{\rm Im} \sin\theta_E ,
\label{delta}
\ee
and the interaction constant $\lambda_d$ at the energy scale $\omega_d$
is defined in Eq.~(\ref{lambda_d}).
We assume that the main contribution to $\Delta$ comes from relatively
high $E$, where $\theta_E (r)$ is small everywhere.
Hence, Eq.~(\ref{usadel}) can be linearized,
and the last term can be neglected once we are interested in
the first-order corrections over $\lambda_d$ to $G_A$.
The result is:
\be
\theta_E(r) = A(E) K_0((1-i)r/L_E) ,
\label{2}
\ee
where $L_E = \sqrt{D/E}$, and $A$ is determined by the boundary condition
\be
  g\left. \frac{d\theta_E(r)}{dr}\right|_{r=d} =
  - \frac{G_T}{2\pi d}\cos\theta_E(r=d) .
\label{boundary}
\ee
The solution for $A(E)$ is conveniently expressed in terms of the
function $\Theta(t)$ defined in Eq.~(\ref{Theta-def}):
$A(E) = \Theta(t_E)/\ln(L_E/d)$, where $t_E = a\ln(\omega_d/E)$.
Substituting $\theta_E(r)$ into Eq.~(\ref{delta}),
using the identity $\int_0^\infty xK_0(x) dx = 1$
and logarithmic slowness of $A(t_E)$ as a
function of $E$, we find for $\Delta(r)$:
\be
  \Delta(r)
  = - \frac{\lambda_d D}{r^2} \frac{\Theta(2a\ln(r/d))}{\ln(r/d)} .
\label{Delta}
\ee
The next step is to solve for $\theta_{E=0}(r)$ with $\Delta(r)$ taken
into account:
\be
  \frac{d^2\theta}{dr^2} + \frac1r\frac{d\theta}{dr} =
  \frac{2\lambda_d}{r^2} \frac{\Theta(2a\ln(r/d))}{\ln(r/d)} \cos\theta .
\label{theta0}
\ee
In terms of the variable $\xi=\ln(r/d)$, this equation reduces to
\be
  \frac{d^2\theta}{d\xi^2}
  = 2\lambda_d \frac{\Theta(2a\xi)}{\xi} \cos\theta ,
\label{theta1}
\ee
with the boundary conditions $\theta(\xi_L)=0$
and $\theta_\xi(0) = -2a \cos\theta(0)$, where $\xi_L=\ln (L/d)$.
To the lowest order in $\lambda_d$ one obtains
\end{multicols}
\top
\bea
  \theta(\xi)
  = \left\{
    \Theta(2a\xi_L) -
    2 \lambda_d \xi_L \int_0^{\xi_L}
    \frac{1+2a\eta \sin\Theta(2a\xi_L)}{1+2a\xi_L \sin\Theta(2a\xi_L)}
    \frac{\Theta(2a\eta)}{\eta}
    \cos \left[ \Theta(2a\xi_L) \left( 1 - \frac{\eta}{\xi_L} \right) \right]
    d\eta
  \right\}
  \left( 1 - \frac{\xi}{\xi_L} \right)   \nonumber
\\ {}
  + 2\lambda_d \int_\xi^{\xi_L} (\eta-\xi) \frac{\Theta(2a\eta)}{\eta}
    \cos \left[ \Theta(2a\xi_L) \left( 1 - \frac{\eta}{\xi_L} \right) \right]
    d\eta .
\label{thetacorr}
\eea
\bottom
\begin{multicols}{2}
Taking into account that $2a\xi_L = a\zeta\equiv t$, and evaluating
expression (\ref{thetacorr}) at $\xi =0$, we obtain
the spectral angle near the interface, $\theta_d \equiv \theta(\xi=0)$:
\bea
  \theta_d
  &=& \Theta(t)
  - \frac{\lambda_d\zeta}{1+t\sin\Theta(t)}
\nonumber \\
  {} && \times
  \int_0^1 \frac{\Theta(xt)}{x} (1-x) \cos \left[ \Theta(t) (1-x) \right] dx .
\label{theta_d}
\eea

\subsection{Distribution function and $G_A$}

To find $G_A$ we use the kinetic equation in the form developed
in Ref.~\cite{StNaz}.
The current density near the N reservoir (at $r=L$) is:
\be
  j_N(R) = g \int dE \left.\frac{df_1(E, r)}{dr}\right|_{r=L},
\label{j}
\ee
where $f_1(E,r)$ is the anomalous (branch imbalance) distribution function,
defined after Eq.~(\ref{Lambda0}).
Note that the Eq.~(\ref{j}) is valid only at the boundary
with the normal lead where $\theta=0$ (otherwise it should be modified
by the presence of the supercurrent ${\bf j}_s \propto \lambda_d$).
The function $f_1(E,r)$ obeys the equation
\be
  \nabla D(r,E)\nabla f_1(E,r) = 2\Delta(r)\sin\theta(E,r) f_1(E,r) .
\label{f1}
\ee
We will assume that the normal reservoir is biased by the voltage $V$,
so that the whole distribution function is shifted:
$F(E,L) = \tanh(E+eV\tau_z)/2T \approx \tanh(E/2T) +
\tau_z (eV/2T) \cosh^{-2}(E/2T)$
and
\be
  f_1(E,L) = eV c(E) = \frac{eV}{2T}\cosh^{-2}\frac{E}{2T}.
\label{ff}
\ee
At $T\to 0$ it is enough to consider Eq.~(\ref{f1}) at zero energy,
where $D(0,r) = D$, and the solution for $\theta(0,r)$ is given by
\be
  \theta(0,r) = \theta_d \frac{\ln(L/r)}{\ln(L/d)} ,
\label{thetaf}
\ee
with $\theta_d$ given by Eq.~(\ref{theta_d}).
Then equation for $f_1(r)$ reduces to
\be
  \frac{d^2f_1}{d\xi^2} =
  - 2\lambda_d \frac{\Theta(2a\xi)}{\xi}
    \sin\left[\theta_d\left(1-\frac{\xi}{\xi_L}\right)\right] f_1 .
\label{f2}
\ee
We seek the solution of Eq.~(\ref{f2}) in the form
$f_1(\xi) = f_1^{(0)}(\xi) + \lambda_d f_1^{(1)}(\xi)$, where
\be
  f_1^{(0)}(\xi) = eV c(E) \left[\frac{\xi}{\xi_L} +
v_d\left(1-\frac{\xi}{\xi_L}\right)\right] .
\label{f3}
\ee
The function $c(E)$ was defined in Eq.~(\ref{ff}),
and the parameter $v_d = V_d/V$
should be determined from the boundary condition
to the Usadel equation:
\be
  \left. \frac{\partial f_1}{\partial\xi} \right|_{\xi=0}
  = 2a f_1 \sin\theta_d ,
\label{b.c.}
\ee
that gives
\be
  v_d = \frac{1}{1+t\sin\theta_d} .
\ee

The general solution of Eq.~(\ref{f2}) for the function $f_1^{(1)}$
can be written as
\be
  f_1^{(1)}(\xi) = \int_0^\xi (\xi-\eta) y(\eta) d\eta + C_1\xi + C_0 ,
\ee
where
$y(\xi) = - 2\lambda_d (\Theta(a\xi)/\xi)
\sin[\Theta(t) (1-\xi/\xi_L) ] f_1^{(0)}(\xi)$.
Using Eq.~(\ref{b.c.}) and the condition $f_1^{(1)}(\xi_L) = 0$,
we find the constants $C_0$ and $C_1$, and obtain
\be
  \left. \frac{\partial f_1^{(1)}}{\partial\xi} \right|_{\xi_L}
  =
  \frac{1}{\xi_L}
    \int_0^{\xi_L} \frac{\xi_L+\eta t\sin\Theta(t)}{1+t\sin\Theta(t)}
      y(\eta) d\eta .
\ee

Calculating the total current as $I_N = 2\pi Lj_N(L)$
where $j_N(L)$ is determined by Eq.~(\ref{j}) and
introducing $x=\eta/\xi_L$, we obtain
\end{multicols}
\top
\be
  I_N = \frac{2\pi g}{\xi_L} V
    \left[
      \frac{t\sin\theta_d}{1+t\sin\theta_d} \nonumber \\
      - 4\lambda_d a \xi_L^2
        \int_0^1 \frac{(1+xt\sin\Theta(t))^2}{(1+t\sin\Theta(t))^2}
          \frac{\Theta(xt)}{xt} \sin [\Theta(t) (1-x)] dx
    \right] .
\label{I_N}
\ee
Finally, the Andreev conductance is given by
\be
  \frac{G_A}{G_D} =
    \frac{t\sin\theta_d}{1+t\sin\theta_d}
    - \lambda_d \zeta
        \int_0^1 \frac{(1+xt\sin\Theta(t))^2}{(1+t\sin\Theta(t))^2}
          \frac{\Theta(xt)}{x} \sin [\Theta(t) (1-x)] dx ,
\label{ga/gn}
\ee
\bottom
\begin{multicols}{2}
\noindent
with $\theta_d$ defined in Eq.~(\ref{theta_d}).

Eq.~(\ref{ga/gn}) is an exact answer in the first order over
$\lambda_d$ (in this approximation supercurrent does not yet mix
equations for $f_1$ and $f$).
In the $t\to\infty$ limit of Eq.~(\ref{ga/gn}), i.~e., at $G_T \gg G_D$,
the interaction-induced correction is
\be
  \frac{G_A}{G_D} =
    1 - 2\frac{\pi-2}{\pi} \lambda_d \ln\frac{L}{d}
\label{Climit}
\ee
and grows with the space scale, contrary to the results of Ref.~\cite{StNaz}
obtained for the 1D geometry.


\end{multicols}


\begin{references}
\bibitem{why?}
Performing averaging over the shadowed diffuson and Cooperons with the help
of Eq.~(\ref{WW}), one obtains an expression similar to Eq.~(\ref{S_nl}).
However, now the trace contains the difference $\ac_3-\ac_1$,
cf.\ Eq.~(\ref{ac}), which vanishes according to the theorem proved
after Eq.~(\ref{ac}).
\end{references}
\end{document}